\documentclass[12pt]{article}
\pdfoutput=1
\usepackage{jheppub} 
\usepackage{epsfig}
\usepackage{amsmath}
\usepackage{amssymb}
\usepackage{amsfonts}
\usepackage{amsxtra}
\usepackage{amsthm}
\usepackage{mathrsfs}
\usepackage{makeidx}
\usepackage{graphics}
\usepackage{dsfont}
\usepackage{mathtools}
\usepackage{graphicx}
\usepackage{subcaption}
\usepackage{placeins}
\usepackage{bm}
\usepackage[capitalise]{cleveref}
\usepackage{empheq}
\usepackage{colortbl}
\usepackage[dvipsnames]{xcolor}
\usepackage{enumerate}
\usepackage{titlesec}
\usepackage{longtable}
\usepackage{float}
\usepackage{color}
\usepackage{tikz}
\usepackage{xfrac}
\usepackage{footnote}
\usepackage{rotating}
\usepackage{lscape}
\usepackage{makecell}
\usepackage{environ}
\usepackage{tabularx}
\usepackage{subfiles}
\usepackage[export]{adjustbox}
\usepackage{ytableau}
\usepackage{tikz-3dplot}
\usepackage{slashed}
\usepackage{pifont}
\usepackage{multirow}
\usepackage{mdframed}
\usepackage{bbm}
\usepackage{ulem}
\usepackage{comment}
\usepackage{placeins}

\titleformat{\paragraph}
{\normalfont\normalsize\bfseries}{\theparagraph}{1em}{}
\titlespacing*{\paragraph}{0pt}{3.25ex plus 1ex minus .2ex}{1.5ex plus .2ex}

\usepackage{tikz}
\usetikzlibrary{positioning}
\usetikzlibrary{calc}

\usetikzlibrary{decorations.pathreplacing,calligraphy}
\usepackage{xstring}
\usetikzlibrary{decorations.pathmorphing} 
\usetikzlibrary{decorations.markings} 
\usetikzlibrary{arrows} 
\usetikzlibrary{shapes} 
\usetikzlibrary{matrix} 
\usetikzlibrary{positioning} 
\usepackage[english]{babel} 
\usepackage[autostyle]{csquotes}

\tikzstyle{brane}=[draw]
\tikzset{roundnode/.style={circle, draw=black, fill=white, inner sep=2.5pt}}
\tikzset{gauge/.style={circle, draw, line width=.2mm, inner sep=2pt}}
\tikzset{gauger/.style={circle, draw,line width=.2mm,fill=red,inner sep=2pt}}
\tikzset{gline/.style={thick}}

\tikzset{bd/.style={circle, draw=black, inner sep=0pt, fill=black, minimum size=1.2mm}}
\tikzset{bld/.style={circle, draw=blue, inner sep=0pt, fill=blue, minimum size=1.2mm}}
\tikzset{wd/.style={circle, draw=black, inner sep=0pt, fill=white, minimum size=1.2mm}}
\tikzset{rd/.style={circle, draw=red, inner sep=0pt, fill=red, minimum size=.9mm}}
\tikzset{wrd/.style={circle, draw=red, inner sep=0pt, fill=white, minimum size=.9mm}}
\tikzstyle{ligne}=[draw, thick] 
\tikzset{doublearrow/.style={ draw=black!75, color=black!75, double distance=3pt, thick}} 
\tikzset{snake it/.style={decorate, decoration={snake, amplitude=.4mm, segment length=2mm,
                       post length=0mm,pre length=0mm}}}

\newcommand{\be}{\begin{equation}}
\newcommand{\ee}{\end{equation}}
\newcommand{\ba}{\begin{aligned}}
\newcommand{\ea}{\end{aligned}}

\newcommand{\cN}{\mathcal{N}}
\newcommand{\Z}{\mathbb{Z}}
\newcommand{\su}{\mathfrak{su}}
\newcommand{\uu}{\mathfrak{u}}

\def\Tr{\mathop{\mathrm{Tr}}\nolimits}
\def\half{{\frac{1}{2}}}

\newcommand{\mvb}[1]{{\bf \color{violet} MvB: #1}}

\title{Connecting 5d Higgs Branches via Fayet-Iliopoulos Deformations}
\author[1]{Marieke van Beest,}
\author[1,2]{~Simone Giacomelli,}
\affiliation[1]{Mathematical Institute, University of Oxford, Woodstock Road, Oxford, \\ OX2 6GG, United Kingdom} 
\affiliation[2]{Dipartimento di Fisica, Universit\`a di Milano-Bicocca, Piazza della Scienza 3, \\ I-20126 Milano, Italy}
\emailAdd{vanbeest@maths.ox.ac.uk}
\emailAdd{simone.giacomelli@unimib.it} 
\abstract{
We describe how the geometry of the Higgs branch of 5d superconformal field theories is transformed under movement along the extended Coulomb branch. Working directly with the (unitary) magnetic quiver, we demonstrate a correspondence between Fayet-Iliopoulos deformations in 3d and 5d mass deformations. 
When the Higgs branch has multiple cones, characterised by a collection of magnetic quivers, the mirror map is not globally well-defined, however we are able to utilize the correspondence to establish a local version of mirror symmetry.
We give several detailed examples of deformations, including decouplings and weak-coupling limits, in $(D_n,D_n)$ conformal matter theories, $T_N$ theory and its parent $P_N$, for which we find new Lagrangian descriptions given by quiver gauge theories with fundamental and anti-symmetric matter.
}

\begin{document}
\maketitle

\section{Introduction}

The intrinsically strongly-coupled nature of 5d $\cN=1$ SCFTs \cite{Seiberg:1996bd,Morrison:1996xf,Intriligator:1997pq} makes the study of them a challenging problem in quantum field theory. New insights into this class of theories has been made possible by the application of methods from string theory, where the 5d SCFTs arise in type IIB on $(p,q)$-5-brane-webs \cite{Hanany:1996ie,Aharony:1997ju,DeWolfe:1999hj,Bergman:2013aca,Zafrir:2014ywa,Hayashi:2015zka,Hayashi:2015fsa,Bergman:2015dpa,Zafrir:2015ftn,Hayashi:2018lyv,Hayashi:2018bkd,Hayashi:2019yxj,Bergman:2020myx,Martone:2021drm,Bertolini:2021cew}, or in M-theory on (non-compact) Calabi-Yau 3-fold singularities \cite{Diaconescu:1998cn,Hayashi:2013lra,Hayashi:2014kca,DelZotto:2017pti,Xie:2017pfl,Jefferson:2017ahm,Jefferson:2018irk,Bhardwaj:2018yhy,Bhardwaj:2018vuu,Apruzzi:2018nre,Closset:2018bjz,Apruzzi:2019vpe,Apruzzi:2019opn,Apruzzi:2019enx,Bhardwaj:2019jtr,Bhardwaj:2019fzv,Bhardwaj:2019ngx,Saxena:2019wuy,Apruzzi:2019kgb,Closset:2019juk,Bhardwaj:2019xeg,Bhardwaj:2020gyu,Eckhard:2020jyr,Bhardwaj:2020kim,Hubner:2020uvb,Bhardwaj:2020ruf,Bhardwaj:2020avz,Bhardwaj:2020phs,Closset:2020afy,Braun:2021lzt,Apruzzi:2021vcu,Closset:2021lhd,Collinucci:2021wty,Collinucci:2021ofd}. The two descriptions are, in principle, related by duality, which is furnished by a one-to-one map between the brane-web and a Generalized Toric Polygon (GTP), also known as a dot diagram \cite{Aharony:1997bh,Benini:2009gi,Bao:2011rc,Taki:2014pba,vanBeest:2020kou,vanBeest:2020civ}. In the toric case, the latter is precisely associated to the Calabi-Yau singularity in M-theory, whereas a geometric interpretation in the general case remains an interesting open problem, on which some headway has been made in \cite{vanBeest:2020kou,vanBeest:2020civ,Closset:2020scj}.

The 5d $\cN=1$ SCFTs have a rich moduli space composed of the Higgs branch (HB) and Coulomb branch (CB), which are respectively parametrized by the vacuum expectation values (vevs) of the scalars in the hyper- and vector-multiplets. In M-theory, the (extended) Coulomb branch is mapped out by resolutions of the Calabi-Yau 3-fold singularity, and the Higgs branch is realized as its deformation space. In the brane-web, the moduli space is described by continuous deformations of the branes, with motion in the plane of the web corresponding to movement on the Coulomb branch, whereas motion in the transverse directions maps out the Higgs branch. The Coulomb branch is well-understood from both an M-theory and brane-web perspective. The Higgs branch, on the other hand, is particularly difficult to characterize, because it receives quantum corrections from instantons in 5d.
Recently, there has been progress in this endeavor, which is due to the description in terms of magnetic quivers (MQ) and Hasse diagrams \cite{Bullimore:2015lsa,Nakajima:2015txa, Braverman:2016pwk, Ferlito:2016grh,Ferlito:2017xdq,Cabrera:2018ann,Cabrera:2018jxt,Cabrera:2019izd,Bourget:2019aer,Bourget:2019rtl,Cabrera:2019dob,Grimminger:2020dmg,Bourget:2020gzi,Bourget:2020asf,Akhond:2020vhc,Bourget:2020mez,Dancer:2020wll,Bourget:2020bxh,Bourget:2020xdz,Akhond:2021knl,Bourget:2021siw,Bourget:2021zyc,Akhond:2021ffo,Bao:2021ohf,Gledhill:2021cbe}. The magnetic quiver represents a 3d $\cN=4$ quiver gauge theory, whose Coulomb branch is identified with the Higgs branch of the 5d theory $\mathcal{T}_{5d}$, 
\be 
\text{CB}[\text{MQ}]=\text{HB}[\mathcal{T}_{5d}]\,,
\ee 
which solves the problem of accounting for instantons, since the Coulomb branch of the magnetic quiver can be examined in detail by computing its Hilbert series \cite{Cremonesi:2015lsa,Hanany:2014dia,Bullimore:2015lsa}.

The focus of this paper is on 5d SCFTs originating, via dimensional reduction, from 6d world-volume theories of M5-branes probing an M9 wall wrapped on an A-type singularity. We concentrate on these models since their Higgs branch can always be described by unitary magnetic quivers (see \cite{Mekareeya:2017jgc}). By successive decoupling of matter in the 5d KK-theories, we can flow to other SCFT fixed points, thus generating a tree of descendant SCFTs. 
The principal motivation for this work is to understand how the geometry of the 5d Higgs branch is transformed, as we move around the extended Coulomb branch. The most literal way to answer this question is to work directly with the magnetic quiver, and understand, for instance, how to decouple matter or take a weak coupling limit explicitly in the magnetic quiver.
We show that this is achieved by Fayet-Iliopoulos (FI) deformations, and argue that, for any extended Coulomb branch deformation of the 5d SCFT, implemented as partial resolutions of the GTP \cite{Apruzzi:2019opn,Eckhard:2020jyr,vanBeest:2020civ}, there is an FI deformation that enacts the corresponding change in the magnetic quiver, thus extrapolating between the Higgs branches emanating from the two points in the extended Coulomb branch. 

Remarkably, although we always start from a star-shaped magnetic quiver describing the Higgs branch of the 6d SCFT \cite{Mekareeya:2017jgc}, FI deformations allow us to generate all the ingredients observed in magnetic quivers of 5d theories, namely edges with nontrivial multiplicity, quivers with loops, adjoint matter and free hypermultiplets. As is well known, many SCFTs are described by a collection of quivers since their Higgs branch is made of multiple components. For such models the mirror map is not well-defined, but we find in several nontrivial cases that each quiver describes the mirror dual of the low-energy effective theory at a singular point in the Coulomb branch of the SCFT dimensionally reduced to 3d. We therefore have in this cases an interpretation in terms of a local mirror symmetry, in which different singular points in the Coulomb branch lead to different quivers.   

In section \ref{sec:FIdef} we explain how to turn on FI parameters in a generic unitary quiver and propagate the non-zero vevs in a consistent (but highly non-unique) way. 
The strategy we implement in section \ref{sec:massdef} for matching 5d mass deformations and FI deformations in the magnetic quiver makes use of the GTP realization of the 5d SCFT. A partial resolution of a GTP $P$ moves us onto the extended Coulomb branch, described by a new polygon $P'$. In any such phase of the 5d theory, we can use the map in \cite{vanBeest:2020kou} to determine the associated magnetic quiver, and identify the FI deformation that connects the two points. The method can be represented diagrammatically as
\be 
\begin{tikzpicture}
    \node at (0,0) {$P$};
    \node at (4,0) {$P'$};
    \node at (0,-2) {$MQ_P$};
    \node at (4,-2) {$MQ_{P'}$};
    
    \draw[->] (1,0)--(3,0) node[midway, above] {\scriptsize resolution};
    \draw[->] (0,-.5)--(0,-1.5); 
    \draw[->] (1,-2)--(3,-2) node[midway, above] {\scriptsize FI deformation};
    \draw[->] (4,-.5)--(4,-1.5); 
\end{tikzpicture}
\ee 
We discuss the concept of local mirror symmetry in the context of SQCD with special unitary gauge group in Section \ref{Seclocmirr3}.
Section \ref{sec:DnDn} illustrates the method in the case of $(D_{N+2},D_{N+2})$ conformal matter, and in section \ref{sec:TN} we analyze the $T_N$ theory. 
Moreover, in many cases, one can use symmetry considerations to constrain the possible FI deformations, allowing us to bypass the construction of $P'$. The method then provides a useful tool for obtaining new quivers, e.g. of descendant SCFTs, or finding new weakly coupled 5d gauge theory descriptions, as we do in section \ref{sec:newlag}.

\section{FI Deformations}
\label{sec:FIdef}

In this paper we are concerned with 5d SCFTs originating from orbi-instanton theories of type A, i.e. the 6d theories on the world-volume of M5-branes probing an M9 wall which wraps a $\mathbb{C}^2/\mathbb{Z}_k$ singularity (transverse to the M5 world-volume) in M-theory. The Higgs branch of all these theories is described by a unitary magnetic quiver, or a collection of them whenever the moduli space of the theory exhibits multiple branches. Upon dimensional reduction of the 6d theory we get a 5d SCFT whose Higgs branch is a single cone described by a star-shaped unitary quiver, which can be identified with the mirror dual of the theory dimensionally reduced to 3d, called $\mathbb{EQ}_{3d}$. By activating mass deformations of the 5d SCFT we can move along the decoupling tree of the theory, and, for the first few decouplings, we can exploit the mirror map dictionary. In particular, we are guaranteed that turning on a mass deformation in the 5d SCFT (which may correspond to a decoupling or to turning on a gauge coupling) results in a mass deformation of the 3d theory, which in turn corresponds to a Fayet-Iliopoulos deformation in the mirror dual quiver.
The duality between the magnetic quiver and $\mathbb{EQ}_{3d}$ breaks down typically as soon as we reach theories whose Higgs branch consists of multiple cones. 
Remarkably, we find that the correspondence between mass deformations of the 5d SCFT and FI deformations of the magnetic quiver remains true even when the magnetic quiver cannot be identified with the mirror dual of $\mathbb{EQ}_{3d}$.

\subsection{FI Deformations of Unitary Quivers}

We would now like to understand FI deformations in a unitary quiver. Let us start by reviewing how FI deformations affect the equations of motion at a given gauge node in the quiver. We turn on a (say complex) FI parameter $\lambda$ in a $U(N)$ gauge theory with $k$ fundamental hypermultiplets $\widetilde{P}_f$ and $P_f$ (in four supercharges notation) and we denote with $\Psi$ the adjoint chiral in the $\mathcal{N}=4$ vector multiplet.
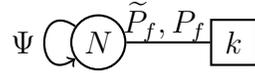
\begin{center}
\begin{tikzpicture}[->,thick, scale=0.4]
\node[](L3) at (-5,0) {$\Psi$};
\node[] (L2) at (-0.3,0.7) {$\widetilde{P}_f,P_f$};
\node[circle, draw, inner sep=2.5](L4) at (-2.5,0){$N$};
\node[rectangle, draw, inner sep=1.7,minimum height=.6cm,minimum width=.6cm](L5) at (2,0){$k$};

  \path[every node/.style={font=\sffamily\small,
  		fill=white,inner sep=1pt}]
(L4) edge [loop, out=145, in=215, looseness=4] (L4);
\draw[-] (L5) -- (L4);
\end{tikzpicture}
\end{center}
After the FI deformation the superpotential reads 
\begin{equation} 
\mathcal{W}= \widetilde{P}_f\Psi P^f + \lambda\Tr\Psi\,,
\end{equation}
and consequently we have to solve the following F-term and D-term equations: 
\be\begin{array}{l} 
P^f\widetilde{P}_f = \lambda I_N\,,\\ 
P^fP_f^{\dagger}-\widetilde{P}^{\dagger f} \widetilde{P}_f=0\,,
\end{array}\ee
where the summation over flavor indices implies a summation over all the bifundamental hypermultiplets charged under the $U(N)$ gauge group, when this is embedded as a node inside a bigger quiver. 
From F-terms we can deduce that
\be 
\label{eq:traceFterm}
\Tr(P^f\widetilde{P}_f)=N\lambda\,.
\ee
Considering each node in the larger quiver in turn, and combining this with the identity $\Tr(P^f\widetilde{P}_f)=\Tr(\widetilde{P}_fP^f)$, we conclude that for any quiver with unitary gauge groups and bifundamental hypermultiplets the FI parameters should obey the relation 
\be\label{constr} \sum_i N_i\lambda^i=0\,,\ee
where $N_i$ denotes the rank of the $i$'th node. We therefore see that we always need to turn on FI parameters at multiple nodes. We will mainly focus on the minimal option with only two FI parameters, although sometimes it is convenient to turn on more than two. We will see examples of this below.

Consider the case of FI parameters turned on at two abelian nodes. Because of (\ref{constr}), the two FI parameters are $\lambda$ and $-\lambda$. The deformation induces a nontrivial expectation value for a chain of bifundamentals connecting the two nodes. This follows from the constraint on the trace of bifundamental bilinears in \eqref{eq:traceFterm}. Finding an explicit solution to the F- and D-term equations in this case is simple. First, we choose a subquiver beginning and ending at the nodes at which we have turned on the FI parameter. All the bifundamentals along the subquiver, which we denote $B_i$ and $\widetilde{B}_i$, acquire a nontrivial vev and modulo gauge transformations we can set
\be\label{solab} \langle B_i\rangle=\sqrt{\lambda}(v_1,0,\dots,0); \quad \langle \widetilde{B}_i\rangle=\langle B_i\rangle^T\,, \ee 
where $v_1$ is the unit vector whose entries are all trivial except for the first. Of course the size of the matrix $\langle B_i\rangle$ is dictated by the rank of the gauge groups under which the bifundamental is charged. If two nodes are connected by $j>1$ bifundamental hypermultiplets, modulo a flavor rotation, we can turn on the vev (\ref{solab}) for one of them, and set the vev of all the others to zero. 
All the unitary gauge groups along the subquiver are spontaneously broken as $U(n_i)\rightarrow U(n_i-1)$, whereas the other nodes in the quiver are unaffected. Among all the broken $U(1)$ factors, the diagonal combination survives and gives rise to a new $U(1)$ node, which is coupled to all the nodes of the quiver connected to those of the subquiver.  Overall, this is equivalent to subtracting \cite{Cabrera:2018ann} an abelian quiver, isomorphic to the subquiver described above, and the new $U(1)$ is identified with the rebalancing node. Notice that if one of the bifundamentals has multiplicity $j$ we should also include $j-1$ adjoints for the rebalancing node. These become $j-1$ free hypermultiplets which are identified with the Goldstone multiplets arising from the global symmetry breaking. Notice that this deformation reduces the dimension of the Higgs branch of (the interacting part of) the quiver. 

Generalizing to the case of FI parameters turned on at two nodes of the same rank $k \geq 1$ is straightforward. The vev for the bifundamentals is obtained from (\ref{solab}) by picking the tensor product with the $k\times k$ identity matrix $I_k$. All the groups along the subquiver are broken as $U(n_i)\rightarrow U(n_i-k)$ and finally we need to add a $U(k)$ node associated with the surviving gauge symmetry.  This operation corresponds to a modified quiver subtraction, where we subtract a quiver with $U(k)$ nodes only and rebalance with a $U(k)$ node. Again, if one of the bifundamentals has multiplicity $j$ the rebalancing node has $j-1$ adjoint hypermultiplets.

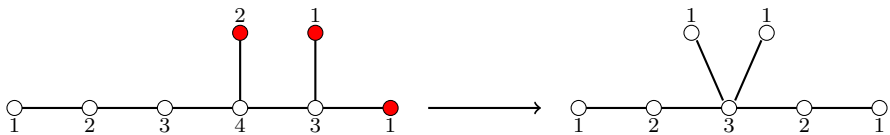
\begin{figure}[ht]
\begin{center}
\begin{tikzpicture}

\filldraw[fill= white] (0,0) circle [radius=0.1] node[below] {\scriptsize 1};
\filldraw[fill= white] (1,0) circle [radius=0.1] node[below] {\scriptsize 2};
\filldraw[fill= white] (2,0) circle [radius=0.1] node[below] {\scriptsize 3};
\filldraw[fill= white] (3,0) circle [radius=0.1] node[below] {\scriptsize 4};
\filldraw[fill= white] (4,0) circle [radius=0.1] node[below] {\scriptsize 3};
\filldraw[fill= red] (5,0) circle [radius=0.1] node[below] {\scriptsize 1};
\filldraw[fill= red] (4,1) circle [radius=0.1] node[above] {\scriptsize 1};
\filldraw[fill= red] (3,1) circle [radius=0.1] node[above] {\scriptsize 2};
\draw [thick] (0.1, 0) -- (0.9,0) ;
\draw [thick] (1.1, 0) -- (1.9,0) ;
\draw [thick] (2.1, 0) -- (2.9,0) ;
\draw [thick] (3.1, 0) -- (3.9,0) ;
\draw [thick] (4.1, 0) -- (4.9,0) ;
\draw [thick] (3, 0.1) -- (3,0.9) ;
\draw [thick] (4,0.1) -- (4,0.9) ;

\draw[->, thick] (5.5,0)--(7,0);

\filldraw[fill= white] (7.5,0) circle [radius=0.1] node[below] {\scriptsize 1};
\filldraw[fill= white] (8.5,0) circle [radius=0.1] node[below] {\scriptsize 2};
\filldraw[fill= white] (9.5,0) circle [radius=0.1] node[below] {\scriptsize 3};
\filldraw[fill= white] (10.5,0) circle [radius=0.1] node[below] {\scriptsize 2};
\filldraw[fill= white] (11.5,0) circle [radius=0.1] node[below] {\scriptsize 1};
\filldraw[fill= white] (9,1) circle [radius=0.1] node[above] {\scriptsize 1};
\filldraw[fill= white] (10,1) circle [radius=0.1] node[above] {\scriptsize 1};
\draw [thick] (7.6, 0) -- (8.4,0) ;
\draw [thick] (8.6, 0) -- (9.4,0) ;
\draw [thick] (9.6, 0) -- (10.4,0) ;
\draw [thick] (10.6, 0) -- (11.4,0) ;
\draw [thick] (9.42, 0.1) -- (9.07,0.9) ;
\draw [thick] (9.58, 0.1) -- (9.93,0.9) ;

\end{tikzpicture} 
\end{center}
\caption{One FI deformation leading from the $SO(12)\times U(1)$ SCFT to its low-energy gauge description, namely $SU(3)$ SQCD with 6 flavors.}\label{su3def}
\end{figure}

We can also write down a more elaborate variant of the above deformation which involves three nodes. Say we turn on FI parameters at the nodes $U(n)$, $U(m)$ and $U(n+m)$. The FI parameters satisfy the relation (\ref{constr}) and we further impose the constraint $\lambda_m=\lambda_n$, so that we still have only one independent parameter. The equations of motion can be solved as follows: We set the vev of all the bifundamentals in the subquiver connecting the nodes $U(m)$ and $U(n+m)$ to be 
\be\label{solab1} \langle B_i\rangle=\sqrt{\lambda_m}(I_m,0,\dots,0); \quad \langle \widetilde{B}_i\rangle=\langle B_i\rangle^T\,, \ee 
and the vev of the bifundamentals in the subquiver connecting the nodes $U(n+m)$ and $U(n)$ to be 
\be\label{solab2} \langle B_i\rangle=\sqrt{\lambda_m}(I_n,0,\dots,0); \quad \langle \widetilde{B}_i\rangle=\langle B_i\rangle^T\,. \ee 
Here we are assuming that the two subquivers meet at the $U(n+m)$ node only\footnote{This assumption can actually be relaxed without any major changes, provided that all the nodes in the intersection have rank at least $n+m$. We will shortly see an example of this in Figure \ref{su3def}.}. 
Overall, the higgsing of the theory can be described in terms of a sequence of two modified quiver subtractions: We first subtract a quiver of $U(n)$ nodes going from $U(n)$ to $U(n+m)$ and, as in the previous case, we rebalance with a $U(n)$ node. Then we subtract from the resulting quiver a quiver of $U(m)$ nodes going from $U(m)$ to $U(n+m)$ and rebalance with a $U(m)$ node. A careful analysis of the higgsing reveals that the $U(n)$ node we introduced at the first step should not be rebalanced when we perform the second subtraction. 
Let us illustrate the procedure for $m=1$ and $n=2$ in the case of the $E_8$ quiver: 
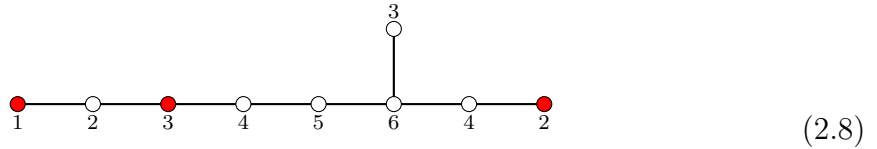
\be\label{E8MQ}
\begin{tikzpicture}
\filldraw[fill= red] (0,0) circle [radius=0.1] node[below] {\scriptsize 1};
\filldraw[fill= white] (1,0) circle [radius=0.1] node[below] {\scriptsize 2};
\filldraw[fill= red] (2,0) circle [radius=0.1] node[below] {\scriptsize 3};
\filldraw[fill= white] (3,0) circle [radius=0.1] node[below] {\scriptsize 4};
\filldraw[fill= white] (4,0) circle [radius=0.1] node[below] {\scriptsize 5};
\filldraw[fill= white] (5,0) circle [radius=0.1] node[below] {\scriptsize 6};
\filldraw[fill= white] (6,0) circle [radius=0.1] node[below] {\scriptsize 4};
\filldraw[fill= white] (5,1) circle [radius=0.1] node[above] {\scriptsize 3};
\filldraw[fill= red] (7,0) circle [radius=0.1] node[below] {\scriptsize 2};
\draw [thick] (0.1, 0) -- (0.9,0) ;
\draw [thick] (1.1, 0) -- (1.9,0) ;
\draw [thick] (2.1, 0) -- (2.9,0) ;
\draw [thick] (3.1, 0) -- (3.9,0) ;
\draw [thick] (4.1, 0) -- (4.9,0) ;
\draw [thick] (5.1, 0) -- (5.9,0) ;
\draw [thick] (5, 0.1) -- (5,0.9) ;
\draw [thick] (6.1, 0) -- (6.9,0) ;
\end{tikzpicture} 
\ee 
We turn on FI parameters at the nodes in red. We first subtract an $A_6$ quiver with $U(2)$ nodes, getting
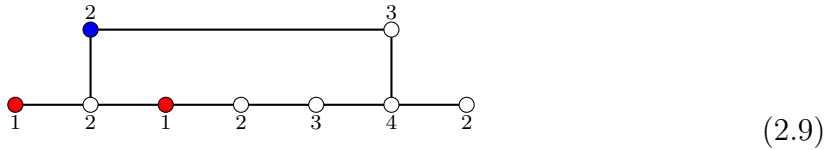
\be\label{E8MQ2}
\begin{tikzpicture}
\filldraw[fill= red] (0,0) circle [radius=0.1] node[below] {\scriptsize 1};
\filldraw[fill= white] (1,0) circle [radius=0.1] node[below] {\scriptsize 2};
\filldraw[fill= red] (2,0) circle [radius=0.1] node[below] {\scriptsize 1};
\filldraw[fill= white] (3,0) circle [radius=0.1] node[below] {\scriptsize 2};
\filldraw[fill= white] (4,0) circle [radius=0.1] node[below] {\scriptsize 3};
\filldraw[fill= white] (5,0) circle [radius=0.1] node[below] {\scriptsize 4};
\filldraw[fill= white] (6,0) circle [radius=0.1] node[below] {\scriptsize 2};
\filldraw[fill= white] (5,1) circle [radius=0.1] node[above] {\scriptsize 3};
\filldraw[fill= blue] (1,1) circle [radius=0.1] node[above] {\scriptsize 2};
\draw [thick] (0.1, 0) -- (0.9,0) ;
\draw [thick] (1.1, 0) -- (1.9,0) ;
\draw [thick] (2.1, 0) -- (2.9,0) ;
\draw [thick] (3.1, 0) -- (3.9,0) ;
\draw [thick] (4.1, 0) -- (4.9,0) ;
\draw [thick] (5.1, 0) -- (5.9,0) ;
\draw [thick] (5, 0.1) -- (5,0.9) ;
\draw [thick] (1.1, 1) -- (4.9,1) ;
\draw [thick] (1, 0.1) -- (1,0.9) ;
\end{tikzpicture} 
\ee 
The node in blue is used to rebalance. Then we subtract an $A_3$ abelian quiver, obtaining 
\be\label{E8MQ3}
\begin{tikzpicture}
\filldraw[fill= blue] (0,0) circle [radius=0.1] node[below] {\scriptsize 1};
\filldraw[fill= white] (1,0) circle [radius=0.1] node[below] {\scriptsize 2};
\filldraw[fill= white] (2,0) circle [radius=0.1] node[below] {\scriptsize 3};
\filldraw[fill= white] (3,0) circle [radius=0.1] node[below] {\scriptsize 4};
\filldraw[fill= white] (4,0) circle [radius=0.1] node[below] {\scriptsize 3};
\filldraw[fill= blue] (5,0) circle [radius=0.1] node[below] {\scriptsize 2};
\filldraw[fill= white] (6,0) circle [radius=0.1] node[below] {\scriptsize 1};
\filldraw[fill= white] (3,1) circle [radius=0.1] node[above] {\scriptsize 2};
\draw [thick] (0.1, 0) -- (0.9,0) ;
\draw [thick] (1.1, 0) -- (1.9,0) ;
\draw [thick] (2.1, 0) -- (2.9,0) ;
\draw [thick] (3.1, 0) -- (3.9,0) ;
\draw [thick] (4.1, 0) -- (4.9,0) ;
\draw [thick] (5.1, 0) -- (5.9,0) ;
\draw [thick] (3, 0.1) -- (3,0.9) ;
\end{tikzpicture} 
\ee 
The $U(1)$ node in blue is introduced to rebalance in the second subtraction, and, as we have explained before, is not connected to the blue $U(2)$ node. Overall, this FI deformation describes the transition from the $E_8$ to the $E_7$ theory.

Finally, let us consider the example of the UV completion of $SU(3)$ SQCD with 6 flavors and Chern-Simons (CS) level 2. The corresponding magnetic quiver is depicted on the left in Figure \ref{su3def}. We want to turn on a finite gauge coupling in 5d and therefore keep a balanced $A_5$ subquiver. This can be done by turning on an FI parameter at the top $U(2)$ node. We then also have to turn on FI parameters at the abelian nodes on the right. The simplest choice is to turn on the same FI parameter $-\lambda$ at both abelian nodes. Because of (\ref{constr}), we then conclude that the $U(2)$ FI parameter is $\lambda$.  
If we denote the $U(1)\times U(3)$ bifundamentals as $\widetilde{P}_{1,2}$ and $P_{1,2}$, all F-terms are solved by setting 
\be \langle\widetilde{P}_1\rangle=(\sqrt{\lambda},0,0);\; \langle P_1\rangle=\langle\widetilde{P}_1\rangle^T;\quad  \langle\widetilde{P}_2\rangle=(0,\sqrt{\lambda},0);\; \langle P_2\rangle=\langle\widetilde{P}_2\rangle^T\,.\ee 
The $U(3)\times U(4)$ and $U(4)\times U(2)$ bifundamentals have instead a nonvanishing $2\times 2$ block equal to $\sqrt{\lambda}I_2$. The result of the higgsing is the quiver on the right of Figure \ref{su3def}. Notice that the same RG flow can be implemented by turning on an FI deformation at the two $U(2)$ nodes. We clearly see here that generically there are multiple choices of FI deformation which implement a given mass deformation. 

\subsection{More General FI Deformations}
\label{sec:genFI}
\begin{figure}[ht]
\begin{center}
\begin{tikzpicture}
\filldraw[fill= red] (0,0) circle [radius=0.1] node[below] {\scriptsize 1};
\filldraw[fill= red] (1,0) circle [radius=0.1] node[below] {\scriptsize 2};
\filldraw[fill= white] (2,0) circle [radius=0.1] node[below] {\scriptsize 3};
\filldraw[fill= white] (3,0) circle [radius=0.1] node[below] {\scriptsize 4};
\filldraw[fill= white] (4,0) circle [radius=0.1] node[below] {\scriptsize 3};
\filldraw[fill= blue] (5,0) circle [radius=0.1] node[below] {\scriptsize 2};
\filldraw[fill= white] (6,0) circle [radius=0.1] node[below] {\scriptsize 1};
\filldraw[fill= white] (3,1) circle [radius=0.1] node[above] {\scriptsize 2};
\draw [thick] (0.1, 0) -- (0.9,0) ;
\draw [thick] (1.1, 0) -- (1.9,0) ;
\draw [thick] (2.1, 0) -- (2.9,0) ;
\draw [thick] (3.1, 0) -- (3.9,0) ;
\draw [thick] (4.1, 0) -- (4.9,0) ;
\draw [thick] (3, 0.1) -- (3,0.9) ;
\draw [thick] (5.1, 0) -- (5.9,0) ;

\draw[->, thick] (6.5,0)--(8,0);

\filldraw[fill= white] (8.5,0) circle [radius=0.1] node[below] {\scriptsize 1};
\filldraw[fill= white] (9.5,0) circle [radius=0.1] node[below] {\scriptsize 2};
\filldraw[fill= white] (10.5,0) circle [radius=0.1] node[below] {\scriptsize 2};
\filldraw[fill= white] (11.5,0) circle [radius=0.1] node[below] {\scriptsize 2};
\filldraw[fill= white] (9.5,1) circle [radius=0.1] node[above] {\scriptsize 1};
\filldraw[fill= white] (12.5,0) circle [radius=0.1] node[below] {\scriptsize 1};
\filldraw[fill= white] (11.5,1) circle [radius=0.1] node[above] {\scriptsize 1};
\draw [thick] (8.6,0) -- (9.4,0) ;
\draw [thick] (9.6,0) -- (10.4,0) ;
\draw [thick] (10.6,0) -- (11.4,0) ;
\draw [thick] (11.6,0) -- (12.4,0) ;
\draw [thick] (9.5,0.1) -- (9.5,0.9) ;
\draw [thick] (11.5,0.1) -- (11.5,0.9) ;

\end{tikzpicture} 
\end{center}
\caption{The $SO(12)$-preserving FI deformation of the rank 1 $E_7$ theory. This corresponds to turning on the gauge coupling in 5d. We turn on FI parameters at the red nodes, and we indicate in blue the node of the $E_7$ quiver which is not higgsed but is connected to a higgsed node. The extra $U(1)$ node is coupled to the blue node only.}\label{e7gaugedef2}
\end{figure}
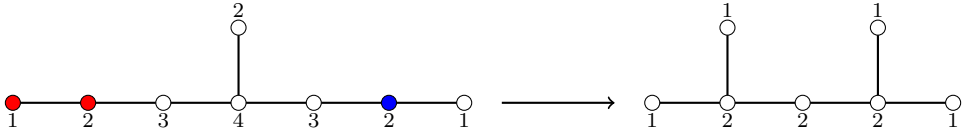

So far we have discussed only a particular set of FI deformations. Most mass deformations in 5d can be understood using the methods described above, however, it is also interesting to identify more general possibilities not considered in \cite{Bourget:2020mez}. When the quiver contains a tail of the form $U(1)-U(2)-\cdots$ we can turn on FI parameters at these $U(1)$ and $U(2)$ nodes only. As before, the generalization to the case $U(k)-U(2k)-\cdots$ is straightforward. Because of (\ref{constr}), we set $\lambda_1=-2\lambda_2\equiv2\lambda$. If we denote the $U(1)\times U(2)$ bifundamentals as $\widetilde{Q}$, $Q$ and the other $U(2)$ fundamentals as $\widetilde{P}_f$, $P^f$, the relevant F-terms are 
\be\label{soldef2} \langle\widetilde{Q}Q\rangle=2\lambda;\quad \langle\widetilde{P}_fP^f\rangle-\langle Q\widetilde{Q}\rangle=\lambda I_2\,,\ee 
where we are summing over flavor indices. These equations are solved by 
\be\label{u1u2}\langle\widetilde{Q}\rangle=\sqrt{\lambda}(1,1);\; \langle Q\rangle=\langle\widetilde{Q}\rangle^T;\; \langle\widetilde{P}\rangle=\sqrt{\lambda}\left(\begin{array}{cccc}1 & 0 & 0 & \dots \\ 0 & 1 & 0 & \dots\\\end{array}\right);\; \langle P\rangle^T=\sqrt{\lambda}\left(\begin{array}{cccc}0 & 1 & 0 & \dots \\ 1 & 0 & 0 & \dots\\\end{array}\right),\ee 
This vev spontaneously breaks $U(1)\times U(2)$ to a diagonal $U(1)$ subgroup. It is easy to check that D-terms are satisfied as well. The vev for the $U(2)$ fundamentals is
\be 
\langle\widetilde{P}_fP^f\rangle=\left(\begin{array}{cc}0 & \lambda \\ \lambda & 0 \\\end{array}\right)=\left(\begin{array}{cc}0 & \lambda \\ 0 & 0 \\\end{array}\right)+\left(\begin{array}{cc}0 & 0 \\ \lambda & 0 \\\end{array}\right)\,.
\ee 
This propagates along the quiver, breaking all the groups as $U(n)\rightarrow U(n-2)$, until we find a junction where we can ``decompose'' the vev as above.
The two nodes connected to the junction are higgsed as $U(n)\rightarrow U(n-1)$ and the vev does not propagate any further. All the nodes connected to the subquiver of nodes which are (partially) higgsed are now coupled to a new $U(1)$ node, which is left unbroken by the vev. 
We give an example of this process in Figure \ref{e7gaugedef2}, which constitutes an alternative way to understand the flow from the $E_7$ theory to $SU(2)$ SQCD with 6 flavors at the level of the 3d mirror theory.

\begin{figure}[ht]
\begin{center}
\begin{tikzpicture}
\filldraw[fill= white] (0,0) circle [radius=0.1] node[below] {\scriptsize 1};
\filldraw[fill= white] (1,0) circle [radius=0.1] node[below] {\scriptsize 2};
\filldraw[fill= white] (2,0) circle [radius=0.1] node[below] {\scriptsize 3};
\filldraw[fill= blue] (3,0) circle [radius=0.1] node[below] {\scriptsize 4};
\filldraw[fill= white] (4,0) circle [radius=0.1] node[below] {\scriptsize 5};
\filldraw[fill= white] (5,0) circle [radius=0.1] node[below] {\scriptsize 6};
\filldraw[fill= red] (6,0) circle [radius=0.1] node[below] {\scriptsize 4};
\filldraw[fill= white] (5,1) circle [radius=0.1] node[above] {\scriptsize 3};
\filldraw[fill= red] (7,0) circle [radius=0.1] node[below] {\scriptsize 2}; 
\filldraw[fill= blue] (8,0) circle [radius=0.1] node[below] {\scriptsize 1};
\draw [thick] (0.1, 0) -- (0.9,0) ;
\draw [thick] (1.1, 0) -- (1.9,0) ;
\draw [thick] (2.1, 0) -- (2.9,0) ;
\draw [thick] (3.1, 0) -- (3.9,0) ;
\draw [thick] (4.1, 0) -- (4.9,0) ;
\draw [thick] (5, 0.1) -- (5,0.9) ;
\draw [thick] (5.1, 0) -- (5.9,0) ;
\draw [thick] (6.1, 0) -- (6.9,0) ;
\draw [thick] (7.1, 0) -- (7.9,0) ;

\draw[->, thick] (8.3,0)--(9.2,0);

\filldraw[fill= white] (9.5,0) circle [radius=0.1] node[below] {\scriptsize 1};
\filldraw[fill= white] (10.5,0) circle [radius=0.1] node[below] {\scriptsize 2};
\filldraw[fill= white] (11.5,0) circle [radius=0.1] node[below] {\scriptsize 3};
\filldraw[fill= white] (12.5,0) circle [radius=0.1] node[below] {\scriptsize 4};
\filldraw[fill= white] (13.5,0) circle [radius=0.1] node[below] {\scriptsize 3};
\filldraw[fill= white] (14.5,0) circle [radius=0.1] node[below] {\scriptsize 2};
\filldraw[fill= white] (15.5,0) circle [radius=0.1] node[below] {\scriptsize 1};
\filldraw[fill= white] (12.5,1) circle [radius=0.1] node[right] {\scriptsize 2};
\filldraw[fill= white] (12.5,2) circle [radius=0.1] node[above] {\scriptsize 1};
\draw [thick] (9.6, 0) -- (10.4,0) ;
\draw [thick] (10.6, 0) -- (11.4,0) ;
\draw [thick] (11.6, 0) -- (12.4,0) ;
\draw [thick] (12.6, 0) -- (13.4,0) ;
\draw [thick] (13.6, 0) -- (14.4,0) ;
\draw [thick] (14.6, 0) -- (15.4,0) ;
\draw [thick] (12.5, 0.1) -- (12.5,0.9) ;
\draw [thick] (12.5, 1.1) -- (12.5,1.9) ;

\end{tikzpicture} 
\end{center}
\caption{The $SU(8)\times SU(2)$ preserving mass deformation of the rank 2 $SO(16)\times SU(2)$ theory. We turn on FI parameters at red nodes. As a result of the higgsing a new $U(2)$ node appears, coupled to the blue nodes.}\label{su3massdef}
\end{figure}
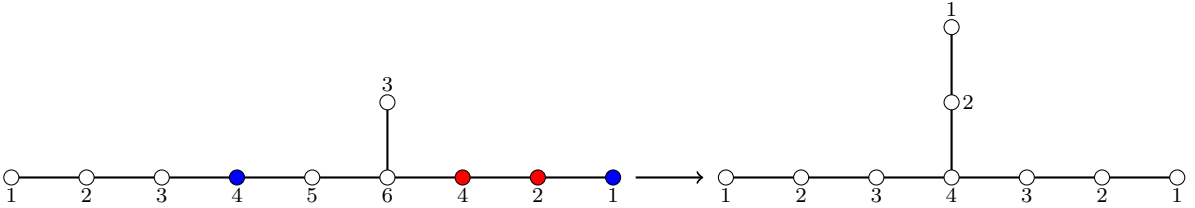

Let us now consider a more complicated example, namely the UV completion of $SU(3)$ SQCD with 8 flavors and CS level 1. The SCFT has global symmetry $SO(16)\times SU(2)$. We wish to turn on a mass deformation which breaks the symmetry to $SU(8)\times SU(2)$, leading to another SCFT whose gauge theory phase is given by $SU(3)$ SQCD with 7 flavors and CS level $1/2$. In other words, we are decoupling by sending a hypermultiplet mass to $-\infty$. In order to preserve an $A_7\times A_1$ balanced subquiver we turn on FI parameters at a $U(2)$ and $U(4)$ node as shown in Figure \ref{su3massdef}. The F-terms are solved as explained above. The only difference is that we have to tensor all the matrices in (\ref{u1u2}) by the $2\times 2$ identity matrix. The central node is higgsed to $U(2)$ and the neighbouring nodes are higgsed as $U(n)\rightarrow U(n-2)$. We further need to add a $U(2)$ node attached to the unhiggsed $U(1)$ and $U(4)$ nodes. 

Finally, let us consider a variant of the above construction involving nodes which are not next to one another. We illustrate this with an example based on the $E_8$ quiver:
\be\label{E8MQ22}
\begin{tikzpicture}
\filldraw[fill= white] (0,0) circle [radius=0.1] node[below] {\scriptsize 1};
\filldraw[fill= red] (1,0) circle [radius=0.1] node[below] {\scriptsize 2};
\filldraw[fill= white] (2,0) circle [radius=0.1] node[below] {\scriptsize 3};
\filldraw[fill= red] (3,0) circle [radius=0.1] node[below] {\scriptsize 4};
\filldraw[fill= white] (4,0) circle [radius=0.1] node[below] {\scriptsize 5};
\filldraw[fill= white] (5,0) circle [radius=0.1] node[below] {\scriptsize 6};
\filldraw[fill= white] (6,0) circle [radius=0.1] node[below] {\scriptsize 4};
\filldraw[fill= white] (5,1) circle [radius=0.1] node[above] {\scriptsize 3};
\filldraw[fill= white] (7,0) circle [radius=0.1] node[below] {\scriptsize 2};
\draw [thick] (0.1, 0) -- (0.9,0) ;
\draw [thick] (1.1, 0) -- (1.9,0) ;
\draw [thick] (2.1, 0) -- (2.9,0) ;
\draw [thick] (3.1, 0) -- (3.9,0) ;
\draw [thick] (4.1, 0) -- (4.9,0) ;
\draw [thick] (5.1, 0) -- (5.9,0) ;
\draw [thick] (5, 0.1) -- (5,0.9) ;
\draw [thick] (6.1, 0) -- (6.9,0) ;
\end{tikzpicture} 
\ee 
If we turn on FI parameters at the $U(2)$ and $U(4)$ nodes, we can solve the equations of motion by setting the expectation value for the $U(2)\times U(3)$ bifundamental to\footnote{Here we have normalized the FI parameter to one at the $U(4)$ node for simplicity.} 
\be \langle B_{23}\rangle=\left(\begin{array}{ccc}\sqrt{2}&0&0\\ 0&\sqrt{2}&0\end{array}\right);\quad \langle\widetilde{B}_{32}\rangle=\left(\begin{array}{cc}\sqrt{2}&0\\0&\sqrt{2}\\ 0&0\\\end{array}\right).\ee 
and the vev of the $U(3)\times U(4)$ bifundamental to
\be \langle B_{34}\rangle=\left(\begin{array}{cccc}1&0&1&0\\ 0&1&0&1\\ 0&0&0&0\end{array}\right);\quad \langle\widetilde{B}_{43}\rangle=\left(\begin{array}{ccc}1&0&0\\0&1&0\\ 1&0&0\\0&1&0\\\end{array}\right).\ee 
The $U(2)\times U(4)$ gauge symmetry is spontaneously broken to a diagonal $U(2)$ and the $U(3)$ in between is broken to $U(1)$.
The vev propagates in the rest of the quiver as in the case of a $U(2)$ node followed by a $U(4)$ node; all the groups along the tail are higgsed as $U(n) \rightarrow U(n-4)$. It turns out that this deformation is equivalent to a weak coupling limit of the 5d theory: 
\be\label{E8MQ33}
\begin{tikzpicture}
\filldraw[fill= white] (0,0) circle [radius=0.1] node[below] {\scriptsize 1};
\filldraw[fill= red] (1,0) circle [radius=0.1] node[below] {\scriptsize 2};
\filldraw[fill= white] (2,0) circle [radius=0.1] node[below] {\scriptsize 3};
\filldraw[fill= red] (3,0) circle [radius=0.1] node[below] {\scriptsize 4};
\filldraw[fill= white] (4,0) circle [radius=0.1] node[below] {\scriptsize 5};
\filldraw[fill= white] (5,0) circle [radius=0.1] node[below] {\scriptsize 6};
\filldraw[fill= white] (6,0) circle [radius=0.1] node[below] {\scriptsize 4};
\filldraw[fill= white] (5,1) circle [radius=0.1] node[above] {\scriptsize 3};
\filldraw[fill= white] (7,0) circle [radius=0.1] node[below] {\scriptsize 2};
\draw [thick] (0.1, 0) -- (0.9,0) ;
\draw [thick] (1.1, 0) -- (1.9,0) ;
\draw [thick] (2.1, 0) -- (2.9,0) ;
\draw [thick] (3.1, 0) -- (3.9,0) ;
\draw [thick] (4.1, 0) -- (4.9,0) ;
\draw [thick] (5.1, 0) -- (5.9,0) ;
\draw [thick] (5, 0.1) -- (5,0.9) ;
\draw [thick] (6.1, 0) -- (6.9,0) ;

\draw [->,thick] (7.5,0) -- (8.5,0); 

\filldraw[fill= white] (9,0) circle [radius=0.1] node[below] {\scriptsize 1};
\filldraw[fill= white] (10,0) circle [radius=0.1] node[below] {\scriptsize 2};
\filldraw[fill= white] (11,0) circle [radius=0.1] node[below] {\scriptsize 2};
\filldraw[fill= white] (12,0) circle [radius=0.1] node[below] {\scriptsize 2};
\filldraw[fill= white] (13,0) circle [radius=0.1] node[below] {\scriptsize 2};
\filldraw[fill= white] (14,0) circle [radius=0.1] node[below] {\scriptsize 1};
\filldraw[fill= white] (10,1) circle [radius=0.1] node[above] {\scriptsize 1};
\filldraw[fill= white] (13,1) circle [radius=0.1] node[above] {\scriptsize 1};
\draw [thick] (9.1, 0) -- (9.9,0) ;
\draw [thick] (10.1, 0) -- (10.9,0) ;
\draw [thick] (11.1, 0) -- (11.9,0) ;
\draw [thick] (12.1, 0) -- (12.9,0) ;
\draw [thick] (13.1, 0) -- (13.9,0) ;
\draw [thick] (10, 0.1) -- (10,0.9) ;
\draw [thick] (13, 0.1) -- (13,0.9) ;
\end{tikzpicture} 
\ee 

\subsection{FI Deformations of the Higher Rank E-String Theory}

As an application of the above discussion, we will now see that using the set of rules we have explained, we can identify the FI deformations which implement decouplings and weak coupling limits of all rank 1 SCFTs, namely the E-string theory and its descendants. The analysis can be easily extended to higher rank E-string theories and therefore we directly discuss this case. The general answer is given in Figure \ref{fig:RankNEn}.
\begin{sidewaysfigure}[ht]
\centering
\begin{tabular}{|c|c|c|}\hline
Rank $N$ $E_n$  & MQ with FI for rank $N$ $E_n$ to rank $N$ $E_{n-1}$ & MQ with FI to get $Sp(N-1) +1\bm{AS} + (n-1) \bm{F}$ \cr \hline\hline 
$E_8$& 
\begin{tikzpicture}
\filldraw[fill= white]    (0,0) circle [radius=0.1] node[below] {\scriptsize 1};
\filldraw[fill= red] (1,0) circle [radius=0.1] node[below] {\scriptsize $N$};
\filldraw[fill= red] (2,0) circle [radius=0.1] node[below] {\scriptsize $2N$};
\filldraw[fill= white] (3,0) circle [radius=0.1] node[below] {\scriptsize $3N$};
\filldraw[fill= white] (4,0) circle [radius=0.1] node[below] {\scriptsize $4N$};
\filldraw[fill= white] (5,0) circle [radius=0.1] node[below] {\scriptsize  $5N$};
\filldraw[fill= white] (6,0) circle [radius=0.1] node[below] {\scriptsize  $6N$ };
\filldraw[fill= white] (7,0) circle [radius=0.1] node[below] {\scriptsize  $4N$ };
\filldraw[fill= white] (8,0) circle [radius=0.1] node[below] {\scriptsize  $2N$};
\filldraw[fill= white] (6,1) circle [radius=0.1] node[above] {\scriptsize $3N$};
\draw [thick] (0.1, 0) -- (0.9,0) ;
\draw [thick] (1.1, 0) -- (1.9,0) ;
\draw [thick] (2.1, 0) -- (2.9,0) ;
\draw [thick] (3.1, 0) -- (3.9,0) ;
\draw [thick] (4.1, 0) -- (4.9,0) ;
\draw [thick] (5.1, 0) -- (5.9,0) ;
\draw [thick] (6.1, 0) -- (6.9,0) ;
\draw [thick] (7.1, 0) -- (7.9,0) ;
\draw [thick] (6, 0.1) -- (6,0.9) ;
\end{tikzpicture}
 &\cr 
&  
\begin{tikzpicture}
\filldraw[fill= white]    (0,0) circle [radius=0.1] node[below] {\scriptsize 1};
\filldraw[fill= white] (1,0) circle [radius=0.1] node[below] {\scriptsize $N$};
\filldraw[fill= red] (2,0) circle [radius=0.1] node[below] {\scriptsize $2N$};
\filldraw[fill= white] (3,0) circle [radius=0.1] node[below] {\scriptsize $3N$};
\filldraw[fill= white] (4,0) circle [radius=0.1] node[below] {\scriptsize $4N$};
\filldraw[fill= white] (5,0) circle [radius=0.1] node[below] {\scriptsize  $5N$};
\filldraw[fill= white] (6,0) circle [radius=0.1] node[below] {\scriptsize  $6N$ };
\filldraw[fill= white] (7,0) circle [radius=0.1] node[below] {\scriptsize  $4N$ };
\filldraw[fill= red] (8,0) circle [radius=0.1] node[below] {\scriptsize  $2N$};
\filldraw[fill= white] (6,1) circle [radius=0.1] node[above] {\scriptsize $3N$};
\draw [thick] (0.1, 0) -- (0.9,0) ;
\draw [thick] (1.1, 0) -- (1.9,0) ;
\draw [thick] (2.1, 0) -- (2.9,0) ;
\draw [thick] (3.1, 0) -- (3.9,0) ;
\draw [thick] (4.1, 0) -- (4.9,0) ;
\draw [thick] (5.1, 0) -- (5.9,0) ;
\draw [thick] (6.1, 0) -- (6.9,0) ;
\draw [thick] (7.1, 0) -- (7.9,0) ;
\draw [thick] (6, 0.1) -- (6,0.9) ;
\end{tikzpicture}
& 
\begin{tikzpicture}
\filldraw[fill= white]    (0,0) circle [radius=0.1] node[below] {\scriptsize 1};
\filldraw[fill= white] (1,0) circle [radius=0.1] node[below] {\scriptsize $N$};
\filldraw[fill= blue] (2,0) circle [radius=0.1] node[below] {\scriptsize $2N$};
\filldraw[fill= white] (3,0) circle [radius=0.1] node[below] {\scriptsize $3N$};
\filldraw[fill= blue] (4,0) circle [radius=0.1] node[below] {\scriptsize $4N$};
\filldraw[fill= white] (5,0) circle [radius=0.1] node[below] {\scriptsize  $5N$};
\filldraw[fill= white] (6,0) circle [radius=0.1] node[below] {\scriptsize  $6N$ };
\filldraw[fill= white] (7,0) circle [radius=0.1] node[below] {\scriptsize  $4N$ };
\filldraw[fill= white] (8,0) circle [radius=0.1] node[below] {\scriptsize  $2N$};
\filldraw[fill= white] (6,1) circle [radius=0.1] node[above] {\scriptsize $3N$};
\draw [thick] (0.1, 0) -- (0.9,0) ;
\draw [thick] (1.1, 0) -- (1.9,0) ;
\draw [thick] (2.1, 0) -- (2.9,0) ;
\draw [thick] (3.1, 0) -- (3.9,0) ;
\draw [thick] (4.1, 0) -- (4.9,0) ;
\draw [thick] (5.1, 0) -- (5.9,0) ;
\draw [thick] (6.1, 0) -- (6.9,0) ;
\draw [thick] (7.1, 0) -- (7.9,0) ;
\draw [thick] (6, 0.1) -- (6,0.9) ;
\end{tikzpicture}
 \cr \hline\hline

$E_7$& 
\begin{tikzpicture}
\filldraw[fill= white]    (0,0) circle [radius=0.1] node[below] {\scriptsize 1};
\filldraw[fill= red] (1,0) circle [radius=0.1] node[below] {\scriptsize $N$};
\filldraw[fill= white] (2,0) circle [radius=0.1] node[below] {\scriptsize $2N$};
\filldraw[fill= white] (3,0) circle [radius=0.1] node[below] {\scriptsize $3N$};
\filldraw[fill= white] (4,0) circle [radius=0.1] node[below] {\scriptsize $4N$};
\filldraw[fill= white] (5,0) circle [radius=0.1] node[below] {\scriptsize  $3N$};
\filldraw[fill= white] (6,0) circle [radius=0.1] node[below] {\scriptsize  $2N$ };
\filldraw[fill= red] (7,0) circle [radius=0.1] node[below] {\scriptsize  $N$ };
\filldraw[fill= white] (4,1) circle [radius=0.1] node[above ] {\scriptsize  $2N$ };
\draw [thick] (0.1, 0) -- (0.9,0) ;
\draw [thick] (1.1, 0) -- (1.9,0) ;
\draw [thick] (2.1, 0) -- (2.9,0) ;
\draw [thick] (3.1, 0) -- (3.9,0) ;
\draw [thick] (4.1, 0) -- (4.9,0) ;
\draw [thick] (5.1, 0) -- (5.9,0) ;
\draw [thick] (6.1, 0) -- (6.9,0) ;
\draw [thick] (4, 0.1) -- (4,0.9) ;
\end{tikzpicture}
&
\begin{tikzpicture}
\filldraw[fill= white]    (0,0) circle [radius=0.1] node[below] {\scriptsize 1};
\filldraw[fill= blue] (1,0) circle [radius=0.1] node[below] {\scriptsize $N$};
\filldraw[fill= blue] (2,0) circle [radius=0.1] node[below] {\scriptsize $2N$};
\filldraw[fill= white] (3,0) circle [radius=0.1] node[below] {\scriptsize $3N$};
\filldraw[fill= white] (4,0) circle [radius=0.1] node[below] {\scriptsize $4N$};
\filldraw[fill= white] (5,0) circle [radius=0.1] node[below] {\scriptsize  $3N$};
\filldraw[fill= white] (6,0) circle [radius=0.1] node[below] {\scriptsize  $2N$ };
\filldraw[fill= white] (7,0) circle [radius=0.1] node[below] {\scriptsize  $N$ };
\filldraw[fill= white] (4,1) circle [radius=0.1] node[above ] {\scriptsize  $2N$ };
\draw [thick] (0.1, 0) -- (0.9,0) ;
\draw [thick] (1.1, 0) -- (1.9,0) ;
\draw [thick] (2.1, 0) -- (2.9,0) ;
\draw [thick] (3.1, 0) -- (3.9,0) ;
\draw [thick] (4.1, 0) -- (4.9,0) ;
\draw [thick] (5.1, 0) -- (5.9,0) ;
\draw [thick] (6.1, 0) -- (6.9,0) ;
\draw [thick] (4, 0.1) -- (4,0.9) ;
\end{tikzpicture}
\cr 
\hline
$E_6$& 
\begin{tikzpicture}
\filldraw[fill= white]    (0,0) circle [radius=0.1] node[below] {\scriptsize 1};
\filldraw[fill= red] (1,0) circle [radius=0.1] node[below] {\scriptsize $N$};
\filldraw[fill= white] (2,0) circle [radius=0.1] node[below] {\scriptsize $2N$};
\filldraw[fill= white] (3,0) circle [radius=0.1] node[below] {\scriptsize $3N$};
\filldraw[fill= white] (4,0) circle [radius=0.1] node[below] {\scriptsize $2N$};
\filldraw[fill= red] (5,0) circle [radius=0.1] node[below] {\scriptsize  $N$};
\filldraw[fill= white] (3,1) circle [radius=0.1] node[right] {\scriptsize  $2N$ };
\filldraw[fill=white] (3,2) circle [radius=0.1] node[right] {\scriptsize  $N$ };
\draw [thick] (0.1, 0) -- (0.9,0) ;
\draw [thick] (1.1, 0) -- (1.9,0) ;
\draw [thick] (2.1, 0) -- (2.9,0) ;
\draw [thick] (3.1, 0) -- (3.9,0) ;
\draw [thick] (4.1, 0) -- (4.9,0) ;
\draw [thick] (3, 0.1) -- (3,0.9) ;
\draw [thick] (3, 1.1) -- (3,1.9) ;
\end{tikzpicture}
&
\begin{tikzpicture}
\filldraw[fill= white]    (0,0) circle [radius=0.1] node[below] {\scriptsize 1};
\filldraw[fill= blue] (1,0) circle [radius=0.1] node[below] {\scriptsize $N$};
\filldraw[fill= white] (2,0) circle [radius=0.1] node[below] {\scriptsize $2N$};
\filldraw[fill= white] (3,0) circle [radius=0.1] node[below] {\scriptsize $3N$};
\filldraw[fill= white] (4,0) circle [radius=0.1] node[below] {\scriptsize $2N$};
\filldraw[fill= blue] (5,0) circle [radius=0.1] node[below] {\scriptsize  $N$};
\filldraw[fill= white] (3,1) circle [radius=0.1] node[right] {\scriptsize  $2N$ };
\filldraw[fill=white] (3,2) circle [radius=0.1] node[right] {\scriptsize  $N$ };
\draw [thick] (0.1, 0) -- (0.9,0) ;
\draw [thick] (1.1, 0) -- (1.9,0) ;
\draw [thick] (2.1, 0) -- (2.9,0) ;
\draw [thick] (3.1, 0) -- (3.9,0) ;
\draw [thick] (4.1, 0) -- (4.9,0) ;
\draw [thick] (3, 0.1) -- (3,0.9) ;
\draw [thick] (3, 1.1) -- (3,1.9) ;
\end{tikzpicture}
\cr 
\hline
$E_5$ & 
\begin{tikzpicture}
\filldraw[fill= white]    (0,0) circle [radius=0.1] node[below] {\scriptsize 1};
\filldraw[fill= red] (1,0) circle [radius=0.1] node[below] {\scriptsize $N$};
\filldraw[fill= white] (2,0) circle [radius=0.1] node[below] {\scriptsize $2N$};
\filldraw[fill= white] (3,0) circle [radius=0.1] node[below] {\scriptsize $2N$};
\filldraw[fill= red] (4,0) circle [radius=0.1] node[below] {\scriptsize $N$};
\filldraw[fill=white] (2,1) circle [radius=0.1] node[right] {\scriptsize  $N$ };
\filldraw[fill=white] (3,1) circle [radius=0.1] node[right] {\scriptsize  $N$ };
\draw [thick] (0.1, 0) -- (0.9,0) ;
\draw [thick] (1.1, 0) -- (1.9,0) ;
\draw [thick] (2.1, 0) -- (2.9,0) ;
\draw [thick] (3.1, 0) -- (3.9,0) ;
\draw [thick] (3, 0.1) -- (3,0.9) ;
\draw [thick] (2, 0.1) -- (2,0.9) ;
\end{tikzpicture}
&
\begin{tikzpicture}
\filldraw[fill= white]    (0,0) circle [radius=0.1] node[below] {\scriptsize 1};
\filldraw[fill= white] (1,0) circle [radius=0.1] node[below] {\scriptsize $N$};
\filldraw[fill= white] (2,0) circle [radius=0.1] node[below] {\scriptsize $2N$};
\filldraw[fill= white] (3,0) circle [radius=0.1] node[below] {\scriptsize $2N$};
\filldraw[fill= blue] (4,0) circle [radius=0.1] node[below] {\scriptsize $N$};
\filldraw[fill=white] (2,1) circle [radius=0.1] node[right] {\scriptsize  $N$ };
\filldraw[fill=blue] (3,1) circle [radius=0.1] node[right] {\scriptsize  $N$ };
\draw [thick] (0.1, 0) -- (0.9,0) ;
\draw [thick] (1.1, 0) -- (1.9,0) ;
\draw [thick] (2.1, 0) -- (2.9,0) ;
\draw [thick] (3.1, 0) -- (3.9,0) ;
\draw [thick] (3, 0.1) -- (3,0.9) ;
\draw [thick] (2, 0.1) -- (2,0.9) ;
\end{tikzpicture}
\cr 

\hline
$E_4$ & 
\begin{tikzpicture}
\filldraw[fill= white]    (0,0) circle [radius=0.1] node[below] {\scriptsize 1};
\filldraw[fill= red] (1,0) circle [radius=0.1] node[below] {\scriptsize $N$};
\filldraw[fill= white] (2,0) circle [radius=0.1] node[below] {\scriptsize $N$};
\filldraw[fill= white] (3,0) circle [radius=0.1] node[below] {\scriptsize $N$};
\filldraw[fill= red] (4,0) circle [radius=0.1] node[below] {\scriptsize $N$};
\filldraw[fill= white] (2.5,1) circle [radius=0.1] node[above] {\scriptsize  $N$ };
\draw [thick] (0.1, 0) -- (0.9,0) ;
\draw [thick] (1.1, 0) -- (1.9,0) ;
\draw [thick] (2.1, 0) -- (2.9,0) ;
\draw [thick] (3.1, 0) -- (3.9,0) ;
\draw [thick] (1.05, 0.1) -- (2.4,0.98) ;
\draw [thick] (3.95, 0.1) -- (2.6,0.98) ;
\end{tikzpicture}
&
\begin{tikzpicture}
\filldraw[fill= white]    (0,0) circle [radius=0.1] node[below] {\scriptsize 1};
\filldraw[fill= blue] (1,0) circle [radius=0.1] node[below] {\scriptsize $N$};
\filldraw[fill= blue] (2,0) circle [radius=0.1] node[below] {\scriptsize $N$};
\filldraw[fill= white] (3,0) circle [radius=0.1] node[below] {\scriptsize $N$};
\filldraw[fill= white] (4,0) circle [radius=0.1] node[below] {\scriptsize $N$};
\filldraw[fill= white] (2.5,1) circle [radius=0.1] node[above] {\scriptsize  $N$ };
\draw [thick] (0.1, 0) -- (0.9,0) ;
\draw [thick] (1.1, 0) -- (1.9,0) ;
\draw [thick] (2.1, 0) -- (2.9,0) ;
\draw [thick] (3.1, 0) -- (3.9,0) ;
\draw [thick] (1.05, 0.1) -- (2.4,0.98) ;
\draw [thick] (3.95, 0.1) -- (2.6,0.98) ;
\end{tikzpicture}
\cr 

\hline
$E_3$ & 
\begin{tikzpicture}
\filldraw[fill= white]    (0,0) circle [radius=0.1] node[below] {\scriptsize 1};
\filldraw[fill= red] (1,0) circle [radius=0.1] node[below] {\scriptsize $N$};
\filldraw[fill= red] (2,0) circle [radius=0.1] node[below] {\scriptsize $N$};
\filldraw[fill= white] (1.5,1) circle [radius=0.1] node[above] {\scriptsize  $N$ };
\draw [thick] (0.1, 0) -- (0.9,0) ;
\draw [thick] (1.1, 0) -- (1.9,0) ;
\draw [thick] (1.05, 0.1) -- (1.4,0.98) ;
\draw [thick] (1.95, 0.1) -- (1.6,0.98) ;
\end{tikzpicture}
\quad 
\begin{tikzpicture}
\filldraw[fill= white]    (0,0) circle [radius=0.1] node[below] {\scriptsize 1};
\filldraw[fill= red] (1,0) circle [radius=0.1] node[below] {\scriptsize $N$};
\filldraw[fill= red] (2,0) circle [radius=0.1] node[below] {\scriptsize $N$};
\draw [thick] (0.1, 0) -- (0.9,0) ;
\draw [thick, double] (1.1, 0) -- (1.9,0) ;
\end{tikzpicture}
&
\begin{tikzpicture}
\filldraw[fill= white]    (0,0) circle [radius=0.1] node[below] {\scriptsize 1};
\filldraw[fill= blue] (1,0) circle [radius=0.1] node[below] {\scriptsize $N$};
\filldraw[fill= blue] (2,0) circle [radius=0.1] node[below] {\scriptsize $N$};
\filldraw[fill= white] (1.5,1) circle [radius=0.1] node[above] {\scriptsize  $N$ };
\draw [thick] (0.1, 0) -- (0.9,0) ;
\draw [thick] (1.1, 0) -- (1.9,0) ;
\draw [thick] (1.05, 0.1) -- (1.4,0.98) ;
\draw [thick] (1.95, 0.1) -- (1.6,0.98) ;
\end{tikzpicture}
\quad 
\begin{tikzpicture}
\filldraw[fill= white]    (0,0) circle [radius=0.1] node[below] {\scriptsize 1};
\filldraw[fill= blue] (1,0) circle [radius=0.1] node[below] {\scriptsize $N$};
\filldraw[fill= blue] (2,0) circle [radius=0.1] node[below] {\scriptsize $N$};
\draw [thick] (0.1, 0) -- (0.9,0) ;
\draw [thick, double] (1.1, 0) -- (1.9,0) ;
\end{tikzpicture}

\cr 

\hline
$E_2$& 
\begin{tikzpicture}
\filldraw[fill= white]    (0,0) circle [radius=0.1] node[below] {\scriptsize 1};
\filldraw[fill= white] (1,0) circle [radius=0.1] node[below] {\scriptsize $N$};
\filldraw[fill= white] (2,0) circle [radius=0.1] node[below] {\scriptsize $N$};
\draw [thick] (0.1, 0) -- (0.9,0) ;
\draw [thick, double] (1.1, 0) -- (1.9,0) ;
\end{tikzpicture}\qquad 
\begin{tikzpicture}
\filldraw[fill= white]    (0,0) circle [radius=0.1] node[below] {\scriptsize 1};
\node[circle, draw, radius=0.1] (A) at (1,0) {}; 
\node[] at (1,-0.35) {\scriptsize $N$};
 \path[every node/.style={font=\sffamily\small,
  		fill=white,inner sep=1pt}]
(A) edge [loop, out=35, in=325, looseness=4] (A); 
\draw [thick] (0.1, 0) -- (0.8,0) ;
\end{tikzpicture}
&
\begin{tikzpicture}
\filldraw[fill= white]    (0,0) circle [radius=0.1] node[below] {\scriptsize 1};
\filldraw[fill= white] (1,0) circle [radius=0.1] node[below] {\scriptsize $N$};
\filldraw[fill= white] (2,0) circle [radius=0.1] node[below] {\scriptsize $N$};
\draw [thick] (0.1, 0) -- (0.9,0) ;
\draw [thick, double] (1.1, 0) -- (1.9,0) ;
\end{tikzpicture}\qquad 
\begin{tikzpicture}
\filldraw[fill= white]    (0,0) circle [radius=0.1] node[below] {\scriptsize 1};
\node[circle, draw, radius=0.1] (A) at (1,0) {}; 
\node[] at (1,-0.35) {\scriptsize $N$};
 \path[every node/.style={font=\sffamily\small,
  		fill=white,inner sep=1pt}]
(A) edge [loop, out=35, in=325, looseness=4] (A); 
\draw [thick] (0.1, 0) -- (0.8,0) ;
\end{tikzpicture}

\cr \hline

\end{tabular}

\caption{Magnetic quivers for the rank $N$ E-strings. The red colored nodes correspond to the FI parameters that need to be turned on to decouple and flow from the rank $N$ $E_n$ to the rank $N$  $E_{n-1}$ theory. The blue nodes denote those at which we have to turn on FI parameters to find the weakly coupled description in terms of $USp(2N)$.  \label{fig:RankNEn}}
\end{sidewaysfigure}

\FloatBarrier

\section{Mass Deformations in 5d via FI Deformations}
\label{sec:massdef}

5d $\cN=1$ SCFTs can be engineered in M-theory on a canonical Calabi-Yau 3-fold singularity 
or, alternatively, one can study the world-volume theories of type IIB 5-brane-webs, where the extended Coulomb branch and Higgs branch are respectively associated with continuous deformations of the branes in and transverse to the plane of the web. Since only supersymmetric and charge-conserving subwebs may move independently, the transverse deformations are determined by the set of maximal subweb decompositions, which in turn encode the magnetic quiver.
The gap between the two descriptions is bridged by the Generalized Toric Polygons, which are the dual diagrams to the brane-webs, and, in the toric case, realize the M-theory compactification space. This is the representation we will work with, as we study the transformation of the Higgs branch under mass deformations and RG flow. The magnetic quiver is determined from the GTP through means of a refined Minkowski sum decomposition, realizing the dual of the subweb decomposition.
Our aim is to describe, using the map between GTP and magnetic quiver, how the Higgs branch emanating from one point in the extended Coulomb branch is changed as we move to a different point.
We focus first on describing transitions in the magnetic quiver associated to decoupling matter, and then discuss gauge theory phases. Finally, we open up general extended Coulomb branch directions that include turning on vevs for the 5d vector-multiplet scalars. 

Because any significant change in the structure of the GTP, i.e. to its shape or distribution of vertices, will result in a major change to the combinatorial data contained in the magnetic quiver, a unified treatment of all 5d SCFTs is not feasible at this point and thus, a case by case analysis is necessary. Therefore, rather than aiming for generality, we demonstrate the correspondence with FI deformations focusing on a representative example. On the other hand, the mechanisms of the correspondence are themselves general, meaning that they carry over from our example to any other 5d SCFT, whose magnetic quivers have only unitary nodes, without any conceptual changes to the analysis. In sections \ref{sec:DnDn} and \ref{sec:TN} we give more examples, including the UV fixed points of higher CS level SQCD theories, the $T_N$ theory and its parent SCFTs $P_N$ and $G_N$, and new lagrangian descriptions of $P_N$ given by quiver gauge theories with fundamental and anti-symmetric matter.

\subsection{Generalized Toric Polygons and Magnetic Quivers}
\label{sec:GTPintro}

For a strictly convex toric Calabi-Yau singularity, there is a beautiful and complete picture relating the M-/string theory realizations of 5d SCFTs by a one-to-one map, and where both branches of the moduli space can be determined in general from either perspective. That is, starting e.g. from a 5-brane-web in type IIB such as 
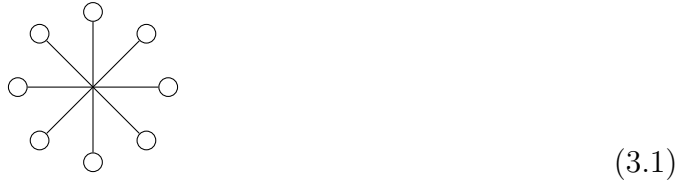
\be
\begin{tikzpicture}
    \node[roundnode] at (-1,0) (1){};
    \node[roundnode] at (1,0) (2){};
    \node[roundnode] at (-.71,.71) (3){};
    \node[roundnode] at (.71,.71) (4){};
    \node[roundnode] at (-.71,-.71) (5){};
    \node[roundnode] at (.71,-.71) (6){};
    \node[roundnode] at (0,1) (7){};
    \node[roundnode] at (0,-1) (8){};
    
    \draw (1)--(2);
    \draw (3)--(6);
    \draw (4)--(5);
    \draw (7)--(8);
\end{tikzpicture}
\ee
the moduli space of the corresponding 5d theory can be studied in terms of its planar and transverse deformations. The brane-web is dual to a strictly convex toric polygon, whose resolutions and deformations are known in general \cite{Altmann,Altmann2} and equivalently map out the moduli space of the 5d theory. In particular, the one-to-one map of the type IIB and M-theory representations associates to each face (compact and non-compact) in the brane-web a vertex in the polygon, and every 5-brane is related to an edge drawn transverse to the brane:
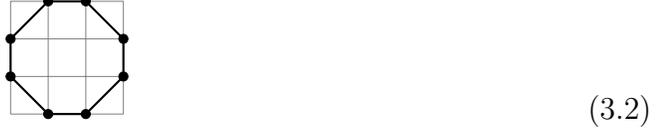
\be 
\begin{tikzpicture}[x=.5cm,y=.5cm]
	\draw[step=.5cm,gray,very thin] (0,-1) grid (3,2);
	
	\draw[ligne,black] (0,0)--(1,-1)--(2,-1)--(3,0)--(3,1)--(2,2)--(1,2)--(0,1)--(0,0);
	
	\node[bd] at (0,0) {};
	\node[bd] at (1,-1) {};
	\node[bd] at (2,-1) {};
	\node[bd] at (3,0) {};
	\node[bd] at (3,1) {};
	\node[bd] at (2,2) {};
	\node[bd] at (1,2) {};
	\node[bd] at (0,1) {};
\end{tikzpicture}
\ee

That being said, many 5d SCFTs have brane-web descriptions whose duals do not fall into the class described above or even the class of toric polygons. In general, the brane-web may contain a sector where more than a single 5-brane ends on the same 7-brane, for which the dual diagram is a Generalized Toric Polygon: a convex polygon, describing an SCFT, that can have lattice points along its edges that are {\it not} vertices.
Where a gauge theory description is available, this phase is represented by a non-convex lattice polygon with a ruling, called a pre-GTP, to which we can apply a (non-unique) series of monodromy-preserving edge-moves \cite{vanBeest:2020civ}, dual to Hanany-Witten brane creation \cite{Hanany:1996ie}, to cure the non-convexity. We then go to the strong coupling point by removing the ruling, and, in this way, arrive at a GTP. 

One such theory is $SU(N)_1+2N\mathbf{F}$ whose pre-GTP and brane-web realizations are
 \be
 
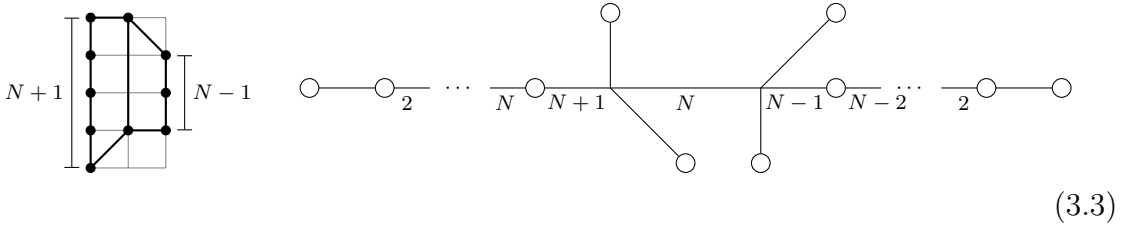
\label{pre-GTPex}
 \begin{tikzpicture}[x=.5cm,y=.5cm]
 	\draw[step=.5cm,gray,very thin] (0,0) grid (2,4);
 	\draw[ligne,black] (0,0)--(1,1)--(2,1)--(2,3)--(1,4)--(0,4)--(0,0);
 	\draw[ligne,black] (1,1)--(1,4);
 	\foreach \x in {0,1,...,4}
 	\node[bd] at (0,\x) {};
 	\foreach \x in {1,2,3}
	\node[bd] at (2,\x) {};
	\node[bd] at (1,4) {};
	\node[bd] at (1,1) {};
	
	\node[] at (-1.5,2) {{\scriptsize $N+1$}};
	\draw[|-|] (-.5,0)--(-.5,4); 				
	\node[] at (3.5,2) {{\scriptsize $N-1$}};
	\draw[|-|] (2.5,1)--(2.5,3); 
 \end{tikzpicture}\hspace{.5cm}
 \begin{tikzpicture}
    \draw (-4,0)--(-2.4,0);
    \draw (-1.6,0)--(3.6,0);
    \node[] at (-2,0) {{\scriptsize $\cdots$}};
    \node[] at (4,0) {{\scriptsize $\cdots$}};
    \draw (4.4,0)--(6,0);
    \foreach \x in {1,3,4}
    \node[roundnode] at (-\x,0) {};
    \foreach \x in {1,3,4}
    \node[roundnode] at (\x+2,0) {};
    \node[roundnode] at (0,1) (1){};
    \node[roundnode] at (3,1) (2){};
    \node[roundnode] at (1,-1) (3){};
    \node[roundnode] at (2,-1) (4){};
    \draw (4)--(2,0)--(2);
    \draw (1)--(0,0)--(3);
    \node[] at (-2.7,-.2) {{\scriptsize $2$}};
    \node[] at (-1.4,-.2) {{\scriptsize $N$}};
    \node[] at (-.45,-.2) {{\scriptsize $N+1$}};
    \node[] at (1,-.2) {{\scriptsize $N$}};
    \node[] at (2.45,-.2) {{\scriptsize $N-1$}};
    \node[] at (3.55,-.2) {{\scriptsize $N-2$}};
    \node[] at (4.7,-.2) {{\scriptsize $2$}};
\end{tikzpicture}
 \ee
where we have drawn the pre-GTP for $N=3$. The label indicates the length of the edge for general $N$, if it is greater than 1. Likewise, in the web, we note the number of individual branes in a stack next to the segment representing it, if it is greater than 1. We take the strong coupling limit of $SU(N)_1+2N\mathbf{F}$ by performing an edge-move to remedy the non-convexity in the pre-GTP and subsequently removing the ruling. Equivalently, we can perform a Hanany-Witten move in the web to arrive at
 \be
 \label{GTPex}
 \begin{tikzpicture}[x=.5cm,y=.5cm]
 	\draw[step=.5cm,gray,very thin] (0,0) grid (2,4);
	
 	\draw[ligne,black] (0,0)--(2,0)--(2,3)--(1,4)--(0,4)--(0,0);

 	\foreach \x in {0,1,...,4}
 	\node[bd] at (0,\x) {};
 	\foreach \x in {0,1,2,3}
	\node[bd] at (2,\x) {};
	\node[bd] at (1,4) {};
	\node[wd] at (1,0) {};
	
	\node[] at (-1.5,2) {{\scriptsize $N+1$}};
	\draw[|-|] (-.5,0)--(-.5,4); 				
	\node[] at (3,2) {{\scriptsize $N$}};
	\draw[|-|] (2.5,0)--(2.5,3); 
	
	\node at (-.5,4.3) {{\scriptsize $\mathbf{v}_0$}};
 \end{tikzpicture}\hspace{1cm}
 \begin{tikzpicture}
    \draw (-4,0)--(-2.4,0);
    \draw (-1.6,0)--(1.6,0);
    \node[] at (-2,0) {{\scriptsize $\cdots$}};
    \node[] at (2,0) {{\scriptsize $\cdots$}};
    \draw (2.4,0)--(4,0);
    \foreach \x in {1,3,4}
    \node[roundnode] at (-\x,0) {};
    \foreach \x in {1,3,4}
    \node[roundnode] at (\x,0) {};
    \node[roundnode] at (0,1) (1){};
    \node[roundnode] at (1,1) (2){};
    \node[roundnode] at (0,-1) (3){};
    \draw (1)--(0,0)--(2);
    \draw (0,0)--(3);

    \node[] at (-2.7,-.2) {{\scriptsize $2$}};
    \node[] at (-1.4,-.2) {{\scriptsize $N$}};
    \node[] at (-.45,-.2) {{\scriptsize $N+1$}};
    \node[] at (.45,-.2) {{\scriptsize $N$}};
    \node[] at (1.55,-.2) {{\scriptsize $N-1$}};
    \node[] at (2.7,-.2) {{\scriptsize $2$}};
    \node[] at (.2,-.6) {{\scriptsize $2$}};
\end{tikzpicture}
 \ee
Notice the empty vertex on the bottom edge of the GTP, dual to the two NS5-branes ending on a single $(0,1)$-7-brane.

A GTP is specified by a set of edges $E_\alpha$ in a $\Z^2$ lattice, which we refer to as having length  $\lambda_\alpha=\text{gcd}(E_\alpha)$. Each edge is equipped with a partition $\mu_\alpha$ of $\lambda_\alpha$ ordered in descending magnitude, which carries information about the filled and empty vertices. For the example in \eqref{GTPex} these are 
\be 
\ba 
\label{eq:GTPexdata}
E_\alpha&=((0,-N-1),(2,0),(0,N),(-1,1),(-1,0))\,, \\ \lambda_\alpha&=(N+1,2,N,1,1)\,, \qquad  \mu_\alpha=(\{1^{N+1}\},\{2\},\{1^N\},\{1\},\{1\})\,,
\ea 
\ee
where we label the edges starting from $\mathbf{v}_0$ and moving counter-clockwise.
Note that toric polygons have only minimal partitions in terms of dominance ordering.

It is not known how or whether one can associate a Calabi-Yau singularity to a generic GTP, and consequently a general geometric analysis of the deformation space remains out of reach. However, the duality with the brane-web extends to the decomposition that determines the magnetic quiver, generalizing 
the work of \cite{Altmann,Altmann2} to include the non-minimal partitions $\mu_\alpha$.
We refer the reader to \cite{vanBeest:2020kou} for details of the algorithm but summarize the main points here. The map from GTP to magnetic quiver relies on an edge coloring, given in terms of a refined Minkowski sum $P_a \oplus P_b$ defined such that the edges agree with those of an ordinary Minkowski sum and 
\be
	\mu_\alpha^{P_a \oplus P_b} = \mu_\alpha^{P_a} + \mu_\alpha^{P_b}\,.
\ee
The edge coloring divides the GTP into closed sub-polygons $S^c$, where $c$ counts the number of colors. Each sub-polygon $S^c$ is some higher multiple of an irreducible polygon that obeys the s-rule minimally. A polygon is said to be irreducible if it cannot be written as a refined Minkowski sum of two other polygons, whereas minimality implies that we cannot remove any vertices, and replace them with a white dot, without violating the s-rule. The colored sub-polygons $S^c$ carry their own partitions $\mu_\alpha^c$ defined analogously to the partitions of the GTP which obey
\be 
\{\mu_\alpha\} \leq \sum_c \{ \mu_\alpha^c\}\,,
\ee 
where the inequality is the dominance partial ordering. 
A GTP may have several consistent edge colorings, each of which will give rise to a magnetic quiver comprising a cone of the Higgs branch.

Let us consider the strong coupling limit of $SU(N)_1+2N\mathbf{F}$ again, which has a unique edge coloring given by
\be
 \label{GTPexcolor}
 \begin{tikzpicture}[x=.5cm,y=.5cm]
 	\draw[step=.5cm,gray,very thin] (0,0) grid (2,4);
	
 	\draw[ligne,ForestGreen] (2,0)--(2,3);
 	\draw[ligne,ForestGreen] (0,4)--(0,1);
 	\draw[ligne,cyan] (0,0)--(1,0);
 	\draw[ligne,cyan] (0,4)--(1,4);
 	\draw[ligne,blue] (1,0)--(2,0);
 	\draw[ligne,blue] (0,0)--(0,1);
 	\draw[ligne,blue] (1,4)--(2,3);

 	\foreach \x in {0,1,...,4}
 	\node[bd] at (0,\x) {};
 	\foreach \x in {0,1,2,3}
	\node[bd] at (2,\x) {};
	\node[bd] at (1,4) {};
	\node[wd] at (1,0) {};
	
	\node[] at (-1.5,2) {{\scriptsize $N+1$}};
	\draw[|-|] (-.5,0)--(-.5,4); 				
	\node[] at (3,2) {{\scriptsize $N$}};
	\draw[|-|] (2.5,0)--(2.5,3); 
	\node at (-.5,4.3) {{\scriptsize $\mathbf{v}_0$}};
	
	\node[] at (4,2) {$<$};
	\draw[ligne,blue] (5,1.5)--(6,1.5)--(5,2.5)--(5,1.5);
	\node[bd] at (5,1.5) {};
	\node[bd] at (6,1.5) {};
	\node[bd] at (5,2.5) {};
	\node[] at (7,2) {$\oplus$};
	\draw[ligne, cyan] (8,2)--(9,2);
	\node[bd] at (8,2) {};
	\node[bd] at (9,2) {};
	\node[] at (10,2) {$\oplus$};
	\draw[ligne,ForestGreen] (11,.5)--(11,3.5);
	\node[bd] at (11,.5) {};
	\node[bd] at (11,3.5) {};
	\node[wd] at (11,1.5) {};
	\node[wd] at (11,2.5) {};
	
	\node[] at (12,2) {{\scriptsize $N$}};
	\draw[|-|] (11.5,.5)--(11.5,3.5); 
 \end{tikzpicture}
 \ee 
where $\oplus$ is the refined Minkowski sum, and the blue $S^b$, turquoise $S^t$ and green $S^g$ sub-polygons are specified by
 \be 
 \ba 
 \lambda_\alpha^b&=(1,1,0,1,0)\,, \qquad \mu_\alpha^b=(\{1\},\{1\},-,\{1\},-)\,,\\
 \lambda_\alpha^t&=(0,1,0,0,1)\,, \qquad \mu_\alpha^t=(-,\{1\},-,-,\{1\})\,,\\
 \lambda_\alpha^g&=(N,0,N,0,0)\,, \qquad \mu_\alpha^g=(\{N\},-,\{N\},-,-)\,.
 \ea 
 \ee
Once all valid edge colorings have been identified, the magnetic quiver is obtained by associating to each sub-polygon $S^c$ a $U(m^c)$ node with $m^c=\text{gcd}(\lambda^c_\alpha)$. Furthermore, vertices along an edge $E_\alpha$ (i.e. non-corner vertices) give rise to tail $U(m_{\alpha,x})$ nodes in the magnetic quiver with
\be 
\label{eq:tailm}
m_{\alpha,x}=\sum_{y=1}^x \left( \sum_c \mu^c_{\alpha,y}-\mu_{\alpha,y} \right)\,.
\ee 
Tail nodes pertaining to the same edge intersect their nearest neighbour once. The color node intersections receive a positive contribution from the mixed volume of the associated sub-polygons and a negative contribution from the colors sharing an edge,
\be 
\label{eq:ccint}
k^{c_1,c_2}=\frac{1}{m^{c_1}m^{c_2}} \left( MV(S^{c_1},S^{c_2})-\sum_{\alpha,x}\mu^{c_1}_{\alpha,x}\mu^{c_2}_{\alpha,x}\right)\,.
\ee 
A color node and tail node intersect as
\be 
\label{eq:ctailint}
k^c_{\alpha,x}=\frac{1}{m^c}\left( \mu_{\alpha,x}^c-\mu_{\alpha,x+1}^c \right)\,.
\ee 
For the example above, 
\be 
\ba
m^b&=1\,, \quad m^t=1\,, \quad m^g=N\,, \\ m_{1,x}&=(N,N-1,\dots,1)\,, \quad m_{3,x}=(N-1,N-2,\dots,1)\,,
\ea 
\ee 
and
\be
k^{tg}=1\,, \quad k^g_{1,1}=k^g_{3,1}=1\,, \quad k^b_{1,1}=1\,,
\ee 
with all other vanishing. Hence, the magnetic quiver is
\be 
\begin{tikzpicture}
	\node (g2) at (4,0) [gauge,label={[yshift=3.2]below:{\scriptsize $1$}}] {};
	\node (space1) at (4.7,0) {};
	\node (dots) at (5,0) {$\cdots$};
	\node (space2) at (5.3,0) {};
	\node (g3) at (6,0) [gauge,label={[yshift=3.2]below:{\scriptsize $N-1$}}] {};
	\node (g7) at (7,0) [gauge,label={[yshift=3.2]below:{\scriptsize $N$}}] {};
	\node (g8) at (8,0) [gauge,draw=ForestGreen,label={[yshift=3.2]below:{\scriptsize $N$}}] {};
	\node (g4) at (9,0) [gauge,label={[yshift=3.2]below:{\scriptsize $N-1$}}] {};
	\node (space3) at (9.7,0) {};
	\node (dots2) at (10,0) {$\cdots$};
	\node (space4) at (10.3,0) {};
	\node (g11) at (11,0) [gauge,label={[yshift=3.2]below:{\scriptsize $1$}}] {};
	\node (g12) at (7,1) [gauge,draw=blue,label={[xshift=3.2]left:{\scriptsize $1$}}] {};
	\node (g13) at (8,1) [gauge,draw=cyan,label={[xshift=-3.2]right:{\scriptsize $1$}}] {};
	\draw[gline] (g2)--(space1);
	\draw[gline] (space2)--(g3)--(g7)--(g8)--(g4)--(space3);
	\draw[gline] (space4)--(g11);
	\draw[gline] (g7)--(g12);
	\draw[gline] (g8)--(g13);
\end{tikzpicture}
\ee 
The flavor symmetry algebra can be obtained from the magnetic quiver as the Dynkin diagram(s) composed of the balanced nodes, with the balance of the $i$'th $U(m_i)$ node given by
\be 
\beta_i=-2m_i+2\ell_i(m_i-1)+\sum_{j\neq i}k_{ij}m_j\,,
\ee 
with $\ell_i$ the number of adjoint loops and $k_{ij}$ bifundamentals between the $i$'th and $j$'th node.
The $U(1)$ factors are given by the number of overbalanced $U(1)$ nodes in the magnetic quiver minus a diagonal $U(1)$. The flavor symmetry algebra of strongly coupled $SU(N)_1+2N\mathbf{F}$ is
\be 
\mathfrak{su}(2N+1) \oplus \mathfrak{u}(1)\,.
\ee

\subsection{Decoupling} 

5d SCFTs can be deformed away from the fixed point either by turning on a gauge coupling or a mass parameter. The latter decouples a matter multiplet from the theory, which will then flow to a new SCFT. In this way, the 5d SCFTs fall into trees of descendants related by mass deformations. In the M-theory realization, decoupling is implemented geometrically by performing flop transitions on curves in the geometry \cite{Apruzzi:2019opn}. In \cite{vanBeest:2020civ} it was shown that the application for toric models in \cite{Eckhard:2020jyr} can be directly generalized: Decoupling can be achieved in the pre-GTP by flopping a curve, which connects an internal vertex along the ruling to an external vertex. Sending the mass to infinity corresponds to removing the section of the polygon that the matter curve separates from the gauge nodes along the ruling after the flop.
As we perform the edge-moves necessary to arrive at the SCFT point, we can keep track of the matter curve, identify the corresponding curve in the GTP, and decouple the associated sector. 
We discuss how a decoupling curve in the GTP is associated to a (non-unique) FI deformation in the magnetic quiver, so that the deformation extrapolates between the Higgs branches before and after decoupling. 

We study the UV completion of $SU(N)$ with CS level $1$ and $2N$ fundamentals, with GTP $P$, and consider the theories resulting from sending a hypermultiplet mass to $m \rightarrow \pm \infty$, namely $SU(N)$ with $2N-1$ fundamentals and CS level $1/2$ or $3/2$. We label the GTPs of the descendants by their CS levels. Let us first consider the transition to $SU(N)_{3/2}+(2N-1)\mathbf{F}$. The decoupling curve is indicated in $P$ in red, and the descendant and magnetic quivers are
\be
\label{GTPSU3_3/2}
 \begin{tikzpicture}[x=.5cm,y=.5cm]
 	\draw[step=.5cm,gray,very thin] (4,0) grid (6,4);
	\node at (.5,2) {$P:$};
 	\draw[ligne,ForestGreen] (6,0)--(6,3);
 	\draw[ligne,ForestGreen] (4,4)--(4,1);
 	\draw[ligne,cyan] (4,0)--(5,0);
 	\draw[ligne,cyan] (4,4)--(5,4);
 	\draw[ligne,blue] (5,0)--(6,0);
 	\draw[ligne,blue] (4,0)--(4,1);
 	\draw[ligne,blue] (5,4)--(6,3);
 	\draw[red] (4,3)--(5,4);

 	\foreach \x in {0,1,...,4}
 	\node[bd] at (4,\x) {};
 	\foreach \x in {0,1,2,3}
	\node[bd] at (6,\x) {};
	\node[bd] at (5,4) {};
	\node[wd] at (5,0) {};
	
	\node[] at (2.5,1.5) {{\scriptsize $N+1$}};
	\draw[|-|] (3.5,0)--(3.5,4); 				
	\node[] at (7,1.5) {{\scriptsize $N$}};
	\draw[|-|] (6.5,0)--(6.5,3); 
	\node at (3.5,4.3) {{\scriptsize $\mathbf{v}_0$}};
	
	\draw[step=.5cm,gray,very thin] (17,0) grid (19,4);
	
	\node at (14,2) {$P_{3/2}:$};
	\node at (16.5,3.3) {{\scriptsize $\mathbf{v}_0$}};
 	\draw[ligne,ForestGreen] (19,0)--(19,2);
 	\draw[ligne,ForestGreen] (17,3)--(17,1);
 	\draw[ligne,Maroon] (17,0)--(18,0);
 	\draw[ligne,Maroon] (19,2)--(19,3);
 	\draw[ligne,blue] (18,0)--(19,0);
 	\draw[ligne,blue] (18,4)--(19,3);
 	\draw[ligne,blue] (17,0)--(17,1);
 	\draw[ligne,Maroon] (17,3)--(18,4);

 	\foreach \x in {0,1,...,3}
 	\node[bd] at (17,\x) {};
 	\foreach \x in {0,1,2,3}
	\node[bd] at (19,\x) {};
	\node[bd] at (18,4) {};
	\node[wd] at (18,0) {};
	
	\node[] at (16,1.5) {{\scriptsize $N$}};
	\draw[|-|] (16.5,0)--(16.5,3); 
	\node[] at (20,1.5) {{\scriptsize $N$}};
	\draw[|-|] (19.5,0)--(19.5,3);

	\node at (0,0) {{\color{white} .}};	
	\node at (24,0) {{\color{white} .}};
	\end{tikzpicture}
 \ee 
\be 
\label{MQSU3_3/2}
\begin{tikzpicture}
	\node (g2) at (0,0) [gauge,label={[yshift=3.2]below:{\scriptsize $1$}}] {};
	\node (space1) at (.7,0) {};
	\node (dots) at (1,0) {$\cdots$};
	\node (space2) at (1.3,0) {};
	\node (g7) at (2,0) [gauge,label={[yshift=3.2]below:{\scriptsize $N$}}] {};
	\node (g8) at (3,0) [gauge,draw=ForestGreen,label={[yshift=3.2]below:{\scriptsize $N$}}] {};
	\node (space3) at (3.7,0) {};
	\node (dots2) at (4,0) {$\cdots$};
	\node (space4) at (4.3,0) {};
	\node (g11) at (5,0) [gauge,label={[yshift=3.2]below:{\scriptsize $1$}}] {};
	\node (g12) at (2,1) [gauge,draw=blue,label={[xshift=3.2]left:{\scriptsize $1$}}] {};
	\node (g13) at (3,1) [gauge,draw=cyan,label={[xshift=-3.2]right:{\scriptsize $1$}}] {};
	\draw[gline] (g2)--(space1);
	\draw[gline] (space2)--(g7)--(g8)--(space3);
	\draw[gline] (space4)--(g11);
	\draw[gline] (g7)--(g12);
	\draw[gline] (g8)--(g13);
	
	\node at (2.5,-1) {{\scriptsize $\su (2N+1) \oplus \uu (1)$}};
	
	\node (g1) at (6,0) [gauge,label={[yshift=3.2]below:{\scriptsize $1$}}] {};
	\node (space1) at (6.7,0) {};
	\node (dots) at (7,0) {$\cdots$};
	\node (space2) at (7.3,0) {};
	\node (g2) at (8,0) [gauge,label={[yshift=3.2]below:{\scriptsize $N-1$}}] {};
	\node (g3) at (9,0) [gauge,draw=ForestGreen,label={[yshift=3.2]below:{\scriptsize $N-1$}}] {};
	\node (g4) at (10,0) [gauge,label={[yshift=3.2]below:{\scriptsize $N-1$}}] {};
	\node (space3) at (10.7,0) {};
	\node (dots) at (11,0) {$\cdots$};
	\node (space4) at (11.3,0) {};
	\node (g5) at (12,0) [gauge,label={[yshift=3.2]below:{\scriptsize $1$}}] {};
	\node (g6) at (8,1) [gauge,draw=blue,label={[xshift=3.2]left:{\scriptsize $1$}}] {};
	\node (g7) at (10,1) [gauge,draw=Maroon,label={[xshift=-3.2]right:{\scriptsize $1$}}] {};
	\draw[gline] (g1)--(space1);
	\draw[gline] (space2)--(g2)--(g3)--(g4)--(space3);
	\draw[gline] (space4)--(g5);
	\draw[gline] (g2)--(g6);
	\draw[gline] (g4)--(g7);
	\draw[gline] (g6)--(g7);
	
	\node at (9,-1) {{\scriptsize $\su (2N) \oplus \uu (1)$}};
\end{tikzpicture}
\ee  
Decoupling the triangle $(\mathbf{v}_0,\mathbf{v}_1,\mathbf{v}_{2N+3})$, defined by the matter curve, changes the edges and partitions from \eqref{eq:GTPexdata} to 
\be 
\ba 
E_\beta&=((0,N),(2,0),(0,N),(-1,1),(-1,-1))\,,\\
\mu_\beta&=(\{1^N\},\{2\},\{1^N\},\{1\},\{1\})\,,
\ea 
\ee 
with $\beta=1,...,4,\text{int}$.
The effect of the decoupling on the map to the magnetic quiver can be described as follows. Shortening $E_1$ by 1 segment reduces the rank of the tail nodes associated to $E_1$ by 1 along the entire tail. 
The intersections of the blue and green nodes with one another and the tail nodes are preserved by the internal edge, but the rank of the green node is reduced by 1. The turquoise sub-polygon does not appear in the Minkowski sum decomposition of $P_{3/2}$. Instead the decomposition of $P_{3/2}$ contains the new triangle shown in maroon in \eqref{GTPSU3_3/2}. The appearance of the maroon triangle gives an new $U(1)$ node, which intersects with the blue triangle and the first tail node from $E_3$. 
The FI deformation that accomplishes these changes in the magnetic quiver is
\be 
\begin{tikzpicture}
\label{MQSU3_decdef+}
	\node (g2) at (0,0) [gauger,label={[yshift=3.2]below:{\scriptsize $1$}}] {};
	\node (space1) at (.7,0) {};
	\node (dots) at (1,0) {$\cdots$};
	\node (space2) at (1.3,0) {};
	\node (g7) at (2,0) [gauge,label={[yshift=3.2]below:{\scriptsize $N$}}] {};
	\node (g8) at (3,0) [gauge,draw=ForestGreen,label={[yshift=3.2]below:{\scriptsize $N$}}] {};
	\node (space3) at (3.7,0) {};
	\node (dots2) at (4,0) {$\cdots$};
	\node (space4) at (4.3,0) {};
	\node (g11) at (5,0) [gauge,label={[yshift=3.2]below:{\scriptsize $1$}}] {};
	\node (g12) at (2,1) [gauge,draw=blue,label={[xshift=3.2]left:{\scriptsize $1$}}] {};
	\node (g13) at (3,1) [gauger,draw=cyan,label={[xshift=-3.2]right:{\scriptsize $1$}}] {};
	\draw[gline] (g2)--(space1);
	\draw[gline] (space2)--(g7)--(g8)--(space3);
	\draw[gline] (space4)--(g11);
	\draw[gline] (g7)--(g12);
	\draw[gline] (g8)--(g13);
\end{tikzpicture}
\ee  
We reach the other descendant SCFT by partially resolving and decoupling as
\be
\label{GTPSU3_1/2}
 \begin{tikzpicture}[x=.5cm,y=.5cm]
 	\draw[step=.5cm,gray,very thin] (0,0) grid (2,4);
	\node at (-3.5,2) {$P:$};
	\node at (-.5,4.3) {{\scriptsize $\mathbf{v}_0$}};
 	\draw[ligne,ForestGreen] (2,0)--(2,3);
 	\draw[ligne,ForestGreen] (0,3)--(0,0);
 	\draw[ligne,cyan] (0,0)--(1,0);
 	\draw[ligne,cyan] (0,4)--(1,4);
 	\draw[ligne,blue] (1,0)--(2,0);
 	\draw[ligne,blue] (0,3)--(0,4);
 	\draw[ligne,blue] (1,4)--(2,3);
 	\draw[red] (0,1)--(2,1);

 	\foreach \x in {0,1,...,4}
 	\node[bd] at (0,\x) {};
 	\foreach \x in {0,1,2,3}
	\node[bd] at (2,\x) {};
	\node[bd] at (1,4) {};
	\node[wd] at (1,0) {};
	\node[rd] at (1,1) {};
	
	\draw[step=.5cm,gray,very thin] (12,1) grid (14,4);
	
	\node at (9,2) {$P_{1/2}:$};
	\node at (11.5,4.3) {{\scriptsize $\mathbf{v}_0$}};
 	\draw[ligne,ForestGreen] (14,1)--(14,3);
 	\draw[ligne,ForestGreen] (12,3)--(12,1);
 	\draw[ligne,cyan] (12,4)--(13,4);
 	\draw[ligne,blue] (13,4)--(14,3);
 	\draw[ligne,blue] (12,3)--(12,4);
 	\draw[ligne,blue] (13,1)--(14,1);
 	\draw[ligne,cyan] (12,1)--(13,1);

 	\foreach \x in {1,2,...,4}
 	\node[bd] at (12,\x) {};
 	\foreach \x in {1,2,3}
	\node[bd] at (14,\x) {};
	\node[bd] at (13,4) {};
	\node[bd] at (13,1) {};
	
	\node[] at (-1.5,2) {{\scriptsize $N+1$}};
	\draw[|-|] (-.5,0)--(-.5,4); 				
	\node[] at (3,2) {{\scriptsize $N$}};
	\draw[|-|] (2.5,0)--(2.5,3); 
	\node[] at (11,2) {{\scriptsize $N$}};
	\draw[|-|] (11.5,1)--(11.5,4); 
	\node[] at (15.5,2) {{\scriptsize $N-1$}};
	\draw[|-|] (14.5,1)--(14.5,3); 
	
	\node at (-4,0) {{\color{white} .}};
	\node at (18,0) {{\color{white} .}};
	\end{tikzpicture}
 \ee 
\be 
\label{MQSU3_1/2}
\begin{tikzpicture}
	\node (g2) at (0,0) [gauge,label={[yshift=3.2]below:{\scriptsize $1$}}] {};
	\node (space1) at (.7,0) {};
	\node (dots) at (1,0) {$\cdots$};
	\node (space2) at (1.3,0) {};
	\node (g7) at (2,0) [gauge,label={[yshift=3.2]below:{\scriptsize $N$}}] {};
	\node (g8) at (3,0) [gauge,draw=ForestGreen,label={[yshift=3.2]below:{\scriptsize $N$}}] {};
	\node (space3) at (3.7,0) {};
	\node (dots2) at (4,0) {$\cdots$};
	\node (space4) at (4.3,0) {};
	\node (g11) at (5,0) [gauge,label={[yshift=3.2]below:{\scriptsize $1$}}] {};
	\node (g12) at (2,1) [gauge,draw=blue,label={[xshift=3.2]left:{\scriptsize $1$}}] {};
	\node (g13) at (3,1) [gauge,draw=cyan,label={[xshift=-3.2]right:{\scriptsize $1$}}] {};
	\draw[gline] (g2)--(space1);
	\draw[gline] (space2)--(g7)--(g8)--(space3);
	\draw[gline] (space4)--(g11);
	\draw[gline] (g7)--(g12);
	\draw[gline] (g8)--(g13);
	\node at (2.5,-1) {{\scriptsize $\su (2N) \oplus \uu (1)$}};
	
	\node (g2) at (6,0) [gauge,label={[yshift=3.2]below:{\scriptsize $1$}}] {};
	\node (space1) at (6.7,0) {};
	\node (dots) at (7,0) {$\cdots$};
	\node (space2) at (7.3,0) {};
	\node (g7) at (8,0) [gauge,label={[yshift=3.2]below:{\scriptsize $N-1$}}] {};
	\node (g8) at (9,0) [gauge,draw=ForestGreen,label={[yshift=3.2]below:{\scriptsize $N-1$}}] {};
	\node (space3) at (9.7,0) {};
	\node (dots2) at (10,0) {$\cdots$};
	\node (space4) at (10.3,0) {};
	\node (g11) at (11,0) [gauge,label={[yshift=3.2]below:{\scriptsize $1$}}] {};
	\node (g12) at (8,1) [gauge,draw=blue,label={[xshift=3.2]left:{\scriptsize $1$}}] {};
	\node (g13) at (9,1) [gauge,draw=cyan,label={[xshift=-3.2]right:{\scriptsize $1$}}] {};
	\node (g14) at (8.5,1.7) [gauge,,label={[yshift=-3.2]above:{\scriptsize $1$}}] {};
	\draw[gline] (g2)--(space1);
	\draw[gline] (space2)--(g7)--(g8)--(space3);
	\draw[gline] (space4)--(g11);
	\draw[gline] (g7)--(g12)--(g14)--(g13)--(g8);
	\node at (8.5,-1) {{\scriptsize $\su (2N-1) \oplus \su (2) \oplus \uu (1)$}};
\end{tikzpicture}
\ee 
The new GTP is characterized by the data
\be 
\ba 
E_\gamma&=((0,-N),(2,0),(0,N-1),(-1,1),(-1,0))\,,\\
\mu_\gamma&=(\{1^N\},\{1^2\},\{1^{N-1}\},\{1\},\{1\})\,,
\ea 
\ee 
with $\gamma=1,\text{int},3,4,5$. Decoupling the sector below the matter curve has the effect of shortening both tails by 1, and decreasing the rank of the green sub-polygon by $1$. Furthermore, the vertex at the bottom of the polygon is filled in, producing an additional length 1 tail, connected to the blue and turqoise nodes. This is precisely the effect of the FI deformation
\be 
\begin{tikzpicture}
\label{MQSU3_decdef-}
	\node (g2) at (0,0) [gauger,label={[yshift=3.2]below:{\scriptsize $1$}}] {};
	\node (space1) at (.7,0) {};
	\node (dots) at (1,0) {$\cdots$};
	\node (space2) at (1.3,0) {};
	\node (g7) at (2,0) [gauge,label={[yshift=3.2]below:{\scriptsize $N$}}] {};
	\node (g8) at (3,0) [gauge,draw=ForestGreen,label={[yshift=3.2]below:{\scriptsize $N$}}] {};
	\node (space3) at (3.7,0) {};
	\node (dots2) at (4,0) {$\cdots$};
	\node (space4) at (4.3,0) {};
	\node (g11) at (5,0) [gauger,label={[yshift=3.2]below:{\scriptsize $1$}}] {};
	\node (g12) at (2,1) [gauge,draw=blue,label={[xshift=3.2]left:{\scriptsize $1$}}] {};
	\node (g13) at (3,1) [gauge,draw=cyan,label={[xshift=-3.2]right:{\scriptsize $1$}}] {};
	\draw[gline] (g2)--(space1);
	\draw[gline] (space2)--(g7)--(g8)--(space3);
	\draw[gline] (space4)--(g11);
	\draw[gline] (g7)--(g12);
	\draw[gline] (g8)--(g13);
\end{tikzpicture}
\ee

\subsection{Weak Coupling}

The SCFTs that we consider admit a weakly coupled gauge theory description. This is reached from the UV by turning on a Super Yang-Mills term for a gauge group, which is a relevant operator and thus triggers an RG flow. In the GTP we go to the gauge theory phase by identifying the ruling, whose vertices correspond to the Cartan subgroups of the gauge group.
A UV fixed point often admits of multiple gauge theory descriptions. When a set of 5d gauge theories have the same SCFT fixed point we say that they are UV dual to each other.

Let us consider the gauge theory descriptions of the example SCFT with GTP $P$. It admits a ruling associated with the $SU(N)_1+2N\mathbf{F}$ gauge theory, as well as another ruling giving rise to $[1]-SU(2)^{N-1}-[3]$.
We denote the GTPs of the two gauge theory phases by $P_{SU(N)}$ and $P_{SU(2)^{N-1}}$, respectively. For the latter the ruling consists of $N-1$ internal edges, corresponding to consecutively turning on a SYM coupling for the $N-1$ gauge groups. The GTPs and magnetic quivers are given by
\be
\label{GTPSUN_weak}
 \begin{tikzpicture}[x=.5cm,y=.5cm]
 	\draw[step=.5cm,gray,very thin] (0,0) grid (2,4);
	\node at (-4,2) {$P_{SU(N)}:$};
 	\draw[ligne,ForestGreen] (2,1)--(2,3);
 	\draw[ligne,ForestGreen] (0,3)--(0,1);
 	\draw[ligne,ForestGreen] (1,1)--(1,3);
 	\draw[ligne,cyan] (0,1)--(0,0)--(1,0);
 	\draw[ligne,cyan] (0,4)--(1,4);
 	\draw[ligne,cyan] (2,0)--(2,1);
 	\draw[ligne,cyan] (0,0)--(1,1);
 	\draw[ligne,blue] (1,0)--(2,0);
 	\draw[ligne,blue] (0,3)--(0,4);
 	\draw[ligne,blue] (2,3)--(1,4)--(1,3);
 	\draw[ligne,blue] (1,1)--(2,1);

 	\foreach \x in {0,1,...,4}
 	\node[bd] at (0,\x) {};
 	\foreach \x in {1,2,3}
	\node[bd] at (1,\x) {};
 	\foreach \x in {0,1,2,3}
	\node[bd] at (2,\x) {};
	\node[bd] at (1,4) {};
	\node[wd] at (1,0) {};
	
	\draw[step=.5cm,gray,very thin] (12,0) grid (14,4);
	
	\node at (7.5,2) {$P_{SU(2)^{N-1}}:$};
 	\draw[ligne,Maroon] (14,2)--(14,3);
 	\draw[ligne,Maroon] (12,3)--(12,2);
 	\draw[ligne,cyan] (12,0)--(13,0);
 	\draw[ligne,cyan] (12,4)--(13,4);
 	\draw[ligne,blue] (13,0)--(14,0);
 	\draw[ligne,blue] (13,4)--(14,3);
 	\draw[ligne,blue] (12,3)--(12,4);
 	
	\foreach \x in {2,3}
 	\draw[ligne,blue] (13,\x)--(14,\x);
	\foreach \x in {2,3}
 	\draw[ligne,cyan] (12,\x)--(13,\x);
 	
 	\draw[ligne,ForestGreen] (12,0)--(12,2);
 	\draw[ligne,ForestGreen] (14,0)--(14,2);

 	\foreach \x in {0,1,...,4}
 	\node[bd] at (12,\x) {};
 	\foreach \x in {0,1,2,3}
	\node[bd] at (14,\x) {};
	\node[bd] at (13,4) {};
	\node[wd] at (13,0) {};
	\foreach \x in {2,3}
	\node[bd] at (13,\x) {};
	
	\node[] at (-1.5,2) {{\scriptsize $N+1$}};
	\draw[|-|] (-.5,0)--(-.5,4); 				
	\node[] at (3,2) {{\scriptsize $N$}};
	\draw[|-|] (2.5,0)--(2.5,3); 
	\node[] at (10.5,2) {{\scriptsize $N+1$}};
	\draw[|-|] (11.5,0)--(11.5,4); 
	\node[] at (15,2) {{\scriptsize $N$}};
	\draw[|-|] (14.5,0)--(14.5,3); 
	
	\node at (-5,0) {{\color{white} .}};
	\node at (16.25,0) {{\color{white} .}};
	\end{tikzpicture}
 \ee 
\be 
\label{MQSUN_weak}
\begin{tikzpicture}
	\node (g1) at (0,0) [gauge,label={[yshift=3.2]below:{\scriptsize $1$}}] {};
	\node (space1) at (.7,0) {};
	\node (dots) at (1,0) {$\cdots$};
	\node (space2) at (1.3,0) {};	
	\node (g2) at (2,0) [gauge,label={[yshift=3.2]below:{\scriptsize $N-1$}}] {};
	\node (g3) at (3,0) [gauge,label={[yshift=3.2]below:{\scriptsize $N$}}] {};
	\node (g4) at (4,0) [gauge,draw=ForestGreen,label={[yshift=3.2]below:{\scriptsize $N-1$}}] {};
	\node (space3) at (4.7,0) {};
	\node (dots) at (5,0) {$\cdots$};
	\node (space4) at (5.3,0) {};
	\node (g5) at (6,0) [gauge,label={[yshift=3.2]below:{\scriptsize $1$}}] {};
	\node (g6) at (2.5,1) [gauge,draw=cyan,label={[xshift=3.2]left:{\scriptsize $1$}}] {};
	\node (g7) at (3.5,1) [gauge,draw=blue,label={[xshift=-3.2]right:{\scriptsize $1$}}] {};
	\draw[gline] (g1)--(space1);
	\draw[gline] (space2)--(g2)--(g3)--(g4)--(space3);
	\draw[gline] (space4)--(g5);
	\draw[gline] (g3)--(g6);
	\draw[gline] (g3)--(g7);
	\node at (3,-1) {{\scriptsize $\su (2N) \oplus \uu (1)$}};

	\node (g1) at (8,0) [gauge,label={[yshift=3.2]below:{\scriptsize $1$}}] {};
	\node (g2) at (9,0) [gauge,draw=ForestGreen,label={[yshift=3.2]below:{\scriptsize $2$}}] {};
	\node (g3) at (10,0) [gauge,label={[yshift=3.2]below:{\scriptsize $1$}}] {};
	\node (g4) at (8.5,1) [gauge,draw=cyan,label={[xshift=3.2]left:{\scriptsize $1$}}] {};
	\node (g5) at (9.5,1) [gauge,draw=blue,label={[xshift=-3.2]right:{\scriptsize $1$}}] {};
	\node (g6) at (9,1) [gauge,draw=Maroon,label={[yshift=-3.2]above:{\scriptsize $1$}}] {};
	\node (g7) at (9,2.3) [gauge,label={[yshift=-3.2]above:{\scriptsize $1$}}] {};
	\draw[gline] (g1)--(g2)--(g3);
	\draw[gline] (g2)--(g4)--(g6)--(g5)--(g2);
	\draw[gline] (g4)--(g7)--(g5);
	
	\node at (9,1.8) {$\vdots$};
	\draw [decorate,decoration={brace,amplitude=7},xshift=-3]
(8.95,.9)--(8.95,2.4) node [midway,xshift=-23] {\scriptsize $N-2$};
	
	\node at (9,-1) {{\scriptsize $\su (4) \oplus \su (2)^{N-2} \oplus \uu (1)$}};
\end{tikzpicture}
\ee 
The FI deformation that gives rise to the magnetic quiver of the weakly coupled $SU(N)$ gauge theory should reduce the length of the short tail and the rank of the green node by 1, and redistribute the intersections of the turqoise node. This is accomplished by
\be 
\begin{tikzpicture}
\label{MQSUN_weakdef}
	\node (g2) at (0,0) [gauge,label={[yshift=3.2]below:{\scriptsize $1$}}] {};
	\node (space1) at (.7,0) {};
	\node (dots) at (1,0) {$\cdots$};
	\node (space2) at (1.3,0) {};
	\node (g7) at (2,0) [gauge,label={[yshift=3.2]below:{\scriptsize $N$}}] {};
	\node (g8) at (3,0) [gauge,draw=ForestGreen,label={[yshift=3.2]below:{\scriptsize $N$}}] {};
	\node (space3) at (3.7,0) {};
	\node (dots2) at (4,0) {$\cdots$};
	\node (space4) at (4.3,0) {};
	\node (g11) at (5,0) [gauger,label={[yshift=3.2]below:{\scriptsize $1$}}] {};
	\node (g12) at (2,1) [gauge,draw=blue,label={[xshift=3.2]left:{\scriptsize $1$}}] {};
	\node (g13) at (3,1) [gauger,draw=cyan,label={[xshift=-3.2]right:{\scriptsize $1$}}] {};
	\draw[gline] (g2)--(space1);
	\draw[gline] (space2)--(g7)--(g8)--(space3);
	\draw[gline] (space4)--(g11);
	\draw[gline] (g7)--(g12);
	\draw[gline] (g8)--(g13);
\end{tikzpicture}
\ee  
To arrive at the $SU(2)$ quiver gauge theory from the UV, we deform the SCFT by turning on $N-1$ independent gauge couplings, each of which corresponds to an FI deformation in the magnetic quiver. The $N-1$ transitions can be implemented as follows. The top ruling in \eqref{GTPSUN_weak} (right) induces a deformation in the magnetic quiver which is equivalent to \eqref{MQSUN_weakdef}. (Indeed this implies that the Higgs branches of the two theories coincide.) The remaining $N-2$ $SU(2)$ rulings will then simply correspond to successively turning on FI parameters at the $U(1)$ nodes in the tails:
\be 
\begin{tikzpicture}
	\node (g1) at (0,0) [gauger,label={[yshift=3.2]below:{\scriptsize $1$}}] {};
	\node (space1) at (.7,0) {};
	\node (dots) at (1,0) {$\cdots$};
	\node (space2) at (1.3,0) {};	
	\node (g2) at (2,0) [gauge,label={[yshift=3.2]below:{\scriptsize $N-1$}}] {};
	\node (g3) at (3,0) [gauge,label={[yshift=3.2]below:{\scriptsize $N$}}] {};
	\node (g4) at (4,0) [gauge,label={[yshift=3.2]below:{\scriptsize $N-1$}}] {};
	\node (space3) at (4.7,0) {};
	\node (dots) at (5,0) {$\cdots$};
	\node (space4) at (5.3,0) {};
	\node (g5) at (6,0) [gauger,label={[yshift=3.2]below:{\scriptsize $1$}}] {};
	\node (g6) at (2.5,1) [gauge,label={[xshift=3.2]left:{\scriptsize $1$}}] {};
	\node (g7) at (3.5,1) [gauge,label={[xshift=-3.2]right:{\scriptsize $1$}}] {};
	\draw[gline] (g1)--(space1);
	\draw[gline] (space2)--(g2)--(g3)--(g4)--(space3);
	\draw[gline] (space4)--(g5);
	\draw[gline] (g3)--(g6);
	\draw[gline] (g3)--(g7);
	
\draw[->,thick] (7,0)--(8,0);

	\node (g1) at (9,0) [gauger,label={[yshift=3.2]below:{\scriptsize $1$}}] {};
	\node (space1) at (9.7,0) {};
	\node (dots) at (10,0) {$\cdots$};
	\node (space2) at (10.3,0) {};	
	\node (g2) at (11,0) [gauge,label={[yshift=3.2]below:{\scriptsize $N-1$}}] {};
	\node (space3) at (11.7,0) {};
	\node (dots) at (12,0) {$\cdots$};
	\node (space4) at (12.3,0) {};
	\node (g5) at (13,0) [gauger,label={[yshift=3.2]below:{\scriptsize $1$}}] {};
	\node (g6) at (10.5,1) [gauge,label={[xshift=3.2]left:{\scriptsize $1$}}] {};
	\node (g7) at (11.5,1) [gauge,label={[xshift=-3.2]right:{\scriptsize $1$}}] {};
	\node (g8) at (11,2) [gauge,label={[yshift=-3.2]above:{\scriptsize $1$}}] {};
	\draw[gline] (g1)--(space1);
	\draw[gline] (space2)--(g2)--(space3);
	\draw[gline] (space4)--(g5);
	\draw[gline] (g6)--(g2)--(g7)--(g8)--(g6);

\draw[->,thick] (11,-.6)--(11,-1.4);

	\node (g1) at (9,-4) [gauger,label={[yshift=3.2]below:{\scriptsize $1$}}] {};
	\node (space1) at (9.7,-4) {};
	\node (dots) at (10,-4) {$\cdots$};
	\node (space2) at (10.3,-4) {};	
	\node (g2) at (11,-4) [gauge,label={[yshift=3.2]below:{\scriptsize $N-2$}}] {};
	\node (space3) at (11.7,-4) {};
	\node (dots) at (12,-4) {$\cdots$};
	\node (space4) at (12.3,-4) {};
	\node (g5) at (13,-4) [gauger,label={[yshift=3.2]below:{\scriptsize $1$}}] {};
	\node (g6) at (10.5,-3) [gauge,label={[xshift=3.2]left:{\scriptsize $1$}}] {};
	\node (g7) at (11.5,-3) [gauge,label={[xshift=-3.2]right:{\scriptsize $1$}}] {};
	\node (g8) at (11,-2) [gauge,label={[yshift=-3.2]above:{\scriptsize $1$}}] {};
	\node (g9) at (11,-3) [gauge,label={[yshift=-3.2]above:{\scriptsize $1$}}] {};
	\draw[gline] (g1)--(space1);
	\draw[gline] (space2)--(g2)--(space3);
	\draw[gline] (space4)--(g5);
	\draw[gline] (g2)--(g7)--(g8)--(g6)--(g2);
	\draw[gline] (g7)--(g9)--(g6);

\draw[-,thick] (8,-4)--(7,-4);
\node at (6.5,-4) {$\cdots$};
\draw[->,thick] (6,-4)--(5,-4);
\node at (6.5,-3.5) {\scriptsize $\times (N-5)$};

\node at (3,-4) {\eqref{MQSUN_weak} \text{ (right)}};

\end{tikzpicture}
\ee

\subsection{General Movement on the Extended Coulomb Branch}

We have seen that the Higgs branches, described by magnetic quivers, of theories that are related by decoupling hypermultiplets or turning on a gauge coupling have a direct connection in terms of an FI deformation. 
For any given magnetic quiver there are, however, many more ways to consistently turn on FI parameters than is captured by decoupling or going to weak coupling in 5d. FI deformations in the magnetic quiver can in general be interpreted in 5d as moving along a sub-manifold of the extended Coulomb branch, adjusting the mass matrix in such a way that a non-trivial Higgs branch emanates there.

Distinct FI deformations may give rise to the same magnetic quiver. In particular, it is common that the more involved FI deformations may equivalently be implemented by composing a set of simpler deformations by applying them successively. It is therefore convenient to introduce an organizing principle that allows us to weed out some of this intrinsic degeneracy. We conjecture that an FI deformation which decreases the rank of the flavor symmetry group of the 5d theory by more than 1 is decomposable into several FI deformations. We call FI deformations which are not decomposable {\it simple}. In the following, we will focus on these simple FI deformations.
However, even within this smaller set of deformations we encounter degeneracies, where inequivalent FI deformations produce the same magnetic quiver.

We now specialize our example further to $SU(3)_1+6\mathbf{F}$, so we can give a complete treatment of the simple FI deformations of its magnetic quiver, without the exposition becoming too tedious. We consider all the FI deformations of the type presented in section \ref{sec:FIdef}, which lower the rank of the flavor symmetry group of the 5d theory by at most 1, and show that they correspond to partial resolutions of the GTP.
We have the following choices for turning on FI parameters at two nodes of the same rank: the two $U(3)$ nodes, the $U(2)$ nodes, and four distinct pairs of $U(1)$ nodes of which we have already considered \eqref{MQSU3_decdef+}, \eqref{MQSU3_decdef-} and \eqref{MQSUN_weakdef}, leaving the two abelian color nodes. The $U(3)$ deformation gives rise to the magnetic quiver of the weakly coupled $SU(3)_1+6\mathbf{F}$ gauge theory in \eqref{MQSUN_weak} (left). Turning on FI parameters at the $U(2)$ nodes we find the magnetic quiver 
\be 
\begin{tikzpicture}
\label{MQSU3_4}
	\node (g1) at (0,0) [gauge,label={[yshift=3.2]below:{\scriptsize $1$}}] {};
	\node (g2) at (1,0) [gauger,label={[yshift=3.2]below:{\scriptsize $2$}}] {};
	\node (g3) at (2,0) [gauge,label={[yshift=3.2]below:{\scriptsize $3$}}] {};
	\node (g4) at (3,0) [gauge,draw=ForestGreen,label={[yshift=3.2]below:{\scriptsize $3$}}] {};
	\node (g5) at (4,0) [gauger,label={[yshift=3.2]below:{\scriptsize $2$}}] {};
	\node (g6) at (5,0) [gauge,label={[yshift=3.2]below:{\scriptsize $1$}}] {};
	\node (g7) at (2,1) [gauge,draw=cyan,label={[xshift=3.2]left:{\scriptsize $1$}}] {};
	\node (g8) at (3,1) [gauge,draw=blue,label={[xshift=-3.2]right:{\scriptsize $1$}}] {};
	\draw[gline] (g1)--(g2)--(g3)--(g4)--(g5)--(g6);
	\draw[gline] (g3)--(g7);
	\draw[gline] (g4)--(g8);

    \draw[->,thick] (5.6,0)--(6.4,0);

	\node (g1) at (7,0) [gauge,label={[yshift=3.2]below:{\scriptsize $1$}}] {};
	\node (g2) at (8,0) [gauge,draw=ForestGreen,label={[yshift=3.2]below:{\scriptsize $2$}}] {};
	\node (g3) at (9,0) [gauge,draw=blue,label={[yshift=3.2]below:{\scriptsize $1$}}] {};
	\node (g4) at (10,0) [gauge,draw=Maroon,label={[yshift=3.2]below:{\scriptsize $1$}}] {};
	\node (g5) at (9,1) [gauge,draw=cyan,label={[xshift=3.2]left:{\scriptsize $1$}}] {};
	\node (g6) at (10,1) [gauge,label={[xshift=-3.2]right:{\scriptsize $1$}}] {};
	\node (g7) at (8,1) [gauge,label={[xshift=3.2]left:{\scriptsize $1$}}] {};
	\draw[gline] (g1)--(g2)--(g3)--(g4)--(g6)--(g5)--(g2)--(g7);
	
	\node at (8.5,-1) {{\scriptsize $\su (4) \oplus \su (3) \oplus \uu (1)$}};
	\end{tikzpicture}
\ee 
which corresponds to inserting only the bottom ruling in $P_{SU(2)^2}$ in \eqref{GTPSUN_weak} (right). The color $U(1)$ nodes give rise to
\be 
\begin{tikzpicture}
	\node (g1) at (0,0) [gauge,label={[yshift=3.2]below:{\scriptsize $1$}}] {};
	\node (g2) at (1,0) [gauge,label={[yshift=3.2]below:{\scriptsize $2$}}] {};
	\node (g3) at (2,0) [gauge,label={[yshift=3.2]below:{\scriptsize $3$}}] {};
	\node (g4) at (3,0) [gauge,draw=ForestGreen,label={[yshift=3.2]below:{\scriptsize $3$}}] {};
	\node (g5) at (4,0) [gauge,label={[yshift=3.2]below:{\scriptsize $2$}}] {};
	\node (g6) at (5,0) [gauge,label={[yshift=3.2]below:{\scriptsize $1$}}] {};
	\node (g7) at (2,1) [gauger,draw=cyan,label={[xshift=3.2]left:{\scriptsize $1$}}] {};
	\node (g8) at (3,1) [gauger,draw=blue,label={[xshift=-3.2]right:{\scriptsize $1$}}] {};
	\draw[gline] (g1)--(g2)--(g3)--(g4)--(g5)--(g6);
	\draw[gline] (g3)--(g7);
	\draw[gline] (g4)--(g8);

    \draw[->,thick] (5.6,0)--(6.4,0);

	\node (g1) at (7,0) [gauge,label={[yshift=3.2]below:{\scriptsize $1$}}] {};
	\node (g2) at (8,0) [gauge,label={[yshift=3.2]below:{\scriptsize $2$}}] {};
	\node (g3) at (9,0) [gauge,label={[yshift=3.2]below:{\scriptsize $2$}}] {};
	\node (g4) at (10,0) [gauge,draw=ForestGreen,label={[yshift=3.2]below:{\scriptsize $2$}}] {};
	\node (g5) at (11,0) [gauge,label={[yshift=3.2]below:{\scriptsize $2$}}] {};
	\node (g6) at (12,0) [gauge,label={[yshift=3.2]below:{\scriptsize $1$}}] {};
	\node (g7) at (9.5,1) [gauge,draw=blue,label={[yshift=-3.2]above:{\scriptsize $1$}}] {};
	\draw[gline] (g1)--(g2)--(g3)--(g4)--(g5)--(g6);
	\draw[gline] (g2)--(g7)--(g5);
	
	\node at (9.5,-1) {{\scriptsize $\su (7)$}};
	\end{tikzpicture}
\ee  
which corresponds to opening the Coulomb branch as
\be
\label{GTPSU3_general}
 \begin{tikzpicture}[x=.5cm,y=.5cm]
 	\draw[step=.5cm,gray,very thin] (0,0) grid (2,4);

 	\draw[ligne,ForestGreen] (0,0)--(0,2);
 	\draw[ligne,ForestGreen] (1,2)--(1,4);
 	\draw[ligne,ForestGreen] (2,0)--(2,2);
 	\draw[ligne,blue] (0,2)--(0,4)--(1,4)--(2,3)--(2,2);
 	\draw[ligne,blue] (0,0)--(1,2)--(2,0)--(0,0);

 	\foreach \x in {0,1,...,4}
 	\node[bd] at (0,\x) {};
 	\foreach \x in {2,3}
	\node[bd] at (1,\x) {};
 	\foreach \x in {0,1,2,3}
	\node[bd] at (2,\x) {};
	\node[bd] at (1,4) {};
	\node[wd] at (1,0) {};
	\end{tikzpicture}
 \ee 
There is furthermore a number of ways to turn on FI parameters at a $U(n)$, $U(m)$ and $U(n+m)$ node, as shown in section \ref{sec:FIdef}. However, no new magnetic quivers are produced in addition to the ones we have already encountered. The same is true of the $U(1)-U(2)-$ deformation which implements a decoupling and subsequent weak coupling limit. 
We note that the set of magnetic quivers we obtain from simple FI deformations of the magnetic quiver of strongly coupled $SU(3)_1+6\mathbf{F}$ can be summarized as comprising the weakly coupled description and
\be 
\begin{tikzpicture}
	\node (g1) at (0,0) [gauge,label={[yshift=3.2]below:{\scriptsize $1$}}] {};
	\node (g2) at (1,0) [gauge,label={[yshift=3.2]below:{\scriptsize $2$}}] {};
	\node (space1) at (1.7,0) {};
	\node (dots) at (2,0) {$\cdots$};
	\node (space2) at (2.3,0) {};
	\node (g3) at (3,0) [gauge,label={[yshift=3.2]below:{\scriptsize $2$}}] {};
	\node (g4) at (4,0) [gauge,label={[yshift=3.2]below:{\scriptsize $1$}}] {};
	\node (g5) at (1,1) [gauge,label={[yshift=-3.2]above:{\scriptsize $1$}}] {};
	\node (space3) at (1.7,1) {};
	\node (dots) at (2,1) {$\cdots$};
	\node (space4) at (2.3,1) {};
	\node (g6) at (3,1) [gauge,label={[yshift=-3.2]above:{\scriptsize $1$}}] {};

	\draw[gline] (g1)--(g2)--(space1);
	\draw[gline] (space2)--(g3)--(g4);
	\draw[gline] (g2)--(g5)--(space3);
	\draw[gline] (space4)--(g6)--(g3);
	
	
	\draw [decorate,decoration={brace,amplitude=7},yshift=11]
(.9,1)--(3.1,1) node [midway,yshift=14] {\scriptsize $k$};
	\draw [decorate,decoration={brace,amplitude=7,mirror},yshift=-12]
(.9,0)--(3.1,0) node [midway,yshift=-14] {\scriptsize $5-k$};
	\end{tikzpicture}
\ee  
for $k=1,2,3,4$.

\subsection{Magnetic Quivers with Loops}

Magnetic quivers with loops are typically associated with low-flavor 5d SCFTs. The presence of a loop implies that to turn on FI parameters at a generic set of nodes, we must specify a path for the propagation of non-zero vevs. 
In the context of decoupling, the path ambiguity is associated to the existence of multiple cones in the 5d Higgs branch of the descendant SCFT. Furthermore, we observe that choosing a longer path, i.e. one that implicates more gauge nodes, usually decreases the rank of the flavor symmetry group by more than 1. 
We also encounter affine Dynkin diagrams, which correspond to a movement onto the extended Coulomb branch where we turn on a mass $m$ and a vev $\langle \phi \rangle$ such that $m-\langle \phi \rangle=0$. 

Let us consider the UV fixed point of $SU(3)_{3/2}+5 \mathbf{F}$ found in \eqref{GTPSU3_3/2}. We repeat it here for convenience:
\be
\label{SU3_3/2_solo}
 \begin{tikzpicture}[x=.5cm,y=.5cm]
	\draw[step=.5cm,gray,very thin] (0,0) grid (2,4);
 	\draw[ligne,ForestGreen] (2,0)--(2,2);
 	\draw[ligne,ForestGreen] (0,3)--(0,0);
 	\draw[ligne,Maroon] (0,0)--(1,0);
 	\draw[ligne,Maroon] (2,2)--(2,3);
 	\draw[ligne,cyan] (1,0)--(2,0);
 	\draw[ligne,cyan] (1,4)--(2,3);
 	\draw[ligne,cyan] (0,2)--(0,3);
 	\draw[ligne,Maroon] (0,3)--(1,4);

 	\foreach \x in {0,1,...,3}
 	\node[bd] at (0,\x) {};
 	\foreach \x in {0,1,2,3}
	\node[bd] at (2,\x) {};
	\node[bd] at (1,4) {};
	\node[wd] at (1,0) {};
	\end{tikzpicture} \hspace{2cm}
\begin{tikzpicture}
	\node (g1) at (0,0) [gauge,label={[yshift=3.2]below:{\scriptsize $1$}}] {};
	\node (g2) at (1,0) [gauge,label={[yshift=3.2]below:{\scriptsize $2$}}] {};
	\node (g3) at (2,0) [gauge,draw=ForestGreen,label={[yshift=3.2]below:{\scriptsize $2$}}] {};
	\node (g4) at (3,0) [gauge,label={[yshift=3.2]below:{\scriptsize $2$}}] {};
	\node (g5) at (4,0) [gauge,label={[yshift=3.2]below:{\scriptsize $1$}}] {};
	\node (g6) at (1,1) [gauge,draw=cyan,label={[xshift=3.2]left:{\scriptsize $1$}}] {};
	\node (g7) at (3,1) [gauge,draw=Maroon,label={[xshift=-3.2]right:{\scriptsize $1$}}] {};
	\draw[gline] (g1)--(g2)--(g3)--(g4)--(g5);
	\draw[gline] (g2)--(g6);
	\draw[gline] (g4)--(g7);
	\draw[gline] (g6)--(g7);
	
	\node at (2,1.5) {{\scriptsize $\su (6) \oplus \uu (1)$}};
\end{tikzpicture}
\ee  
Decoupling a hypermultiplet generates a theory whose Higgs branch is described by two cones. In the GTP the two cones arise as two inequivalent edge colorings, whereas in the magnetic quiver the two cones appear as we turn on FI parameters at the associated set of gauge nodes and propagate the non-zero vevs around the loop along the two possible paths. In other words, a single matter curve can still be associated with turning on FI parameters at a single set of nodes, and the multiple cones arise from the ambiguity in connecting these nodes. 
Sending a hypermultiplet mass to $+\infty$ we obtain the UV fixed point of $SU(3)_{2}+4\mathbf{F}$, whose Higgs branch is described by
\be
\label{GTPSU3_2}
 \begin{tikzpicture}[x=.5cm,y=.5cm]
 	\draw[step=.5cm,gray,very thin] (0,0) grid (2,4);
 	
 	\draw[ligne,ForestGreen] (0,0)--(0,2);
 	\draw[ligne,ForestGreen] (2,0)--(2,2);
 	\draw[ligne,blue] (0,0)--(2,0);
 	\draw[ligne,blue] (0,2)--(0,3)--(1,4)--(2,2);

 	\foreach \x in {0,1,...,3}
 	\node[bd] at (0,\x) {};
 	\foreach \x in {0,1,2}
	\node[bd] at (2,\x) {};
	\node[bd] at (1,4) {};
	\node[wd] at (1,0) {};
	
	\draw[step=.5cm,gray,very thin] (10,0) grid (12,4);
	
 	\draw[ligne,ForestGreen] (10,0)--(10,1);
 	\draw[ligne,ForestGreen] (12,0)--(12,1);
 	\draw[ligne,blue] (10,0)--(11,0);
 	\draw[ligne,blue] (10,1)--(10,3);
 	\draw[ligne,blue] (11,4)--(12,2);
 	\draw[ligne,Maroon] (11,0)--(12,0);
 	\draw[ligne,Maroon] (12,1)--(12,2);
 	\draw[ligne,Maroon] (10,3)--(11,4);

 	\foreach \x in {0,1,...,3}
 	\node[bd] at (10,\x) {};
 	\foreach \x in {0,1,2}
	\node[bd] at (12,\x) {};
	\node[bd] at (11,4) {};
	\node[wd] at (11,0) {};
	
	\node at (-2,0) {{\color{white} .}};
	\node at (14,0) {{\color{white} .}};
	\end{tikzpicture}
 \ee 
\be 
\label{MQSU3_2}
\begin{tikzpicture}
	\node (g1) at (0,0) [gauge,label={[yshift=3.2]below:{\scriptsize $1$}}] {};
	\node (g2) at (1,0) [gauge,label={[yshift=3.2]below:{\scriptsize $2$}}] {};
	\node (g3) at (2,0) [gauge,draw=ForestGreen,label={[yshift=3.2]below:{\scriptsize $2$}}] {};
	\node (g4) at (3,0) [gauge,label={[yshift=3.2]below:{\scriptsize $1$}}] {};
	\node (g5) at (1.5,1) [gauge,draw=blue,label={[yshift=-3.2]above:{\scriptsize $1$}}] {};
	\draw[gline] (g1)--(g2)--(g3)--(g4);
	\draw[gline] (g2)--(g5)--(g3);
	\node at (1.5,-1) {{\scriptsize $\su (5)$}};
	
	\node (g1) at (5,0) [gauge,label={[yshift=3.2]below:{\scriptsize $1$}}] {};
	\node (g2) at (6,0) [gauge,draw=ForestGreen,label={[yshift=3.2]below:{\scriptsize $1$}}] {};
	\node (g3) at (7,0) [gauge,label={[yshift=3.2]below:{\scriptsize $1$}}] {};
	\node (g4) at (8,0) [gauge,label={[yshift=3.2]below:{\scriptsize $1$}}] {};
	\node (g5) at (6,1) [gauge,draw=Maroon,label={[xshift=3.2]left:{\scriptsize $1$}}] {};
	\node (g6) at (7,1) [gauge,draw=blue,label={[xshift=-3.2]right:{\scriptsize $1$}}] {};
	\draw[gline] (g5)--(g1)--(g2)--(g3)--(g4)--(g6);
	\draw[doublearrow] (g5)--(g6);
	\node at (6.5,-1) {{\scriptsize $\su (5) \oplus \uu (1)$}};
\end{tikzpicture}
\ee 
If the global symmetry is not simple, it is possible that only certain simple components act on each cone of the Higgs branch. Here we see that the $\uu(1)$ only acts on the cone described by the righthand magnetic quiver, whereas the lefthand magnetic quiver does not see it.
The corresponding FI deformation is given by moving along the top, respectively the bottom, of the loop in
\be
\label{FISU3_2}
\begin{tikzpicture}
	\node (g1) at (0,0) [gauger,label={[yshift=3.2]below:{\scriptsize $1$}}] {};
	\node (g2) at (1,0) [gauge,label={[yshift=3.2]below:{\scriptsize $2$}}] {};
	\node (g3) at (2,0) [gauge,draw=ForestGreen,label={[yshift=3.2]below:{\scriptsize $2$}}] {};
	\node (g4) at (3,0) [gauge,label={[yshift=3.2]below:{\scriptsize $2$}}] {};
	\node (g5) at (4,0) [gauge,label={[yshift=3.2]below:{\scriptsize $1$}}] {};
	\node (g6) at (1,1) [gauge,draw=cyan,label={[xshift=3.2]left:{\scriptsize $1$}}] {};
	\node (g7) at (3,1) [gauger,draw=Maroon,label={[xshift=-3.2]right:{\scriptsize $1$}}] {};
	\draw[gline] (g1)--(g2)--(g3)--(g4)--(g5);
	\draw[gline] (g2)--(g6);
	\draw[gline] (g4)--(g7);
	\draw[gline] (g6)--(g7);
\end{tikzpicture}
\ee  
We note that, in correspondence with our conjecture, the lefthand magnetic quiver can also be obtained from two separate FI deformations, e.g. by identifying two adjacent $U(2)$ nodes in \eqref{SU3_3/2_solo} (right) and subsequently identifying the two adjacent $U(1)$ nodes at the top.
Sending a hypermultiplet mass to $-\infty$ we obtain strongly coupled $SU(3)_{1}+4\mathbf{F}$, whose Higgs branch is described by
\be
\label{GTPSU3_1}
 \begin{tikzpicture}[x=.5cm,y=.5cm]
 	\draw[step=.5cm,gray,very thin] (0,0) grid (2,3);
 	
 	\draw[ligne,ForestGreen] (0,0)--(0,2);
 	\draw[ligne,ForestGreen] (2,0)--(2,2);
 	\draw[ligne,blue] (0,0)--(2,0);
 	\draw[ligne,blue] (0,2)--(1,3)--(2,2);

 	\foreach \x in {0,1,2}
 	\node[bd] at (0,\x) {};
 	\foreach \x in {0,1,2}
	\node[bd] at (2,\x) {};
	\node[bd] at (1,3) {};
	\node[bd] at (1,0) {};
	
	\draw[step=.5cm,gray,very thin] (10,0) grid (12,3);
	
 	\draw[ligne,ForestGreen] (10,0)--(10,1);
 	\draw[ligne,ForestGreen] (12,0)--(12,1);
 	\draw[ligne,blue] (10,0)--(11,0);
 	\draw[ligne,blue] (10,1)--(10,2);
 	\draw[ligne,blue] (11,3)--(12,2);
 	\draw[ligne,Maroon] (11,0)--(12,0);
 	\draw[ligne,Maroon] (12,1)--(12,2);
 	\draw[ligne,Maroon] (10,2)--(11,3);

 	\foreach \x in {0,1,2}
 	\node[bd] at (10,\x) {};
 	\foreach \x in {0,1,2}
	\node[bd] at (12,\x) {};
	\node[bd] at (11,3) {};
	\node[bd] at (11,0) {};
	
	\node at (-1,0) {{\color{white} .}};
	\node at (13.5,0) {{\color{white} .}};
	\end{tikzpicture}
 \ee 
\be 
\label{MQSU3_1}
\begin{tikzpicture}
	\node (g1) at (0,0) [gauge,label={[yshift=3.2]below:{\scriptsize $1$}}] {};
	\node (g2) at (1,0) [gauge,draw=ForestGreen,label={[yshift=3.2]below:{\scriptsize $2$}}] {};
	\node (g4) at (2,0) [gauge,label={[yshift=3.2]below:{\scriptsize $1$}}] {};
	\node (g5) at (1,1) [gauge,draw=blue,label={[yshift=-3.2]above:{\scriptsize $1$}}] {};
	\draw[gline] (g1)--(g2)--(g4);
	\draw[doublearrow] (g2)--(g5);
	\node at (1,-1) {{\scriptsize $\su (4)$}};
	
	\node (g1) at (5,0) [gauge,label={[yshift=3.2]below:{\scriptsize $1$}}] {};
	\node (g2) at (6,0) [gauge,draw=ForestGreen,label={[yshift=3.2]below:{\scriptsize $1$}}] {};
	\node (g3) at (7,0) [gauge,label={[yshift=3.2]below:{\scriptsize $1$}}] {};
	\node (g4) at (5.5,1) [gauge,draw=Maroon,label={[xshift=3.2]left:{\scriptsize $1$}}] {};
	\node (g5) at (6.5,1) [gauge,draw=blue,label={[xshift=-3.2]right:{\scriptsize $1$}}] {};
	\node (g6) at (6,2) [gauge,label={[yshift=-3.2]above:{\scriptsize $1$}}] {};
	\draw[gline] (g4)--(g1)--(g2)--(g3)--(g5);
	\draw[gline] (g5)--(g6)--(g4)--(g5);
	\node at (6,-1) {{\scriptsize $\su (4) \oplus \su(2) \oplus \uu (1)$}};
\end{tikzpicture}
\ee 
For this descendant, only the cone characterized by the righthand quiver sees the full flavor symmetry, whereas the lefthand quiver only knows about the $\su(4)$ subalgebra. 
The FI deformation that implements this decoupling in the magnetic quiver is given by turning on FI parameters at
\be
\label{FISU3_1}
\begin{tikzpicture}
	\node (g1) at (0,0) [gauger,label={[yshift=3.2]below:{\scriptsize $1$}}] {};
	\node (g2) at (1,0) [gauge,label={[yshift=3.2]below:{\scriptsize $2$}}] {};
	\node (g3) at (2,0) [gauge,draw=ForestGreen,label={[yshift=3.2]below:{\scriptsize $2$}}] {};
	\node (g4) at (3,0) [gauge,label={[yshift=3.2]below:{\scriptsize $2$}}] {};
	\node (g5) at (4,0) [gauger,label={[yshift=3.2]below:{\scriptsize $1$}}] {};
	\node (g6) at (1,1) [gauge,draw=cyan,label={[xshift=3.2]left:{\scriptsize $1$}}] {};
	\node (g7) at (3,1) [gauge,draw=Maroon,label={[xshift=-3.2]right:{\scriptsize $1$}}] {};
	\draw[gline] (g1)--(g2)--(g3)--(g4)--(g5);
	\draw[gline] (g2)--(g6);
	\draw[gline] (g4)--(g7);
	\draw[gline] (g6)--(g7);
\end{tikzpicture}
\ee  
and propagating the non-zero vevs along the top/bottom of the loop. The lefthand quiver in \eqref{MQSU3_1} can also be obtained by composing three FI deformations, e.g. by merging the $U(2)$ nodes in \eqref{MQSU3_2} (left).

Like its parent, the 5d SCFT given by \eqref{SU3_3/2_solo} admits two weakly coupled gauge theory descriptions, namely $SU(3)_{3/2}+5\mathbf{F}$ and $SU(2)_\pi-SU(2)-[3]$.
We denote the GTPs of the two gauge theory phases by $Q_{SU(3)}$ and $Q_{SU(2)^2}$, respectively. The GTPs and associated magnetic quivers are 
\be
\label{GTPSU3_3/2weak}
 \begin{tikzpicture}[x=.5cm,y=.5cm]
 	\draw[step=.5cm,gray,very thin] (0,0) grid (2,4);
	\node at (-2,2) {$Q_{SU(3)}:$};
 	\draw[ligne,ForestGreen] (2,1)--(2,3);
 	\draw[ligne,ForestGreen] (0,0)--(0,2);
 	\draw[ligne,ForestGreen] (1,1)--(1,3);
 	\draw[ligne,blue] (0,3)--(1,4);
 	\draw[ligne,blue] (2,0)--(2,1);
 	\draw[ligne,blue] (1,0)--(0,0)--(1,1);
 	\draw[ligne,cyan] (1,0)--(2,0);
 	\draw[ligne,cyan] (2,3)--(1,4)--(1,3);
 	\draw[ligne,cyan] (1,1)--(2,1);
 	\draw[ligne,cyan] (0,2)--(0,3);

 	\foreach \x in {0,1,2,3}
 	\node[bd] at (0,\x) {};
 	\foreach \x in {1,2,3}
	\node[bd] at (1,\x) {};
 	\foreach \x in {0,1,2,3}
	\node[bd] at (2,\x) {};
	\node[bd] at (1,4) {};
	\node[wd] at (1,0) {};
	
	\draw[step=.5cm,gray,very thin] (10,0) grid (12,4);
	
	\node at (7,2) {$Q_{SU(2)^2}:$};
 	\draw[ligne,Maroon] (12,2)--(12,3);
 	\draw[ligne,Maroon] (10,3)--(10,2);
 	\draw[ligne,blue] (10,1)--(10,0)--(12,0)--(12,1);
 	\draw[ligne,blue] (10,3)--(11,4)--(12,3)--(10,3);

	\foreach \x in {2,3}
 	\draw[ligne,blue] (10,\x)--(12,\x);
 	
 	\draw[ligne,ForestGreen] (10,1)--(10,2);
 	\draw[ligne,ForestGreen] (12,1)--(12,2);

 	\foreach \x in {0,1,2,3}
 	\node[bd] at (10,\x) {};
 	\foreach \x in {0,1,2,3}
	\node[bd] at (12,\x) {};
	\node[bd] at (11,4) {};
	\node[wd] at (11,0) {};
	\foreach \x in {2,3}
	\node[bd] at (11,\x) {};

	\node at (-2,0) {{\color{white} .}};
	\node at (15.2,0) {{\color{white} .}};
	\end{tikzpicture}
 \ee 
\be 
\label{MQSU3_weak}
\begin{tikzpicture}
	\node (g1) at (0,0) [gauge,label={[yshift=3.2]below:{\scriptsize $1$}}] {};
	\node (g2) at (1,0) [gauge,label={[yshift=3.2]below:{\scriptsize $2$}}] {};
	\node (g3) at (2,0) [gauge,draw=ForestGreen,label={[yshift=3.2]below:{\scriptsize $2$}}] {};
	\node (g4) at (3,0) [gauge,label={[yshift=3.2]below:{\scriptsize $1$}}] {};
	\node (g5) at (1,1) [gauge,draw=cyan,label={[xshift=3.2]left:{\scriptsize $1$}}] {};
	\node (g6) at (2,1) [gauge,draw=blue,label={[xshift=-3.2]right:{\scriptsize $1$}}] {};
	\draw[gline] (g1)--(g2)--(g3)--(g4);
	\draw[gline] (g2)--(g5)--(g6)--(g3);
	\node at (1.5,-1) {{\scriptsize $\su (5) \oplus \uu (1)$}};

	\node (g1) at (6,1) [gauge,label={[yshift=-3.2]above:{\scriptsize $1$}}] {};
	\node (g2) at (5,0) [gauge,draw=ForestGreen,label={[yshift=3.2]below:{\scriptsize $1$}}] {};
	\node (g3) at (6,0) [gauge,label={[yshift=3.2]below:{\scriptsize $1$}}] {};
	\node (g4) at (7,0) [gauge,draw=blue,label={[yshift=3.2]below:{\scriptsize $1$}}] {};
	\node (g5) at (8,0) [gauge,draw=Maroon,label={[yshift=3.2]below:{\scriptsize $1$}}] {};
	\draw[gline] (g4)--(g1)--(g2)--(g3)--(g4);
	\draw[doublearrow] (g4)--(g5);
	
	\node at (6.5,-1) {{\scriptsize $\su (4) \oplus \su (2)$}};
\end{tikzpicture}
\ee 
To reach the magnetic quiver of the weakly coupled $SU(3)$ gauge theory, we turn on FI parameters at
\be
\label{FISU3_3/2weak}
\begin{tikzpicture}
	\node (g1) at (0,0) [gauger,label={[yshift=3.2]below:{\scriptsize $1$}}] {};
	\node (g2) at (1,0) [gauge,label={[yshift=3.2]below:{\scriptsize $2$}}] {};
	\node (g3) at (2,0) [gauge,draw=ForestGreen,label={[yshift=3.2]below:{\scriptsize $2$}}] {};
	\node (g4) at (3,0) [gauge,label={[yshift=3.2]below:{\scriptsize $2$}}] {};
	\node (g5) at (4,0) [gauge,label={[yshift=3.2]below:{\scriptsize $1$}}] {};
	\node (g6) at (1,1) [gauger,draw=cyan,label={[xshift=3.2]left:{\scriptsize $1$}}] {};
	\node (g7) at (3,1) [gauge,draw=Maroon,label={[xshift=-3.2]right:{\scriptsize $1$}}] {};
	\draw[gline] (g2)--(g3)--(g4)--(g5);
	\draw[gline] (g4)--(g7);
	\draw[gline] (g6)--(g7);
	\draw[red] (g1)--(g2)--(g6);
\end{tikzpicture}
\ee  
taking the shortest path around the loop. The long path gives $\widetilde A_4$, which does not correspond to a decoupling or gauge theory description. This transition can be decomposed into two separate FI deformations e.g. identify the $U(1)$ nodes connected by a double edge in \eqref{MQSU3_2} (right).
The magnetic quiver of the $SU(2)$ quiver gauge theory is obtained by composing the two steps
\be 
\begin{tikzpicture}
	\node (g1) at (0,0) [gauger,label={[yshift=3.2]below:{\scriptsize $1$}}] {};
	\node (g2) at (1,0) [gauge,label={[yshift=3.2]below:{\scriptsize $2$}}] {};
	\node (g3) at (2,0) [gauge,draw=ForestGreen,label={[yshift=3.2]below:{\scriptsize $2$}}] {};
	\node (g4) at (3,0) [gauge,label={[yshift=3.2]below:{\scriptsize $2$}}] {};
	\node (g5) at (4,0) [gauger,label={[yshift=3.2]below:{\scriptsize $1$}}] {};
	\node (g6) at (1,1) [gauge,draw=cyan,label={[xshift=3.2]left:{\scriptsize $1$}}] {};
	\node (g7) at (3,1) [gauge,draw=Maroon,label={[xshift=-3.2]right:{\scriptsize $1$}}] {};
	\draw[red] (g1)--(g2)--(g3)--(g4)--(g5);
	\draw[gline] (g2)--(g6);
	\draw[gline] (g4)--(g7);
	\draw[gline] (g6)--(g7);

    \draw[->,thick] (4.6,0)--(5.4,0);

	\node (g1) at (6,0) [gauge,label={[yshift=3.2]below:{\scriptsize $1$}}] {};
	\node (g2) at (7,0) [gauge,draw=ForestGreen,label={[yshift=3.2]below:{\scriptsize $1$}}] {};
	\node (g3) at (8,0) [gauge,label={[yshift=3.2]below:{\scriptsize $1$}}] {};
	\node (g4) at (6.5,1) [gauger,draw=Maroon,label={[xshift=3.2]left:{\scriptsize $1$}}] {};
	\node (g5) at (7.5,1) [gauger,draw=blue,label={[xshift=-3.2]right:{\scriptsize $1$}}] {};
	\node (g6) at (7,2) [gauge,label={[yshift=-3.2]above:{\scriptsize $1$}}] {};
	\draw[gline] (g4)--(g1)--(g2)--(g3)--(g5);
	\draw[gline] (g5)--(g6)--(g4);
	\draw[red] (g4)--(g5);
	\end{tikzpicture}
\ee  
corresponding to first inserting the bottom ruling and then the top in $Q_{SU(2)^2}$ of \eqref{GTPSU3_3/2weak}.

Which other FI deformations can we perform on the magnetic quiver of \eqref{SU3_3/2_solo} such that the rank of the flavor group is lowered by at most 1? There are four distinct ways of turning on FI deformations at two $U(1)$ nodes, and two ways for the $U(2)$ nodes. 
We have already studied three of the deformations using $U(1)$ pairs, namely $m \rightarrow \pm \infty$ in \eqref{FISU3_2} and \eqref{FISU3_1}, and the weak coupling limit in \eqref{FISU3_3/2weak}. The remaning set of $U(1)$s gives
\be 
\begin{tikzpicture}
	\node (g1) at (0,0) [gauge,label={[yshift=3.2]below:{\scriptsize $1$}}] {};
	\node (g2) at (1,0) [gauge,label={[yshift=3.2]below:{\scriptsize $2$}}] {};
	\node (g3) at (2,0) [gauge,draw=ForestGreen,label={[yshift=3.2]below:{\scriptsize $2$}}] {};
	\node (g4) at (3,0) [gauge,label={[yshift=3.2]below:{\scriptsize $2$}}] {};
	\node (g5) at (4,0) [gauge,label={[yshift=3.2]below:{\scriptsize $1$}}] {};
	\node (g6) at (1,1) [gauger,draw=cyan,label={[xshift=3.2]left:{\scriptsize $1$}}] {};
	\node (g7) at (3,1) [gauger,draw=Maroon,label={[xshift=-3.2]right:{\scriptsize $1$}}] {};
	\draw[gline] (g1)--(g2)--(g3)--(g4)--(g5);
	\draw[gline] (g2)--(g6);
	\draw[gline] (g4)--(g7);
	\draw[red] (g6)--(g7);

    \draw[->,thick] (4.6,0)--(5.4,0);

	\node (g1) at (6,0) [gauge,label={[yshift=3.2]below:{\scriptsize $1$}}] {};
	\node (g2) at (7,0) [gauge,label={[yshift=3.2]below:{\scriptsize $2$}}] {};
	\node (g3) at (8,0) [gauge,draw=ForestGreen,label={[yshift=3.2]below:{\scriptsize $2$}}] {};
	\node (g4) at (9,0) [gauge,label={[yshift=3.2]below:{\scriptsize $2$}}] {};
	\node (g5) at (10,0) [gauge,label={[yshift=3.2]below:{\scriptsize $1$}}] {};
	\node (g6) at (8,1) [gauge,draw=blue,label={[yshift=-3.2]above:{\scriptsize $1$}}] {};
	\draw[gline] (g1)--(g2)--(g3)--(g4)--(g5);
	\draw[gline] (g2)--(g6)--(g4);	
	\node at (8,-1) {{\scriptsize $\su (6)$}};
	\end{tikzpicture}
\ee  
The long way around the loop gives $\widetilde A_5$. The above transition corresponds to going onto the Coulomb branch as
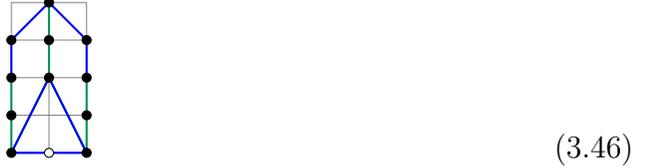
\be
 \begin{tikzpicture}[x=.5cm,y=.5cm]
	\draw[step=.5cm,gray,very thin] (0,0) grid (2,4);
 	\foreach \x in {0,2}
 	\draw[ligne,ForestGreen] (\x,0)--(\x,2);
 	\draw[ligne,ForestGreen] (0,0)--(1,2)--(2,0);
 	\draw[ligne,blue] (0,0)--(2,0)--(1,2)--(0,0);
 	\draw[ligne,blue] (0,2)--(0,3)--(1,4)--(2,3)--(2,2);
 	\draw[ligne,ForestGreen] (1,4)--(1,2);

 	\foreach \x in {0,1,2,3}
 	\node[bd] at (0,\x) {};
 	\foreach \x in {2,3}
 	\node[bd] at (1,\x) {};
 	\foreach \x in {0,1,2,3}
	\node[bd] at (2,\x) {};
	\node[bd] at (1,4) {};
	\node[wd] at (1,0) {};
\end{tikzpicture}
\ee  
For the $U(2)$ nodes we can turn on FI parameters at two adjacent nodes or at the two tail nodes. This time it is only possible to propagate through the bottom of the loop, since a vev for the bifundamental of $U(2) \times U(1)$ does not contain enough degrees of freedom to solve the equations of motion. Turning on FI parameters at adjacent $U(2)$ nodes is another way to obtain the magnetic quiver of weakly coupled $SU(3)_{3/2}+5\mathbf{F}$. The FI deformation of the other set of $U(2)$ nodes is decomposable into two applications of the former transition.

Finally, we can also turn on FI parameters at two $U(1)$ nodes and a $U(2)$ node. There are ten distinct ways of choosing this set of nodes, and three of them allow for multiple paths.
A single new magnetic quiver is obtained, which can be generated in two ways
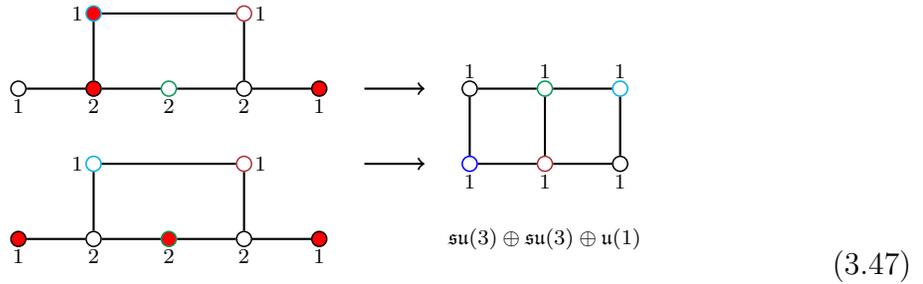
\be\label{F33} 
\begin{tikzpicture}
	\node (g1) at (0,0) [gauge,label={[yshift=3.2]below:{\scriptsize $1$}}] {};
	\node (g2) at (1,0) [gauger,label={[yshift=3.2]below:{\scriptsize $2$}}] {};
	\node (g3) at (2,0) [gauge,draw=ForestGreen,label={[yshift=3.2]below:{\scriptsize $2$}}] {};
	\node (g4) at (3,0) [gauge,label={[yshift=3.2]below:{\scriptsize $2$}}] {};
	\node (g5) at (4,0) [gauger,label={[yshift=3.2]below:{\scriptsize $1$}}] {};
	\node (g6) at (1,1) [gauger,draw=cyan,label={[xshift=3.2]left:{\scriptsize $1$}}] {};
	\node (g7) at (3,1) [gauge,draw=Maroon,label={[xshift=-3.2]right:{\scriptsize $1$}}] {};
	\draw[gline] (g1)--(g2)--(g3)--(g4)--(g5);
	\draw[gline] (g2)--(g6);
	\draw[gline] (g4)--(g7);
	\draw[gline] (g6)--(g7);
	
	\node (g1) at (0,-2) [gauger,label={[yshift=3.2]below:{\scriptsize $1$}}] {};
	\node (g2) at (1,-2) [gauge,label={[yshift=3.2]below:{\scriptsize $2$}}] {};
	\node (g3) at (2,-2) [gauger,draw=ForestGreen,label={[yshift=3.2]below:{\scriptsize $2$}}] {};
	\node (g4) at (3,-2) [gauge,label={[yshift=3.2]below:{\scriptsize $2$}}] {};
	\node (g5) at (4,-2) [gauger,label={[yshift=3.2]below:{\scriptsize $1$}}] {};
	\node (g6) at (1,-1) [gauge,draw=cyan,label={[xshift=3.2]left:{\scriptsize $1$}}] {};
	\node (g7) at (3,-1) [gauge,draw=Maroon,label={[xshift=-3.2]right:{\scriptsize $1$}}] {};
	\draw[gline] (g1)--(g2)--(g3)--(g4)--(g5);
	\draw[gline] (g2)--(g6);
	\draw[gline] (g4)--(g7);
	\draw[gline] (g6)--(g7);

    \draw[->,thick] (4.6,0)--(5.4,0);
    \draw[->,thick] (4.6,-1)--(5.4,-1);
    
	\node (g1) at (6,-1) [gauge,draw=blue,label={[yshift=3.2]below:{\scriptsize $1$}}] {};
	\node (g2) at (7,-1) [gauge,draw=Maroon,label={[yshift=3.2]below:{\scriptsize $1$}}] {};
	\node (g3) at (8,-1) [gauge,label={[yshift=3.2]below:{\scriptsize $1$}}] {};
	\node (g4) at (6,0) [gauge,label={[yshift=-3.2]above:{\scriptsize $1$}}] {};
	\node (g5) at (7,0) [gauge,draw=ForestGreen,label={[yshift=-3.2]above:{\scriptsize $1$}}] {};
	\node (g6) at (8,0) [gauge,draw=cyan,label={[yshift=-3.2]above:{\scriptsize $1$}}] {};
	\draw[gline] (g1)--(g2)--(g3)--(g6)--(g5)--(g4)--(g1);
	\draw[gline] (g2)--(g5);
	\node at (7,-2) {{\scriptsize $\su (3) \oplus \su (3) \oplus \uu (1)$}};
	\end{tikzpicture}
\ee 
Note that there is no choice of path in either of these deformations.
The resulting magnetic quiver characterizes the Higgs branch emanating from the point on the Coulomb branch described by
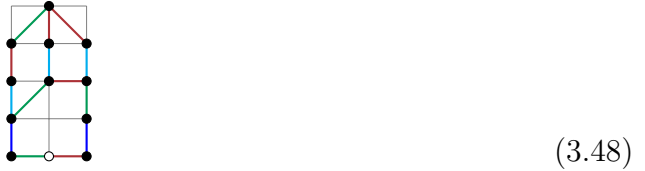
\be
 \begin{tikzpicture}[x=.5cm,y=.5cm]
	\draw[step=.5cm,gray,very thin] (0,0) grid (2,4);
 	\draw[ligne,ForestGreen] (0,3)--(1,4);
 	\draw[ligne,ForestGreen] (0,1)--(1,2);
 	\draw[ligne,ForestGreen] (0,0)--(1,0);
 	\draw[ligne,ForestGreen] (2,1)--(2,2);
 	\draw[ligne,blue] (0,0)--(0,1);
 	\draw[ligne,blue] (2,0)--(2,1);
 	 	\foreach \x in {1,2}
 	\draw[ligne,cyan] (\x,2)--(\x,3);
 	\draw[ligne,cyan] (0,1)--(0,2);
 	\draw[ligne,Maroon] (0,2)--(0,3);
 	\draw[ligne,Maroon] (1,3)--(1,4)--(2,3);
 	\draw[ligne,Maroon] (1,2)--(2,2);
 	\draw[ligne,Maroon] (1,0)--(2,0);

 	\foreach \x in {0,1,2,3}
 	\node[bd] at (0,\x) {};
 	\foreach \x in {2,3}
 	\node[bd] at (1,\x) {};
 	\foreach \x in {0,1,2,3}
	\node[bd] at (2,\x) {};
	\node[bd] at (1,4) {};
	\node[wd] at (1,0) {};
\end{tikzpicture}
\ee 

\subsection{Mirror Symmetry and FI Deformations}\label{Seclocmirr3} 

Let us now discuss the 3d interpretation of the mass deformations we have just studied. We focus for simplicity on the rank 2 example discussed above, although all we do in this section can be generalized to higher rank models.  

We consider again the magnetic quiver for the SCFT UV completing $SU(3)_{3/2}$ with $5\mathbf{F}$, whose mirror dual is a $U(2)$ gauge theory with six fundamentals and one hypermultiplet with charge 2 under the central $U(1)$ subgroup. This latter theory can therefore be interpreted as the dimensional reduction to 3d of the SCFT in five dimensions. 
\be \label{35F}
\begin{tikzpicture}
\filldraw[fill= white] (0,0) circle [radius=0.1] node[below] {\scriptsize 1};
\filldraw[fill= white] (1,0) circle [radius=0.1] node[below] {\scriptsize 2};
\filldraw[fill= white] (2,0) circle [radius=0.1] node[below] {\scriptsize 2};
\filldraw[fill= white] (3,0) circle [radius=0.1] node[below] {\scriptsize 2};
\filldraw[fill= white] (4,0) circle [radius=0.1] node[below] {\scriptsize 1};
\filldraw[fill= white] (1,1) circle [radius=0.1] node[above] {\scriptsize 1};
\filldraw[fill= white] (3,1) circle [radius=0.1] node[above] {\scriptsize 1};
\draw [thick] (0.1, 0) -- (0.9,0) ;
\draw [thick] (1.1, 0) -- (1.9,0) ;
\draw [thick] (2.1, 0) -- (2.9,0) ;
\draw [thick] (3.1, 0) -- (3.9,0) ;
\draw [thick] (1.1, 1) -- (2.9,1) ;
\draw [thick] (1, 0.1) -- (1,0.9) ;
\draw [thick] (3, 0.1) -- (3,0.9) ; 

\draw [<->,thick] (4.5,0) -- (6.5,0); 

\node[] (A) at (5.5,0.2) {Mirror}; 
\node[rectangle, draw, inner sep=1.7,minimum height=.6cm,minimum width=.6cm] (B) at (7.5,0) {6}; 
\node[] (C) at (9,0) {$U(2)$}; 
\node[rectangle, draw, inner sep=1.7,minimum height=.6cm,minimum width=.6cm] (D) at (10.5,0) {1}; 

\draw[-] (B) -- (C);
\draw[solid, thick, snake it] (D) to (C);
\end{tikzpicture} 
\ee 
The statement \eqref{35F} can be explained as follows: We first notice that by ungauging one abelian node, the magnetic quiver can be equivalently described as 
\be\begin{tikzpicture} 
\filldraw[fill= white] (7,0) circle [radius=0.1] node[below] {\scriptsize 1};
\filldraw[fill= white] (8,0) circle [radius=0.1] node[below] {\scriptsize 2};
\filldraw[fill= white] (9,0) circle [radius=0.1] node[below] {\scriptsize 2};
\filldraw[fill= white] (10,0) circle [radius=0.1] node[below] {\scriptsize 2};
\filldraw[fill= white] (11,0) circle [radius=0.1] node[below] {\scriptsize 1};
\node[rectangle, draw, minimum height=0.1,minimum width=0.1] (D) at (8,1) {}; 
\node[] at (8,1.3) {\scriptsize 1};
\node[rectangle, draw, minimum height=0.1,minimum width=0.1] (D) at (9,1) {}; 
\node[] at (9,1.3) {\scriptsize 1};
\filldraw[fill= white] (10,1) circle [radius=0.1] node[above] {\scriptsize 1};
\draw [thick] (7.1, 0) -- (7.9,0) ;
\draw [thick] (8.1, 0) -- (8.9,0) ;
\draw [thick] (9.1, 0) -- (9.9,0) ;
\draw [thick] (10.1, 0) -- (10.9,0) ;
\draw [thick] (9.12, 1) -- (9.9,1) ;
\draw [thick] (8, 0.1) -- (8,0.88) ;
\draw [thick] (10, 0.1) -- (10,0.9) ; 
\end{tikzpicture}\ee
Next we notice that this quiver can be obtained by ``fusing'' two smaller quivers, 
\be\label{mirfuse}
\begin{tikzpicture} 
\filldraw[fill= white] (-1,0) circle [radius=0.1] node[below] {\scriptsize 1};
\filldraw[fill= white] (0,0) circle [radius=0.1] node[below] {\scriptsize 2};
\filldraw[fill= white] (1,0) circle [radius=0.1] node[below] {\scriptsize 2};
\filldraw[fill= white] (2,0) circle [radius=0.1] node[below] {\scriptsize 2};
\filldraw[fill= white] (3,0) circle [radius=0.1] node[below] {\scriptsize 1};
\node[rectangle, draw, minimum height=0.1,minimum width=0.1] (D) at (0,1) {}; 
\node[] at (0,1.3) {\scriptsize 1};
\filldraw[fill= blue] (2,1) circle [radius=0.1] node[above] {\scriptsize 1};
\draw [thick] (-0.1, 0) -- (-0.9,0) ;
\draw [thick] (0.1, 0) -- (0.9,0) ;
\draw [thick] (1.1, 0) -- (1.9,0) ;
\draw [thick] (2.1, 0) -- (2.9,0) ;
\draw [thick] (0, 0.1) -- (0,0.9) ;
\draw [thick] (2, 0.1) -- (2,0.9) ; 

\node[] at (3.75,0) {+};

\filldraw[fill= blue] (4.5,0) circle [radius=0.1] node[below] {\scriptsize 1};
\node[rectangle, draw, minimum height=0.1,minimum width=0.1] (D) at (4.5,1) {}; 
\node[] at (4.5,1.3) {\scriptsize 1};
\draw [thick] (4.5, 0.1) -- (4.5,0.88) ;

\draw [->,thick] (5,0) -- (6.5,0); 

\filldraw[fill= white] (7,0) circle [radius=0.1] node[below] {\scriptsize 1};
\filldraw[fill= white] (8,0) circle [radius=0.1] node[below] {\scriptsize 2};
\filldraw[fill= white] (9,0) circle [radius=0.1] node[below] {\scriptsize 2};
\filldraw[fill= white] (10,0) circle [radius=0.1] node[below] {\scriptsize 2};
\filldraw[fill= white] (11,0) circle [radius=0.1] node[below] {\scriptsize 1};
\node[rectangle, draw, minimum height=0.1,minimum width=0.1] (D) at (8,1) {}; 
\node[] at (8,1.3) {\scriptsize 1};
\node[rectangle, draw, minimum height=0.1,minimum width=0.1] (D) at (9,1) {}; 
\node[] at (9,1.3) {\scriptsize 1};
\filldraw[fill= white] (10,1) circle [radius=0.1] node[above] {\scriptsize 1};
\draw [thick] (7.1, 0) -- (7.9,0) ;
\draw [thick] (8.1, 0) -- (8.9,0) ;
\draw [thick] (9.1, 0) -- (9.9,0) ;
\draw [thick] (10.1, 0) -- (10.9,0) ;
\draw [thick] (9.12, 1) -- (9.9,1) ;
\draw [thick] (8, 0.1) -- (8,0.88) ;
\draw [thick] (10, 0.1) -- (10,0.9) ; 
\end{tikzpicture}\ee
where fusing means gauging the diagonal combination of the topological symmetries associated with the nodes colored in blue. 
On the left-hand side of \eqref{mirfuse} one can now recognize the mirror duals of $SU(2)$ SQCD with $6\mathbf{F}$ and a free hypermultiplet and the topological symmetries associated with the nodes in blue are mapped to the baryonic symmetry of SQCD and the $U(1)$ acting on the hypermultiplet respectively. We therefore conclude that by gauging the diagonal combination of these symmetries we find the model on the right-hand side of \eqref{35F}.

Let us now discuss all the transitions which decrease the rank of the flavor symmetry by 1. The easiest to describe involves FI parameters at the two abelian nodes on top, which reduces the magnetic quiver to the mirror dual of $U(2)$ SQCD with $6\mathbf{F}$. 
\be \label{356F}
\begin{tikzpicture}
\filldraw[fill= white] (0,0) circle [radius=0.1] node[below] {\scriptsize 1};
\filldraw[fill= white] (1,0) circle [radius=0.1] node[below] {\scriptsize 2};
\filldraw[fill= white] (2,0) circle [radius=0.1] node[below] {\scriptsize 2};
\filldraw[fill= white] (3,0) circle [radius=0.1] node[below] {\scriptsize 2};
\filldraw[fill= white] (4,0) circle [radius=0.1] node[below] {\scriptsize 1};
\filldraw[fill= red] (1,1) circle [radius=0.1] node[above] {\scriptsize 1};
\filldraw[fill= red] (3,1) circle [radius=0.1] node[above] {\scriptsize 1};
\draw [thick] (0.1, 0) -- (0.9,0) ;
\draw [thick] (1.1, 0) -- (1.9,0) ;
\draw [thick] (2.1, 0) -- (2.9,0) ;
\draw [thick] (3.1, 0) -- (3.9,0) ;
\draw [red]   (1.1, 1) -- (2.9,1) ;
\draw [thick] (1, 0.1) -- (1,0.9) ;
\draw [thick] (3, 0.1) -- (3,0.9) ; 

\draw [->,thick] (4.5,0) -- (6.5,0); 

\filldraw[fill= white] (7,0) circle [radius=0.1] node[below] {\scriptsize 1};
\filldraw[fill= white] (8,0) circle [radius=0.1] node[below] {\scriptsize 2};
\filldraw[fill= white] (9,0) circle [radius=0.1] node[below] {\scriptsize 2};
\filldraw[fill= white] (10,0) circle [radius=0.1] node[below] {\scriptsize 2};
\filldraw[fill= white] (11,0) circle [radius=0.1] node[below] {\scriptsize 1};
\filldraw[fill= white] (9,1) circle [radius=0.1] node[above] {\scriptsize 1};
\draw [thick] (7.1, 0) -- (7.9,0) ;
\draw [thick] (8.1, 0) -- (8.9,0) ;
\draw [thick] (9.1, 0) -- (9.9,0) ;
\draw [thick] (10.1, 0) -- (10.9,0) ;
\draw [thick] (8.05, 0.05) -- (8.95,0.95) ;
\draw [thick] (9.05, 0.95) -- (9.95,0.05) ;
\end{tikzpicture} 
\ee 
In the dual $U(2)$ theory this simply corresponds to turning on a mass for the hypermultiplet charged under the central $U(1)$ only. 

It is also easy to describe the deformation to the weakly-coupled theory:
\be \label{355F}
\begin{tikzpicture}
\filldraw[fill= red] (0,0) circle [radius=0.1] node[below] {\scriptsize 1};
\filldraw[fill= white] (1,0) circle [radius=0.1] node[below] {\scriptsize 2};
\filldraw[fill= white] (2,0) circle [radius=0.1] node[below] {\scriptsize 2};
\filldraw[fill= white] (3,0) circle [radius=0.1] node[below] {\scriptsize 2};
\filldraw[fill= white] (4,0) circle [radius=0.1] node[below] {\scriptsize 1};
\filldraw[fill= red] (1,1) circle [radius=0.1] node[above] {\scriptsize 1};
\filldraw[fill= white] (3,1) circle [radius=0.1] node[above] {\scriptsize 1};
\draw [red]   (0.1, 0) -- (0.9,0) ;
\draw [thick] (1.1, 0) -- (1.9,0) ;
\draw [thick] (2.1, 0) -- (2.9,0) ;
\draw [thick] (3.1, 0) -- (3.9,0) ;
\draw [thick] (1.1, 1) -- (2.9,1) ;
\draw [red]   (1, 0.1) -- (1,0.9) ;
\draw [thick] (3, 0.1) -- (3,0.9) ; 

\draw [->,thick] (4.5,0) -- (6.5,0); 

\filldraw[fill= white] (7,0) circle [radius=0.1] node[below] {\scriptsize 1};
\filldraw[fill= white] (8,0) circle [radius=0.1] node[below] {\scriptsize 2};
\filldraw[fill= white] (9,0) circle [radius=0.1] node[below] {\scriptsize 2};
\filldraw[fill= white] (10,0) circle [radius=0.1] node[below] {\scriptsize 1};
\filldraw[fill= white] (8,1) circle [radius=0.1] node[above] {\scriptsize 1};
\filldraw[fill= white] (9,1) circle [radius=0.1] node[above] {\scriptsize 1};
\draw [thick] (7.1, 0) -- (7.9,0) ;
\draw [thick] (8.1, 0) -- (8.9,0) ;
\draw [thick] (9.1, 0) -- (9.9,0) ;
\draw [thick] (8.1, 1) -- (8.9,1) ;
\draw [thick] (8, 0.1) -- (8,0.9) ;
\draw [thick] (9, 0.1) -- (9,0.9) ; 
\end{tikzpicture} 
\ee 
where on the right of \eqref{355F} we recognize the mirror dual of $SU(3)$ SQCD with $5\mathbf{F}$ (see e.g. \cite{Bourget:2019rtl}). In the dual $U(2)$ theory of \eqref{35F} this deformation corresponds to turning on a mass for one of the fundamentals. This reduces the theory to the dual of $SU(3)$ SQCD with $5\mathbf{F}$ recently discussed in \cite{Dey:2021rxw}. This represents a nice consistency check of our claim \eqref{35F}.

Let us now turn our attention to the decouplings. As we have reviewed before, the Higgs branch of the UV completion of $SU(3)_2$ with $4\mathbf{F}$ is described by two quivers. One of them is the mirror dual of
$U(2)$ SQCD with $5\mathbf{F}$. Let us now analyze in detail this deformation from the perspective of the $U(2)$ theory: We should turn on an $SU(5)$-preserving mass for the six fundamentals, i.e. five of them are given the same mass, which we denote as $m$, and $-5m$ for the sixth. We can make the first five fields massless again by tuning the vev for the scalar in the central $U(1)$ vector multiplet. This also has the effect of making the hypermultiplet charged under the central $U(1)$ massive. At low-energy we are therefore left with a $U(2)$ theory with $5\mathbf{F}$. Here we see that we should combine a mass deformation with a motion in the Coulomb branch. Said differently, the FI deformation describes a motion in the extended Coulomb branch. 

The magnetic quiver describing the second cone of the Higgs branch of the UV completion of $SU(3)_2$ with $4\mathbf{F}$ arises by deforming the $U(2)$ theory to the model 
\be\label{higgsed1} \boxed{5}-U(1)-U(1)-\boxed{1}\ee 
which is achieved by activating a vev for the $SU(2)$ scalar as well. This can of course be diagonalized thus breaking the gauge symmetry to the Cartan subgroup. For ease of presentation we collect the charges of the various matter fields in Table \ref{ccharge}.
\begin{table}[ht]
    \centering
\begin{equation*}
    \begin{array}{|c|c|c|c|}
\hline
\text{Matter Fields} & I_3 & U(1) & \text{Mass} \\
\hline
B_1 & \frac{1}{2} & -1 & M_{su(6)}+\phi-\varphi \\
\hline
B_2 & -\frac{1}{2} & -1 & M_{su(6)}-\phi-\varphi \\
\hline
q & 0 & 2 & 2\varphi-M_{u(1)} \\ 
\hline
\end{array}
\end{equation*}
    \caption{Masses and charges of matter fields. $B_{1,2}$ denote the $SU(2)$ color components of the fundamentals and $q$ the $U(1)$ flavor. $I_3$ stands for the charge under the Cartan of $SU(2)$. We also denote with $\phi$ the eigenvalue of the $SU(2)$ scalar and with $\varphi$ the vev of the $U(1)$ field.}
    \label{ccharge}
\end{table}
\\
Choosing again the mass matrix for the flavors of the form
\be 
M_{su(6)}=\text{diag}(m,m,m,m,m,-5m)\,, \quad \phi=-3m\,, \quad \varphi=M_{u(1)}/2=-2m\,,
\ee 
the theory reduces precisely to \eqref{higgsed1}, where the two abelian gauge groups correspond to the sum and difference of $I_3$ and $U(1)/2$ in the table. 

From the above discussion we clearly see that in 3d the two cones of the Higgs branch of the 5d theory arise by deforming \eqref{35F} with the $SU(6)$ mass matrix $M_{su(6)}=\text{diag}(m,m,m,m,m,-5m)$ and then restricting to different singular loci of the Coulomb branch of the resulting theory. The corresponding magnetic quivers are simply the mirror duals of the low-energy effective theories at the singular loci. In particular the Higgs branch of the magnetic quiver describes the local structure of the Coulomb branch of the theory (dimensionally reduced to 3d) in a neighbourhood of the singular loci. 

With the choice 
\be 
M_{su(6)}=\text{diag}(m,m,m,-m,-m,-m)\,, \quad \phi=-m\,, \quad \varphi=M_{U(1)}=0\,,
\ee 
we can similarly obtain the theory \be\label{kk33}\boxed{3}-U(1)-U(1)-\boxed{3}\ee 
This corresponds to the transition we have discussed before (see \eqref{F33})
\be \label{3F3F}
\begin{tikzpicture}
\filldraw[fill= red] (0,0) circle [radius=0.1] node[below] {\scriptsize 1};
\filldraw[fill= white] (1,0) circle [radius=0.1] node[below] {\scriptsize 2};
\filldraw[fill= red] (2,0) circle [radius=0.1] node[below] {\scriptsize 2};
\filldraw[fill= white] (3,0) circle [radius=0.1] node[below] {\scriptsize 2};
\filldraw[fill= red] (4,0) circle [radius=0.1] node[below] {\scriptsize 1};
\filldraw[fill= white] (1,1) circle [radius=0.1] node[above] {\scriptsize 1};
\filldraw[fill= white] (3,1) circle [radius=0.1] node[above] {\scriptsize 1};
\draw [thick] (0.1, 0) -- (0.9,0) ;
\draw [thick] (1.1, 0) -- (1.9,0) ;
\draw [thick] (2.1, 0) -- (2.9,0) ;
\draw [thick] (3.1, 0) -- (3.9,0) ;
\draw [thick] (1.1, 1) -- (2.9,1) ;
\draw [thick] (1, 0.1) -- (1,0.9) ;
\draw [thick] (3, 0.1) -- (3,0.9) ; 

\draw [->,thick] (4.5,0) -- (6.5,0); 

\filldraw[fill= white] (7,0) circle [radius=0.1] node[below] {\scriptsize 1};
\filldraw[fill= white] (8,0) circle [radius=0.1] node[below] {\scriptsize 1};
\filldraw[fill= white] (9,0) circle [radius=0.1] node[below] {\scriptsize 1};
\filldraw[fill= white] (7,1) circle [radius=0.1] node[above] {\scriptsize 1};
\filldraw[fill= white] (8,1) circle [radius=0.1] node[above] {\scriptsize 1};
\filldraw[fill= white] (9,1) circle [radius=0.1] node[above] {\scriptsize 1};
\draw [thick] (7.1, 0) -- (7.9,0) ;
\draw [thick] (8.1, 0) -- (8.9,0) ;
\draw [thick] (7.1, 1) -- (7.9,1) ;
\draw [thick] (8.1, 1) -- (8.9,1) ;
\draw [thick] (7, 0.1) -- (7,0.9) ;
\draw [thick] (8, 0.1) -- (8,0.9) ;
\draw [thick] (9, 0.1) -- (9,0.9) ; 
\end{tikzpicture} 
\ee 
If we ungauge one of the nodes at the center of the second quiver we get equivalently 
\be\begin{tikzpicture}
\filldraw[fill= white] (7,0) circle [radius=0.1] node[below] {\scriptsize 1};
\filldraw[fill= white] (8,0) circle [radius=0.1] node[below] {\scriptsize 1};
\filldraw[fill= white] (9,0) circle [radius=0.1] node[below] {\scriptsize 1};
\filldraw[fill= white] (10,0) circle [radius=0.1] node[below] {\scriptsize 1};
\filldraw[fill= white] (11,0) circle [radius=0.1] node[below] {\scriptsize 1};
\node[rectangle, draw, minimum height=0.1,minimum width=0.1] (D) at (9,1) {}; 
\node[] at (9,1.3) {\scriptsize 1};
\node[rectangle, draw, minimum height=0.1,minimum width=0.1] (D) at (6,0) {}; 
\node[] at (5.7,0) {\scriptsize 1};
\node[rectangle, draw, minimum height=0.1,minimum width=0.1] (D) at (12,0) {}; 
\node[] at (12.3,0) {\scriptsize 1};
\draw [thick] (6.12, 0) -- (6.9,0) ;
\draw [thick] (7.1, 0) -- (7.9,0) ;
\draw [thick] (8.1, 0) -- (8.9,0) ;
\draw [thick] (9.1, 0) -- (9.9,0) ;
\draw [thick] (10.1, 0) -- (10.9,0) ;
\draw [thick] (9, 0.1) -- (9,0.88) ;
\draw [thick] (11.1, 0) -- (11.88,0) ; 
\end{tikzpicture}\ee
which is indeed the mirror dual of \eqref{kk33}. 

Similarly we can discuss the decoupling to the SCFT which UV completes $SU(3)_1$ with $4\mathbf{F}$. The corresponding magnetic quivers are given in \eqref{MQSU3_1}.
This process involves an $SU(6)$ mass deformation of the form 
\be 
M_{su(6)}=(m,m,m,m,-2m,-2m)\,.
\ee
Again we find the two cones by zooming in around different loci in the Coulomb branch. If we set 
\be 
\phi=M_{U(1)}=0\,, \quad \varphi=m\,,
\ee
the $U(2)$ theory reduces to $U(2)$ SQCD with $4\mathbf{F}$, whose mirror dual is the quiver on the left of \eqref{MQSU3_1}. The other quiver is obtained by choosing the locus 
\be 
\phi=-3m/2\,, \quad \varphi=-m/2\,, \quad M_{U(1)}=-m\,,
\ee 
which reduces the $U(2)$ theory to 
\be \boxed{4}-U(1)-U(1)-\boxed{2}\ee
Again we see that the magnetic quivers of the 5d theory can be interpreted as the mirror duals of the low-energy theories at special loci in the Coulomb branch. 

Finally, let us briefly revisit the case of trivial $M_{su(6)}$. We have already seen that choosing one path around the closed loop of the magnetic quiver \eqref{SU3_3/2_solo} we find the mirror dual of $U(2)$ SQCD with $6\mathbf{F}$. Choosing the other path leads instead to the $\widetilde{A}_5$ quiver plus one free hypermultiplet, which is the mirror dual of SQED with $6\mathbf{F}$ plus one twisted free hyper. This is the low-energy theory on the locus $\phi=\varphi=a$ with $a$ arbitrary. The twisted free hyper is identified with the variation of the modulus $a$. We therefore see that we have a flat direction in the Coulomb branch and this explains why the magnetic quiver has a one dimensional Higgs branch although the theory we are discussing has rank two. 

In conclusion, this analysis illustrates the fact that the magnetic quivers describe the structure of the theory at the singular points in the moduli space of the SCFT, dimensionally reduced to 3d. When the Coulomb branch is a cone we have a single magnetic quiver which is the mirror dual of the SCFT. When instead there are multiple singular loci in the Coulomb branch, we have for each one of them a quiver which is the mirror dual of the low-energy theory at the corresponding singular locus. When the singular locus is not a single point but rather a submanifold the dimension of the Higgs branch of the magnetic quiver is lower than the rank of the SCFT and is accompanied by free hypermultiplets, reflecting the presence in the low-energy theory of free twisted hypermultiplets whose moduli parametrize the motion along the singular locus.

\section{\texorpdfstring{$(D_{N+2},D_{N+2})$}{(D{N+2},D{N+2})} Conformal Matter Descendant Tree} 
\label{sec:DnDn}

In this section we study the descendants of the $(D_{N+2},D_{N+2})$ conformal matter theory (for the definition of six-dimensional conformal matter see \cite{DelZotto:2014hpa}). 
In the polygon description, we can successively decouple fundamental matter multiplets by flopping out the corresponding curves. These curves are easily identified in the pre-GTP, which describes the gauge theory phase, and can be traced through the edge-moves to the corresponding curves in the GTP. 
A matter curve in the pre-GTP connects a vertex bounding the ruling to a vertex along an external edge. Sending a fundamental mass to $+ \infty$ or $- \infty$ corresponds to a matter curve emanating from either end of the ruling, and respectively increases or decreases the CS by $\half$ in the decoupling.

The pre-GTP representing the gauge theory phase of the first descendant $SU(N)_\half +(2N+3)\mathbf{F}$ is (drawn for $N=4$)
\be
\begin{tikzpicture}[x=.5cm,y=.5cm]
	\draw[step=.5cm,gray,very thin] (0,0) grid (2,6);
	
	\draw[ligne,black] (0,0)--(1,1)--(2,0)--(2,6)--(1,5)--(0,5)--(0,0);
	\draw[ligne,black] (1,1)--(1,5);
	
	\foreach \x in {0,1,...,5}
	\node[bd] at (0,\x) {};
	\node[bd] at (1,1) {};
	\foreach \x in {0,1,...,6}
	\node[bd] at (2,\x) {};
	\node[bd] at (1,5) {};
	
	\node[] at (3.5,3) {{\scriptsize $N+2$}};
	\node[] at (-1.5,2.5) {{\scriptsize $N+1$}};
	\node at (-.5,-.5) {{\scriptsize $\mathbf{v}_0$}};
	
	\draw[|-|] (-.5,0)--(-.5,5);
	\draw[|-|] (2.5,0)--(2.5,6);
\end{tikzpicture}
\ee
To reach the SCFT point, we must remove the non-convexities by applying edge-moves. This can be achieved by first exchanging the $(-1,-1)$- and $(-1,0)$-edge at the top, and subsequently pruning the $(1,1)$- and $(1,-1)$-edges at the bottom of the polygon:
\be
\label{tikz:simplemove}
\begin{tikzpicture}[x=.5cm,y=.5cm]
	\draw[step=.5cm,gray,very thin] (0,0) grid (2,6);
	
	\draw[ligne,black] (0,0)--(1,1)--(2,0)--(2,6)--(0,6)--(0,0);
	\draw[ligne,black] (1,1)--(1,5)--(2,6);
	\draw[ligne,black] (1,5)--(0,5);
	
	\foreach \x in {0,1,...,6}
	\node[bd] at (0,\x) {};
	\node[bd] at (1,1) {};
	\foreach \x in {0,1,...,6}
	\node[bd] at (2,\x) {};
	\node[wd] at (1,6) {};

	\node[] at (-1.6,3) {{\scriptsize $N+2$}};
	\draw[|-|] (-.5,0)--(-.5,6);
	\node at (-.5,-.5) {{\scriptsize $\mathbf{v}_0$}};
	
	\draw[->,thick] (3,3)--(5,3);
	
	\draw[step=.5cm,gray,very thin] (6,1) grid (18,6);

	\draw[ligne,black] (6,1)--(18,1)--(13,6)--(11,6)--(6,1);
	
	\foreach \x in {0,1,...,12}
	\node[bd] at (6+\x,1) {};
	\foreach \x in {1,2,3,4}
	\node[wd] at (18-\x,\x+1) {};
	\node[bd] at (13,6) {};
	\node[wd] at (12,6) {};
	\node[bd] at (11,6) {};
	\foreach \x in {1,2,3,4}
	\node[wd] at (11-\x,6-\x) {};
	
	\node[] at (8.5,-.2) {{\scriptsize $N+1$}};	
	\node[] at (12,-.2) {{\scriptsize $2$}};	
	\node[] at (15.5,-.2) {{\scriptsize $N+1$}};
	\node[] at (19.5,3.5) {{\scriptsize $N+1$}};
	\node at (5.5,.5) {{\scriptsize $\mathbf{v}_0$}};
	
	\draw[|-|] (18.5,1)--(18.5,6);
	\draw[|-] (6,.5)--(11,.5);
	\draw[|-] (11,.5)--(13,.5);
	\draw[|-|] (13,.5)--(18,.5);
\end{tikzpicture}
\ee
Taking the bottom left vertex as the origin and using the definitions in \cite{vanBeest:2020civ} the transformations are summarised as
\be 
\label{eq:transform}
\mathfrak{T}=\mathfrak{M}_{N+5}^+ \mathfrak{P}_1^+ \mathfrak{P}_1^-\,.
\ee 
For the UV completion of $SU(N)_\half +(2N+3)\mathbf{F}$ the unique edge coloring is 
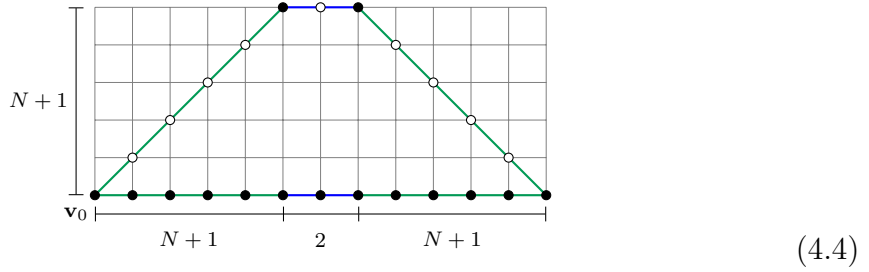
\be
\begin{tikzpicture}[x=.5cm,y=.5cm]
	\draw[step=.5cm,gray,very thin] (0,0) grid (12,5);
	
	\draw[ligne,ForestGreen] (5,5)--(0,0)--(5,0);
	\draw[ligne,blue] (5,0)--(7,0);
	\draw[ligne,ForestGreen] (7,0)--(12,0)--(7,5);
	\draw[ligne,blue] (5,5)--(7,5);
	
	\foreach \x in {0,1,...,12}
	\node[bd] at (\x,0) {};
	\foreach \x in {1,2,3,4}
	\node[wd] at (12-\x,\x) {};
	\node[bd] at (7,5) {};
	\node[wd] at (6,5) {};
	\node[bd] at (5,5) {};
	\foreach \x in {1,2,3,4}
	\node[wd] at (5-\x,5-\x) {};
	
	\node[] at (2.5,-1.2) {{\scriptsize $N+1$}};	
	\node[] at (6,-1.2) {{\scriptsize $2$}};	
	\node[] at (9.5,-1.2) {{\scriptsize $N+1$}};
	\node[] at (-1.5,2.5) {{\scriptsize $N+1$}};
	
	\draw[|-|] (-.5,0)--(-.5,5);
	\draw[|-] (0,-.5)--(5,-.5);
	\draw[|-] (5,-.5)--(7,-.5);
	\draw[|-|] (7,-.5)--(12,-.5);
	\node at (-.5,-.5) {{\scriptsize $\mathbf{v}_0$}};
\end{tikzpicture}
\ee
The data of the decomposition is
\be 
\ba 
\lambda_\alpha&=(2N+4,N+1,2,N+1)\,, \qquad && \mu_{\alpha }=(\{1^{2N+4}\},\{N+1\},\{2\},\{N+1\})\,, \\
\lambda^b_\alpha&=(2,0,2,0)\,, \qquad && \mu^b_{\alpha }=(\{2\},-,\{2\},-)\,, \\ \lambda^g_\alpha&=(2N+2,N+1,0,N+1)\,, \qquad && \mu^g_{\alpha }=(\{N+1,N+1\},\{N+1\},-,\{N+1\})\,,
\ea 
\ee 
which results in the nodes
\be 
m^b=2\,, \qquad \ m^g=N+1\,, \qquad m_{1, x}=(N+2, 2N+2,2N+1,\dots,1)\,,
\ee 
and non-zero intersections
\be 
k^g_{1 ,2}=1\,, \qquad k^b_{1 ,1}=1\,,
\ee 
as well as nearest neighbours intersecting once in the tail. Thus, the magnetic quiver is
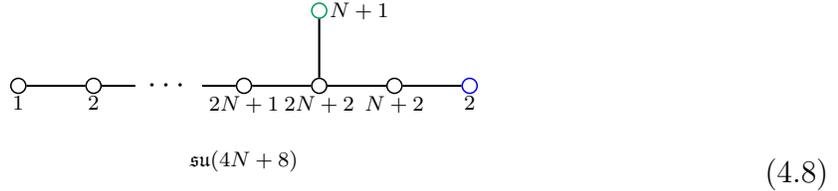
\be 
\begin{tikzpicture}
	\node (g1) at (4,0) [gauge,label={[yshift=3.2]below:{\scriptsize $1$}}] {};
	\node (g2) at (5,0) [gauge,label={[yshift=3.2]below:{\scriptsize $2$}}] {};
	\node (space1) at (5.7,0) {};
	\node (dots) at (6,0) {$\cdots$};
	\node (space2) at (6.3,0) {};
	\node (g7) at (7,0) [gauge,label={[yshift=3.2]below:{\scriptsize $2N+1$}}] {};
	\node (g8) at (8,0) [gauge,label={[yshift=3.2]below:{\scriptsize $2N+2$}}] {};
	\node (g9) at (9,0) [gauge,label={[yshift=3.2]below:{\scriptsize $N+2$}}] {};
	\node (g10) at (10,0) [gauge,draw=blue,label={[yshift=3.2]below:{\scriptsize $2$}}] {};
	\node (g11) at (8,1) [gauge,draw=ForestGreen,label={[xshift=-3.2]right:{\scriptsize $N+1$}}] {};
	\draw[gline] (g1)--(g2)--(space1);
	\draw[gline] (space2)--(g7)--(g8)--(g9)--(g10);
	\draw[gline] (g8)--(g11);
	
    \node at (7,-1) {{\scriptsize $\su (4N+8)$}};
\end{tikzpicture}
\ee  

Using this theory as a starting point and with a decoupling we get either $SU(N)_1$ or $SU(N)_0$ with $2N+2$ flavors. By further decoupling (always with positive sign) we get the two sequences $SU(N)_{\frac{k+1}{2}}$ with $2N+3-k$ flavors or  $SU(N)_{\frac{k-1}{2}}$ with $2N+3-k$ flavors. These two sequences are described by the FI deformations in Figure \ref{dndn}.
We note that the magnetic quivers of the descendants of $(D_{N+2},D_{N+2})$ conformal matter are star shaped for $N_f > 2N-1$. At $N_f = 2N-1$ a loop appears, which results in multiple cones for $N_f < 2N-1$. The corresponding descendant tree of GTPs is given in Figure \ref{fig:dndnGTP}.

\begin{figure}[ht]
\begin{center}
\begin{tikzpicture}[]

\filldraw[fill= white] (0,0) circle [radius=0.1] node[below] {\scriptsize 1};
\filldraw[fill= red] (1,0) circle [radius=0.1] node[below] {\scriptsize 2};
\node[] at (2,0){...};
\filldraw[fill= white] (3,0) circle [radius=0.1] node[below] {\scriptsize 2N};
\filldraw[fill= white] (4,0) circle [radius=0.1] node[below] {\scriptsize 2N+1};
\filldraw[fill= white] (5,0) circle [radius=0.1] node[below] {\scriptsize 2N+2};
\filldraw[fill= white] (6,0) circle [radius=0.1] node[below] {\scriptsize N+2};
\filldraw[fill= white] (5,1) circle [radius=0.1] node[above] {\scriptsize N+1};
\filldraw[fill= red] (7,0) circle [radius=0.1] node[below] {\scriptsize 2};
\draw [thick] (0.1, 0) -- (0.9,0) ;
\draw [thick] (1.1, 0) -- (1.75,0) ;
\draw [thick] (2.25, 0) -- (2.9,0) ;
\draw [thick] (3.1, 0) -- (3.9,0) ;
\draw [thick] (4.1, 0) -- (4.9,0) ;
\draw [thick] (5.1, 0) -- (5.9,0) ;
\draw [thick] (6.1,0) -- (6.9,0) ;
\draw [thick] (5,0.1) -- (5,0.9) ;

\draw[->] (3,-0.5)--(3,-2);

\filldraw[fill= red] (0,-3) circle [radius=0.1] node[below] {\scriptsize 1};
\filldraw[fill= white] (1,-3) circle [radius=0.1] node[below] {\scriptsize 2};
\node[] at (2,-3){...};
\filldraw[fill= white] (3,-3) circle [radius=0.1] node[below] {\scriptsize 2N-1};
\filldraw[fill= white] (4,-3) circle [radius=0.1] node[below] {\scriptsize 2N};
\filldraw[fill= white] (5,-3) circle [radius=0.1] node[below] {\scriptsize N+1};
\filldraw[fill= white] (6,-3) circle [radius=0.1] node[below] {\scriptsize 2};
\filldraw[fill= white] (4,-2) circle [radius=0.1] node[above] {\scriptsize N};
\filldraw[fill= red] (7,-3) circle [radius=0.1] node[below] {\scriptsize 1};
\draw [thick] (0.1, -3) -- (0.9,-3) ;
\draw [thick] (1.1, -3) -- (1.75,-3) ;
\draw [thick] (2.25, -3) -- (2.9,-3) ;
\draw [thick] (3.1, -3) -- (3.9,-3) ;
\draw [thick] (4.1, -3) -- (4.9,-3) ;
\draw [thick] (5.1, -3) -- (5.9,-3) ;
\draw [thick] (6.1,-3) -- (6.9,-3) ;
\draw [thick] (4,-2.1) -- (4,-2.9) ;

\draw[->] (3,-3.5)--(3,-5);

\filldraw[fill= red] (0,-6) circle [radius=0.1] node[below] {\scriptsize 1};
\filldraw[fill= white] (1,-6) circle [radius=0.1] node[below] {\scriptsize 2};
\node[] at (2,-6){...};
\filldraw[fill= white] (3,-6) circle [radius=0.1] node[below] {\scriptsize 2N-2};
\filldraw[fill= white] (4,-6) circle [radius=0.1] node[below] {\scriptsize 2N-1};
\filldraw[fill= white] (5,-6) circle [radius=0.1] node[below] {\scriptsize N};
\filldraw[fill= white] (6,-6) circle [radius=0.1] node[below] {\scriptsize 1};
\filldraw[fill= white] (4,-5) circle [radius=0.1] node[above] {\scriptsize N};
\filldraw[fill= red] (5,-5) circle [radius=0.1] node[above] {\scriptsize 1};
\draw [thick] (0.1, -6) -- (0.9,-6);
\draw [thick] (1.1, -6) -- (1.75,-6) ;
\draw [thick] (2.25, -6) -- (2.9,-6) ;
\draw [thick] (3.1, -6) -- (3.9,-6) ;
\draw [thick] (4.1, -6) -- (4.9,-6) ;
\draw [thick] (5.1, -6) -- (5.9,-6) ;
\draw [thick] (4.1,-5) -- (4.9,-5) ;
\draw [thick] (4,-5.1) -- (4,-5.9) ;

\draw[->] (3,-6.5)--(3,-8);

\filldraw[fill= red] (0,-9) circle [radius=0.1] node[below] {\scriptsize 1};
\filldraw[fill= white] (1,-9) circle [radius=0.1] node[below] {\scriptsize 2};
\node[] at (2,-9){...};
\filldraw[fill= white] (3,-9) circle [radius=0.1] node[below] {\scriptsize 2N-3};
\filldraw[fill= white] (4,-9) circle [radius=0.1] node[below] {\scriptsize 2N-2};
\filldraw[fill= white] (5,-9) circle [radius=0.1] node[below] {\scriptsize N};
\filldraw[fill= white] (6,-9) circle [radius=0.1] node[below] {\scriptsize 1};
\filldraw[fill= white] (4,-8) circle [radius=0.1] node[above] {\scriptsize N-1};
\filldraw[fill= red] (5,-8) circle [radius=0.1] node[above] {\scriptsize 1};
\draw [thick] (0.1, -9) -- (0.9,-9);
\draw [thick] (1.1, -9) -- (1.75,-9) ;
\draw [thick] (2.25, -9) -- (2.9,-9) ;
\draw [thick] (3.1, -9) -- (3.9,-9) ;
\draw [thick] (4.1, -9) -- (4.9,-9) ;
\draw [thick] (5.1, -9) -- (5.9,-9) ;
\draw [thick] (5,-8.1) -- (5,-8.9) ;
\draw [thick] (4,-8.1) -- (4,-8.9) ;

\draw[->] (3,-9.5)--(3,-11);

\filldraw[fill= red] (0,-12) circle [radius=0.1] node[below] {\scriptsize 1};
\filldraw[fill= white] (1,-12) circle [radius=0.1] node[below] {\scriptsize 2};
\node[] at (2,-12){...};
\filldraw[fill= white] (3,-12) circle [radius=0.1] node[below] {\scriptsize 2N-4};
\filldraw[fill= white] (4,-12) circle [radius=0.1] node[below] {\scriptsize 2N-3};
\filldraw[fill= white] (5,-12) circle [radius=0.1] node[below] {\scriptsize N-1};
\filldraw[fill= white] (6,-12) circle [radius=0.1] node[below] {\scriptsize 1};
\filldraw[fill= white] (4,-11) circle [radius=0.1] node[above] {\scriptsize N-1};
\filldraw[fill= red] (5,-11) circle [radius=0.1] node[above] {\scriptsize 1};
\draw [thick] (0.1, -12) -- (0.9,-12);
\draw [thick] (1.1, -12) -- (1.75,-12) ;
\draw [thick] (2.25, -12) -- (2.9,-12) ;
\draw [thick] (3.1, -12) -- (3.9,-12) ;
\draw [thick] (4.1, -12) -- (4.9,-12) ;
\draw [thick] (5.1, -12) -- (5.9,-12) ;
\draw [thick] (4.1,-11) -- (4.9,-11) ;
\draw [thick] (4,-11.1) -- (4,-11.9) ;
\draw [thick] (5.05,-11.05) -- (5.95,-11.95) ;

\filldraw[fill= white] (8,0) circle [radius=0.1] node[below] {\scriptsize 1};
\filldraw[fill= red] (10,0) circle [radius=0.1] node[below] {\scriptsize N+1};
\node[] at (9,0){...};
\node[] at (11,0){...};
\filldraw[fill= white] (12,0) circle [radius=0.1] node[below] {\scriptsize 2N+1};
\filldraw[fill= white] (13,0) circle [radius=0.1] node[below] {\scriptsize 2N+2};
\filldraw[fill= white] (14,0) circle [radius=0.1] node[below] {\scriptsize N+2};
\filldraw[fill= red] (13,1) circle [radius=0.1] node[above] {\scriptsize N+1};
\filldraw[fill= white] (15,0) circle [radius=0.1] node[below] {\scriptsize 2};
\draw [thick] (8.1, 0) -- (8.75,0) ;
\draw [thick] (9.25, 0) -- (9.9,0) ;
\draw [thick] (10.1, 0) -- (10.75,0) ;
\draw [thick] (11.25, 0) -- (11.9,0) ;
\draw [thick] (12.1, 0) -- (12.9,0) ;
\draw [thick] (13.1, 0) -- (13.9,0) ;
\draw [thick] (14.1,0) -- (14.9,0) ;
\draw [thick] (13,0.1) -- (13,0.9) ;

\draw[->] (11,-0.5)--(11,-2);

\filldraw[fill= red] (9,-3) circle [radius=0.1] node[below] {\scriptsize 1};
\filldraw[fill= white] (11,-3) circle [radius=0.1] node[below] {\scriptsize N+1};
\node[] at (10,-3){...};
\filldraw[fill= white] (12,-3) circle [radius=0.1] node[below] {\scriptsize N+2};
\filldraw[fill= white] (13,-3) circle [radius=0.1] node[below] {\scriptsize N+1};
\filldraw[fill= white] (12,-2) circle [radius=0.1] node[above] {\scriptsize 2};
\node[] at (14,-3) {...};
\filldraw[fill= red] (15,-3) circle [radius=0.1] node[below] {\scriptsize 1};
\draw [thick] (9.1, -3) -- (9.75,-3) ;
\draw [thick] (10.25, -3) -- (10.9,-3) ;
\draw [thick] (11.1, -3) -- (11.9,-3) ;
\draw [thick] (12.1, -3) -- (12.9,-3) ;
\draw [thick] (13.1, -3) -- (13.75,-3) ;
\draw [thick] (14.25, -3) -- (14.9,-3) ;
\draw [thick] (12,-2.1) -- (12,-2.9) ;

\draw[->] (11,-3.5)--(11,-5);

\filldraw[fill= red] (9,-6) circle [radius=0.1] node[below] {\scriptsize 1};
\filldraw[fill= white] (11,-6) circle [radius=0.1] node[below] {\scriptsize N};
\node[] at (10,-6){...};
\filldraw[fill= white] (12,-6) circle [radius=0.1] node[below] {\scriptsize N+1};
\filldraw[fill= white] (13,-6) circle [radius=0.1] node[below] {\scriptsize N};
\filldraw[fill= white] (12,-5) circle [radius=0.1] node[above] {\scriptsize 2};
\node[] at (14,-6) {...};
\filldraw[fill= white] (15,-6) circle [radius=0.1] node[below] {\scriptsize 1};
\filldraw[fill= red] (13,-5) circle [radius=0.1] node[above] {\scriptsize 1};
\draw [thick] (9.1, -6) -- (9.75,-6) ;
\draw [thick] (10.25, -6) -- (10.9,-6) ;
\draw [thick] (11.1, -6) -- (11.9,-6) ;
\draw [thick] (12.1, -6) -- (12.9,-6) ;
\draw [thick] (13.1, -6) -- (13.75,-6) ;
\draw [thick] (14.25, -6) -- (14.9,-6) ;
\draw [thick] (12,-5.1) -- (12,-5.9) ;
\draw [thick] (12.1, -5) -- (12.9,-5) ;

\draw[->] (11,-6.5)--(11,-8);

\filldraw[fill= red] (9,-9) circle [radius=0.1] node[below] {\scriptsize 1};
\filldraw[fill= white] (11,-9) circle [radius=0.1] node[below] {\scriptsize N-1};
\node[] at (10,-9){...};
\filldraw[fill= white] (12,-9) circle [radius=0.1] node[below] {\scriptsize N};
\filldraw[fill= white] (13,-9) circle [radius=0.1] node[below] {\scriptsize N};
\filldraw[fill= white] (12,-8) circle [radius=0.1] node[above] {\scriptsize 1};
\node[] at (14,-9) {...};
\filldraw[fill= white] (15,-9) circle [radius=0.1] node[below] {\scriptsize 1};
\filldraw[fill= red] (13,-8) circle [radius=0.1] node[above] {\scriptsize 1};
\draw [thick] (9.1, -9) -- (9.75,-9) ;
\draw [thick] (10.25, -9) -- (10.9,-9) ;
\draw [thick] (11.1, -9) -- (11.9,-9) ;
\draw [thick] (12.1, -9) -- (12.9,-9) ;
\draw [thick] (13.1, -9) -- (13.75,-9) ;
\draw [thick] (14.25, -9) -- (14.9,-9) ;
\draw [thick] (12,-8.1) -- (12,-8.9) ;
\draw [thick] (13, -8.1) -- (13,-8.9) ;

\draw[->] (11,-9.5)--(11,-11);

\filldraw[fill= red] (8,-12) circle [radius=0.1] node[below] {\scriptsize 1};
\filldraw[fill= white] (10,-12) circle [radius=0.1] node[below] {\scriptsize N-1};
\node[] at (9,-12){...};
\filldraw[fill= white] (11,-12) circle [radius=0.1] node[below] {\scriptsize N-1};
\filldraw[fill= white] (12,-12) circle [radius=0.1] node[below] {\scriptsize N-1};
\filldraw[fill= white] (14,-12) circle [radius=0.1] node[below] {\scriptsize 1};
\node[] at (13,-12) {...};
\filldraw[fill= white] (10,-11) circle [radius=0.1] node[above] {\scriptsize 1};
\filldraw[fill= red] (12,-11) circle [radius=0.1] node [above] {\scriptsize 1};
\draw [thick] (8.1, -12) -- (8.75,-12) ;
\draw [thick] (9.25, -12) -- (9.9,-12) ;
\draw [thick] (10.1, -12) -- (10.9,-12) ;
\draw [thick] (11.1, -12) -- (11.9,-12) ;
\draw [thick] (12.1, -12) -- (12.75,-12) ;
\draw [thick] (13.25, -12) -- (13.9,-12) ;
\draw [thick] (12,-11.1) -- (12,-11.9) ;
\draw [thick] (10, -11.1) -- (10,-11.9) ;
\draw [thick] (10.1,-11)--(11.9,-11);

\end{tikzpicture} 
\end{center}
\caption{The decoupling trees for $SU(N)_{\frac{k+1}{2}}$ with $2N+3-k$ flavors (on the left) and $SU(N)_{\frac{k-1}{2}}$ with $2N+3-k$ flavors (on the right). We exhibit explicitly all the deformations from $k=0$ to $k=4$. Notice that apart from the first decoupling we always turn on FI parameters at two abelian nodes and in this case the deformation is equivalent to a standard quiver subtraction. If we proceed further with the decoupling  the Higgs branch becomes the union of multiple cones, each described by a quiver. With our procedure this arises because the magnetic quivers for $k=4$ contain closed loops and therefore we have multiple choices for the set of bifundamentals for which we turn on a vev. The FI deformation leading to the $k=5$ magnetic quivers is given explicitly.}\label{dndn}
\end{figure}
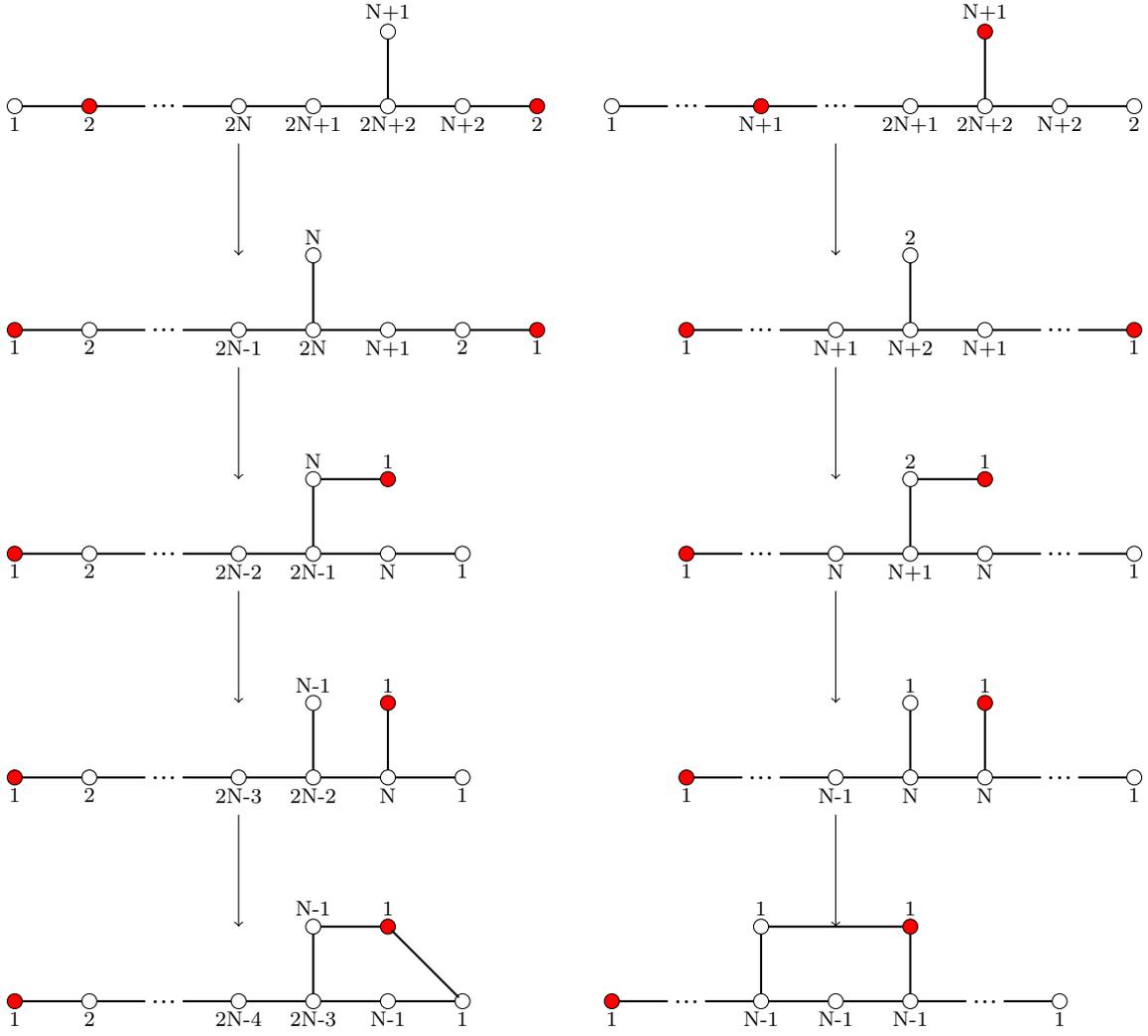 

\begin{figure}
    \centering
\begin{tikzpicture}[x=.5cm,y=.5cm]
	\draw[step=.5cm,gray,very thin] (1,8) grid (11,12);
	
	\draw[ligne,ForestGreen] (5,12)--(1,8)--(5,8);
	\draw[ligne,blue] (5,8)--(7,8);
	\draw[ligne,ForestGreen] (7,8)--(11,8)--(7,12);
	\draw[ligne,blue] (5,12)--(7,12);
	\draw[red] (2,8)--(5,11)--(7,11)--(10,8);
	\draw[red] (5,12)--(5,11)--(7,11)--(7,12);
	
	\foreach \x in {1,2,...,11}
	\node[bd] at (\x,8) {};
	\foreach \x in {1,2,3}
	\node[wd] at (11-\x,8+\x) {};
	\node[bd] at (7,12) {};
	\node[wd] at (6,12) {};
	\node[bd] at (5,12) {};
	\foreach \x in {1,2,3}
	\node[wd] at (5-\x,12-\x) {};
	\foreach \x in {1,2}
	\node[wrd] at (2+\x,8+\x) {};
	\foreach \x in {1,2}
	\node[wrd] at (10-\x,8+\x) {};
	\foreach \x in {0,1,2}
	\node[rd] at (5+\x,11) {};
	
	\node[] at (3,7) {{\scriptsize $N+1$}};	
	\node[] at (6,7) {{\scriptsize $2$}};	
	\node[] at (9,7) {{\scriptsize $N+1$}};
	\node[] at (-.5,10) {{\scriptsize $N+1$}};
	\node[] at (12,9.5) {{\scriptsize $N$}};
	\node[] at (12,11.5) {{\scriptsize $1$}};
	
	\draw[|-|] (.5,8)--(.5,12);
	\draw[|-] (1,7.5)--(5,7.5);
	\draw[|-] (5,7.5)--(7,7.5);
	\draw[|-|] (7,7.5)--(11,7.5);
	\draw[|-|] (11.5,8)--(11.5,11);
	\draw[-|] (11.5,11)--(11.5,12);

	\draw[step=.5cm,gray,very thin] (2,0) grid (10,3);
	\draw[step=.5cm,gray,very thin] (2.5,-8) grid (9.5,-5);
	
	\foreach \x in {16,24}
	\draw[step=.5cm,gray,very thin] (3,-\x) grid (9,-\x+3);
	\draw[ligne,ForestGreen] (5,0)--(2,0)--(5,3);
	\draw[ligne,blue] (5,3)--(7,3);
	\draw[ligne,blue] (5,0)--(7,0);
	\draw[ligne,ForestGreen] (7,3)--(10,0)--(7,0);
	\draw[red] (6,3)--(9,0);
	
	\foreach \x in {2,3,...,10}
	\node[bd] at (\x,0) {};
	\foreach \x in {1,2}
	\node[wd] at (2+\x,\x) {};
	\foreach \x in {1,2}
	\node[wd] at (7+\x,3-\x) {};
	\foreach \x in {0,1,2}
	\node[bd] at (5+\x,3) {};
	\node[wrd] at (7,2) {};
	\node[rd] at (8,1) {};
	
	\node[] at (3.5,-1) {{\scriptsize $N$}};	
	\node[] at (6,-1) {{\scriptsize $2$}};	
	\node[] at (8.5,-1) {{\scriptsize $N$}};
	
	\node[] at (1,1.5) {{\scriptsize $N$}};
	\node[] at (1.5,-6.5) {{\scriptsize $N$}};
	
	\node[] at (11.5,2) {{\scriptsize $N-1$}};
	\node[] at (11,.5) {{\scriptsize $1$}};
	
	\node[] at (11,-6) {{\scriptsize $N-1$}};
	\node[] at (10.5,-7.5) {{\scriptsize $1$}};
	
	\foreach \x in {16,24}
	\node[] at (2,1.5-\x) {{\scriptsize $N$}};
	\foreach \x in {16,24}
	\node[] at (10.5,2-\x) {{\scriptsize $N-1$}};
	\foreach \x in {16,24}
	\node[] at (10,-15.5) {{\scriptsize $1$}};
	
	\draw[|-|] (1.5,0)--(1.5,3);
	\draw[|-|] (2,-8)--(2,-5);
	
	\draw[|-|] (10.5,1)--(10.5,3);
	\draw[|-] (10.5,0)--(10.5,1);

	\draw[|-|] (10,-7)--(10,-5);
	\draw[|-] (10,-8)--(10,-7);
	
	\foreach \x in {16,24}
	\draw[|-|] (2.5,-\x)--(2.5,-\x+3);
	\foreach \x in {16,24}
	\draw[|-|] (9.5,-\x+1)--(9.5,-\x+3);
	\foreach \x in {16,24}
	\draw[|-] (9.5,-\x)--(9.5,-\x+1);
	
	\draw[|-] (2,-.5)--(5,-.5);
	\draw[|-] (5,-.5)--(7,-.5);
	\draw[|-|] (7,-.5)--(10,-.5);
	
	\foreach \x in {6,-2,-10,-18}
	\draw[->] (6,\x)--(6,\x-2);
	
	\draw[ligne,ForestGreen] (5.5,-8)--(2.5,-8)--(5.5,-5);
	\draw[ligne,blue] (5.5,-5)--(6.5,-5);
	\draw[ligne,blue] (5.5,-8)--(6.5,-8);
	\draw[ligne,ForestGreen] (6.5,-5)--(9.5,-8)--(6.5,-8);
	\draw[red] (3.5,-8)--(6.5,-5);
	
	\foreach \x in {2,3,...,9}
	\node[bd] at (.5+\x,-8) {};
	\foreach \x in {2,3}
	\node[wd] at (1.5+\x,\x-9) {};
	\foreach \x in {0,1}
	\node[bd] at (5.5+\x,-5) {};
	\node[wrd] at (5.5,-6) {};
	\node[wd] at (7.5,-6) {};
	\node[rd] at (4.5,-7) {};
	\node[bd] at (8.5,-7) {};
	
	\node[] at (4,-9) {{\scriptsize $N$}};	
	\node[] at (6,-9) {{\scriptsize $1$}};	
	\node[] at (8,-9) {{\scriptsize $N$}};
	
	\draw[|-] (2.5,-8.5)--(5.5,-8.5);
	\draw[|-] (5.5,-8.5)--(6.5,-8.5);
	\draw[|-|] (6.5,-8.5)--(9.5,-8.5);

	\draw[ligne,ForestGreen] (3,-16)--(6,-13)--(9,-16)--(3,-16);
	\draw[red] (4,-16)--(4,-15);
	
	\foreach \x in {3,4,...,9}
	\node[bd] at (\x,-16) {};
	\node[bd] at (6,-13) {};
	\foreach \x in {0,2}
	\node[wd] at (5+\x,-14) {};
	\foreach \x in {0,4}
	\node[bd] at (4+\x,-15) {};
	
	\node[] at (4.5,-17) {{\scriptsize $N$}};
	\node[] at (7.5,-17) {{\scriptsize $N$}};
	
	\draw[|-] (3,-16.5)--(6,-16.5);
	\draw[|-|] (6,-16.5)--(9,-16.5);

	\draw[ligne,ForestGreen] (4,-24)--(8,-24);
	\draw[ligne,ForestGreen] (4,-23)--(6,-21)--(8,-23);
	\draw[ligne,blue] (4,-24)--(4,-23);
	\draw[ligne,blue] (8,-24)--(9,-24)--(8,-23);
	\draw[red] (8,-24)--(8,-23);
	
	\foreach \x in {4,5,...,9}
	\node[bd] at (\x,-24) {};
	\node[bd] at (6,-21) {};
	\foreach \x in {0,2}
	\node[wd] at (5+\x,-22) {};
	\foreach \x in {0,4}
	\node[bd] at (4+\x,-23) {};
	
	\node[] at (5,-25) {{\scriptsize $N-1$}};
	\node[] at (7.5,-25) {{\scriptsize $N$}};
	
	\draw[|-] (4,-24.5)--(6,-24.5);
	\draw[|-|] (6,-24.5)--(9,-24.5);
	
	
	\draw[step=.5cm,gray,very thin] (16,8) grid (26,12);
	
	\draw[ligne,ForestGreen] (20,12)--(16,8)--(20,8);
	\draw[ligne,blue] (20,8)--(22,8);
	\draw[ligne,ForestGreen] (22,8)--(26,8)--(22,12);
	\draw[ligne,blue] (20,12)--(22,12);
	\draw[red] (22,12)--(22,8);
	
	\foreach \x in {1,2,...,11}
	\node[bd] at (15+\x,8) {};
	\foreach \x in {1,2,3}
	\node[wd] at (26-\x,8+\x) {};
	\node[bd] at (22,12) {};
	\node[wd] at (21,12) {};
	\node[bd] at (20,12) {};
	\foreach \x in {1,2,3}
	\node[wd] at (20-\x,12-\x) {};
	\foreach \x in {0,1,2}
	\node[rd] at (22,9+\x) {};
	
	\node[] at (18,7) {{\scriptsize $N+1$}};	
	\node[] at (21,7) {{\scriptsize $2$}};	
	\node[] at (24,7) {{\scriptsize $N+1$}};
	\node[] at (14.5,10) {{\scriptsize $N+1$}};
	\node[] at (27,9.5) {{\scriptsize $N$}};
	\node[] at (27,11.5) {{\scriptsize $1$}};
	
	\draw[|-|] (15.5,8)--(15.5,12);
	\draw[|-] (16,7.5)--(20,7.5);
	\draw[|-] (20,7.5)--(22,7.5);
	\draw[|-|] (22,7.5)--(26,7.5);
	\draw[|-|] (26.5,8)--(26.5,11);
	\draw[-|] (26.5,11)--(26.5,12);

	\foreach \x in {1,2,3}
	\draw[step=.5cm,gray,very thin] (20,-8*\x) grid (22,4-8*\x);
	
%
%
%

	\draw[step=.5cm,gray,very thin] (18,0) grid (24,4);
	
	\foreach \x in {0,4}
	\draw[ligne,blue] (22,\x)--(24,\x);
	\draw[ligne,ForestGreen] (22,4)--(18,0)--(22,0);
	\draw[ligne,ForestGreen] (24,0)--(24,4);
	\draw[red] (22,0)--(22,4);	
	
	\foreach \x in {0,1,...,6}
	\node[bd] at (18+\x,0) {};				
	\foreach \x in {1,2,...,4}
	\node[bd] at (24,\x) {};
	\foreach \x in {1,2,3}
	\node[wd] at (18+\x,\x) {};
	\foreach \x in {1,2,3}
	\node[rd] at (22,\x) {};
	\node[bd] at (22,4) {};		
	\node[wd] at (23,4) {};		
	
	\node[] at (23,-1) {{\scriptsize $2$}};
	\draw[|-|] (22,-.5)--(24,-.5);	
	
	\node[] at (25.5,2) {{\scriptsize $N+1$}};	
	\draw[|-|] (24.5,0)--(24.5,4);
	
	\node[] at (20,-1) {{\scriptsize $N+1$}};
	\draw[|-] (18,-.5)--(22,-.5);

	\foreach \x in {9,17,25}
	\node[] at (23,4.5-\x) {{\scriptsize $1$}};
	\foreach \x in {9,17,25}
	\draw[-|] (22.5,4-\x)--(22.5,5-\x); 
	
	\foreach \x in {-6.5,1.5,9.5,17.5}
	\draw[->] (21,-\x)--(21,-\x-2);
	
	\foreach \x in {8,16,24}
	\node[] at (21,-\x-1) {{\scriptsize $2$}};
	\foreach \x in {8,16,24}
	\draw[|-|] (20,-\x-.5)--(22,-\x-.5);

	\foreach \x in {0,4}
	\draw[ligne,blue] (20,\x-8)--(22,\x-8);
	\foreach \x in {0,2}
	\draw[ligne,ForestGreen] (\x+20,-8)--(\x+20,-4);
	\draw[red] (21,-4)--(22,-5);	
	
	\foreach \x in {0,1,...,4}
	\node[bd] at (20,\x-8) {};				
	\foreach \x in {0,1,...,4}
	\node[bd] at (22,\x-8) {};
	\node[wd] at (21,-8) {};
	\node[bd] at (21,-4) {};
	
	\node[] at (18.5,-6) {{\scriptsize $N+1$}};
	\draw[|-|] (19.5,-8)--(19.5,-4);

	\foreach \x in {0,4}
	\draw[ligne,blue] (20,\x-16)--(21,\x-16);
	\foreach \x in {0,2}
	\draw[ligne,ForestGreen] (\x+20,-16)--(\x+20,-13);
	\draw[ligne,cyan] (20,-12)--(20,-13);
	\draw[ligne,cyan] (21,-12)--(22,-13);
	\draw[ligne,cyan] (21,-16)--(22,-16);
	\draw[red] (21,-12)--(20,-13);	
	
	\foreach \x in {0,1,...,4}
	\node[bd] at (20,\x-16) {};				
	\foreach \x in {0,1,...,3}
	\node[bd] at (22,\x-16) {};
	\node[wd] at (21,-16) {};
	\node[bd] at (21,-12) {};
	
	\node[] at (18.5,-14) {{\scriptsize $N+1$}};
	\draw[|-|] (19.5,-16)--(19.5,-12); 	
	\foreach \x in {0,8}			
	\node[] at (23,-14.5+\x) {{\scriptsize $N$}};	
	\foreach \x in {0,8}
	\draw[|-|] (22.5,-16+\x)--(22.5,-13+\x);

	\draw[ligne,blue] (20,-21)--(21,-20);
	\draw[ligne,blue] (22,-22)--(22,-21);
	\draw[ligne,blue] (20,-24)--(21,-24);
	\foreach \x in {0,2}
	\draw[ligne,ForestGreen] (\x+20,-22)--(\x+20,-24);
	\draw[ligne,cyan] (20,-22)--(20,-21);
	\draw[ligne,cyan] (21,-20)--(22,-21);
	\draw[ligne,cyan] (21,-24)--(22,-24);
	\draw[red] (21,-20)--(20,-22);	
	
	\foreach \x in {0,1,2,3}
	\node[bd] at (20,\x-24) {};				
	\foreach \x in {0,1,2,3}
	\node[bd] at (22,\x-24) {};
	\node[wd] at (21,-24) {};
	\node[bd] at (21,-20) {};
	
	\node[] at (23,-22.5) {{\scriptsize $N$}};
	\draw[|-|] (22.5,-24)--(22.5,-21);
	\node[] at (18.5,-22) {{\scriptsize $N+1$}};
	\draw[|-|] (19.5,-24)--(19.5,-20);
\end{tikzpicture}
    \caption{The GTP decoupling trees for $SU(N)_{\frac{k+1}{2}}$ with $2N+3-k$ flavors (on the left) and $SU(N)_{\frac{k-1}{2}}$ with $2N+3-k$ flavors (on the right), drawn for $N=3$, corresponding to the FI deformations shown in Figure \ref{dndn}.}
    \label{fig:dndnGTP}
\end{figure}
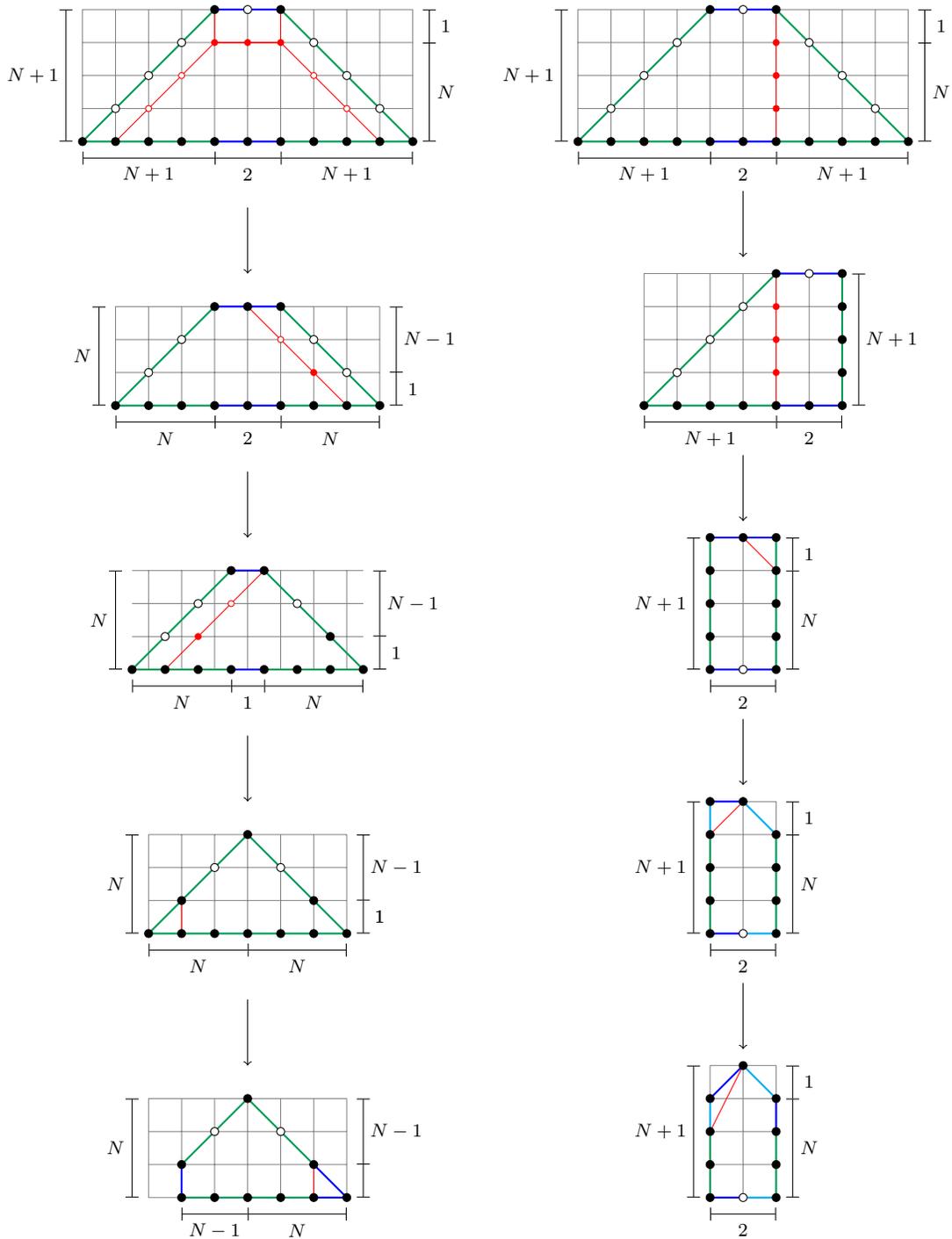

\FloatBarrier

\section{\texorpdfstring{$T_N$}{TN} Theory} 
\label{sec:TN}

We will now study the $T_N$ theory, its parent SCFTs and its Lagrangian descriptions. In particular we are interested in relating the magnetic quivers describing the Higgs branches of these 5d theories through FI deformations. The GTP of $T_N$ is (drawn for $N=5$)
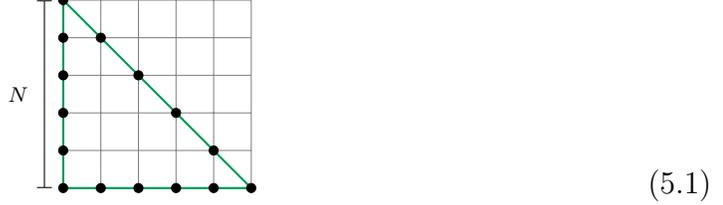
\be
\begin{tikzpicture}[x=.5cm,y=.5cm]
	\draw[step=.5cm,gray,very thin] (0,0) grid (5,5);
	
	\draw[ligne,ForestGreen] (0,0)--(5,0)--(0,5)--(0,0);
	
	\foreach \x in {0,1,...,5}
	\node[bd] at (0,\x) {};
	\foreach \x in {1,2,...,5}
	\node[bd] at (\x,0) {};
	\foreach \x in {1,2,...,4}
	\node[bd] at (0+\x,5-\x) {};
		
	\node[] at (-1.2,2.5) {{\scriptsize $N$}};
	
	\draw[|-|] (-.5,0)--(-.5,5);
\end{tikzpicture}
\ee
where we have also given the unique edge coloring with
\be 
\ba 
\lambda_\alpha&=(N,N,N)\,, \qquad && \mu_{\alpha }=(\{1^N\},\{1^N\},\{1^N\})\,, \\
\lambda^g_\alpha&=(N,N,N)\,, \qquad && \mu^g_{\alpha }=(\{N\},\{N\},\{N\})\,,
\ea 
\ee 
which results in the nodes
\be 
m^g=N\,, \qquad m_{\alpha ,x}=(N-1,N-2,\dots,1) \ \forall \alpha
\ee 
and non-zero intersections
\be 
k^g_{\alpha ,1}=1 \ \forall \alpha
\ee 
along with nearest neighbours intersecting once in the tails. Hence, the magnetic quiver is
\be 
\label{MQTN}
\begin{tikzpicture}
	\node (g1) at (0,0) [gauge,label={[yshift=3.2]below:{\scriptsize $1$}}] {};
	\node (g2) at (1,0) [gauge,label={[yshift=3.2]below:{\scriptsize $2$}}] {};
	\node (space1) at (1.7,0) {};
	\node (dots) at (2,0) {$\cdots$};
	\node (space2) at (2.3,0) {};
	\node (g3) at (3,0) [gauge,label={[yshift=3.2]below:{\scriptsize $N-1$}}] {};
	\node (g4) at (4,0) [gauge,draw=ForestGreen,label={[yshift=3.2]below:{\scriptsize $N$}}] {};
	\node (g5) at (5,0) [gauge,label={[yshift=3.2]below:{\scriptsize $N-1$}}] {};
	\node (space3) at (5.7,0) {};
	\node (dots) at (6,0) {$\cdots$};
	\node (space4) at (6.3,0) {};
	\node (g6) at (7,0) [gauge,label={[yshift=3.2]below:{\scriptsize $2$}}] {};
	\node (g7) at (8,0) [gauge,label={[yshift=3.2]below:{\scriptsize $1$}}] {};
	\node (g8) at (4,1) [gauge,label={[xshift=3.2]left:{\scriptsize $N-1$}}] {};
	\node (space5) at (4,1.7) {};
	\node (dots) at (4,2.1) {$\vdots$};
	\node (space6) at (4,2.3) {};
	\node (g9) at (4,3) [gauge,label={[xshift=3.2]left:{\scriptsize $2$}}] {};
	\node (g10) at (4,4) [gauge,label={[xshift=3.2]left:{\scriptsize $1$}}] {};
	\draw[gline] (g1)--(g2)--(space1);
	\draw[gline] (space2)--(g3)--(g4)--(g5)--(space3);
	\draw[gline] (space4)--(g6)--(g7);
	\draw[gline] (g4)--(g8)--(space5);
	\draw[gline] (space6)--(g9)--(g10);
	
	\node at (4,-1) {{\scriptsize $\su (N)^3$}};	
\end{tikzpicture}
\ee

\subsection{Parent SCFTs}
The respective parent and ``grandparent'' SCFTs of $T_N$ are the so-called $P_N$ and $G_N$ theories. The GTPs of these theories can be obtained by constructing the pre-GTP representing their low energy Lagrangian descriptions 
\be 
\ba 
G_N\,: \ &[N+1]-SU(N-1)_\half-\cdots-SU(2)-[3]\\
&[N+2]-SU(N-1)_0-\cdots-SU(2)-[2]\\\\
P_N\,: \ &[N]-SU(N-1)_0-\cdots-SU(2)-[3]\\
&[N+1]-SU(N-1)_\half-\cdots-SU(2)-[2]
\ea
\ee 
and performing edge-moves to reach the SCFT point. For $G_N$ we find
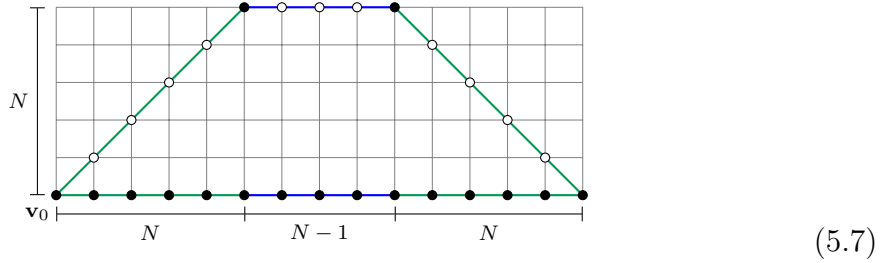
\be
\label{GTPGN}
\begin{tikzpicture}[x=.5cm,y=.5cm]
	\draw[step=.5cm,gray,very thin] (0,0) grid (14,5);
	
	\draw[ligne,ForestGreen] (5,5)--(0,0)--(5,0);
	\draw[ligne,blue] (5,0)--(9,0);
	\draw[ligne,ForestGreen] (9,0)--(14,0)--(9,5);
	\draw[ligne,blue] (9,5)--(5,5);
	
	\foreach \x in {0,1,...,14}
	\node[bd] at (\x,0) {};
	\foreach \x in {1,2,3,4}
	\node[wd] at (14-\x,\x) {};
	\node[bd] at (9,5) {};
	\foreach \x in {1,2,3}
	\node[wd] at (9-\x,5) {};
	\node[bd] at (5,5) {};
	\foreach \x in {1,2,3,4}
	\node[wd] at (5-\x,5-\x) {};
	
	\node[] at (2.5,-1) {{\scriptsize $N$}};	
	\node[] at (7,-1) {{\scriptsize $N-1$}};	
	\node[] at (11.5,-1) {{\scriptsize $N$}};
	\node[] at (-1,2.5) {{\scriptsize $N$}};
	\node[] at (-.5,-.5) {{\scriptsize $\mathbf{v}_0$}};
	
	\draw[|-|] (-.5,0)--(-.5,5);
	\draw[|-] (0,-.5)--(5,-.5);
	\draw[|-] (5,-.5)--(9,-.5);
	\draw[|-|] (9,-.5)--(14,-.5);
\end{tikzpicture}
\ee
with 
\be 
\ba 
\lambda_\alpha&=(3N-1,N,N-1,N)\,, \qquad && \mu_{\alpha }=(\{1^{3N-1}\},\{N\},\{N-1\},\{N\})\,, \\
\lambda^b_\alpha&=(N-1,0,N-1,0)\,, \qquad && \mu^b_{\alpha }=(\{N-1\},-,\{N-1\},-)\,, \\ \lambda^g_\alpha&=(2N,N,0,N)\,, \qquad && \mu^g_{\alpha }=(\{N,N\},\{N\},-,\{N\})\,,
\ea 
\ee 
which results in the nodes
\be 
m^b=N-1\,, \qquad \ m^g=N\,, \qquad m_{1 x}=(2N-2, 3N-3,3N-4,\dots,1)\,,
\ee 
and intersections
\be 
k^g_{1 ,2}=1\,, \qquad k^b_{1,1}=1\,,
\ee 
as well as nearest neighbours intersecting once in the tail. Thus, the magnetic quiver is
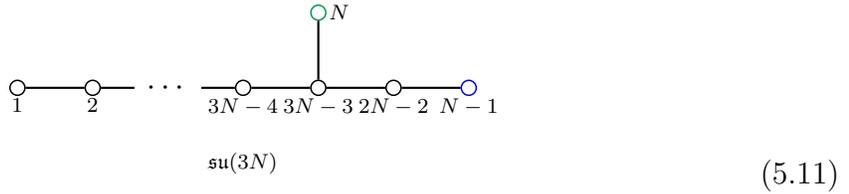
\be 
\begin{tikzpicture}
	\node (g1) at (4,0) [gauge,label={[yshift=3.2]below:{\scriptsize $1$}}] {};
	\node (g2) at (5,0) [gauge,label={[yshift=3.2]below:{\scriptsize $2$}}] {};
	\node (space1) at (5.7,0) {};
	\node (dots) at (6,0) {$\cdots$};
	\node (space2) at (6.3,0) {};
	\node (g7) at (7,0) [gauge,label={[yshift=3.2]below:{\scriptsize $3N-4$}}] {};
	\node (g8) at (8,0) [gauge,label={[yshift=3.2]below:{\scriptsize $3N-3$}}] {};
	\node (g9) at (9,0) [gauge,label={[yshift=3.2]below:{\scriptsize $2N-2$}}] {};
	\node (g10) at (10,0) [gauge,draw=blue,label={[yshift=3.2]below:{\scriptsize $N-1$}}] {};
	\node (g11) at (8,1) [gauge,draw=ForestGreen,label={[xshift=-3.2]right:{\scriptsize $N$}}] {};
	\draw[gline] (g1)--(g2)--(space1);
	\draw[gline] (space2)--(g7)--(g8)--(g9)--(g10);
	\draw[gline] (g8)--(g11);
	\node at (7,-1) {{\scriptsize $\su (3N)$}};
\end{tikzpicture}
\ee  
The matter curves which give $P_N$ and $T_N$ are
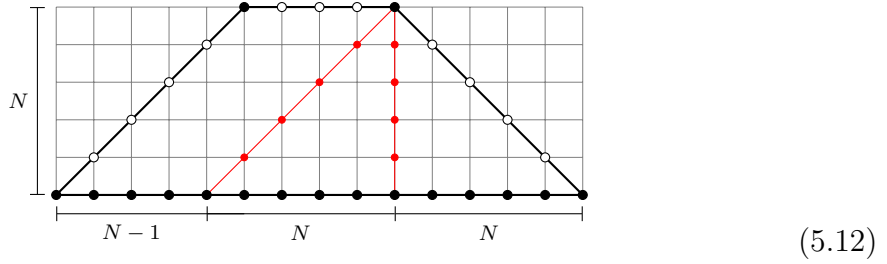
\be
\label{GNGTP}
\begin{tikzpicture}[x=.5cm,y=.5cm]
	\draw[step=.5cm,gray,very thin] (0,0) grid (14,5);
	
	\draw[ligne] (0,0)--(14,0)--(9,5)--(5,5)--(0,0);
	\draw[red] (9,0)--(9,5)--(4,0);
	
	\foreach \x in {0,1,...,14}
	\node[bd] at (\x,0) {};
	\foreach \x in {1,2,3,4}
	\node[wd] at (14-\x,\x) {};
	\node[bd] at (9,5) {};
	\foreach \x in {1,2,3}
	\node[wd] at (9-\x,5) {};
	\node[bd] at (5,5) {};
	\foreach \x in {1,2,3,4}
	\node[wd] at (5-\x,5-\x) {};
	
	\foreach \x in {1,2,3,4}
	\node[rd] at (4+\x,\x) {};
	\foreach \x in {1,2,3,4}
	\node[rd] at (9,\x) {};
	
	\node[] at (2,-1) {{\scriptsize $N-1$}};	
	\node[] at (6.5,-1) {{\scriptsize $N$}};	
	\node[] at (11.5,-1) {{\scriptsize $N$}};
	\node[] at (-1,2.5) {{\scriptsize $N$}};

	\draw[|-|] (-.5,0)--(-.5,5);
	\draw[|-] (0,-.5)--(5,-.5);
	\draw[|-] (4,-.5)--(9,-.5);
	\draw[|-|] (9,-.5)--(14,-.5);
\end{tikzpicture}
\ee
Resolving by either curve on its own and decoupling gives a representation of $P_N$:
\be
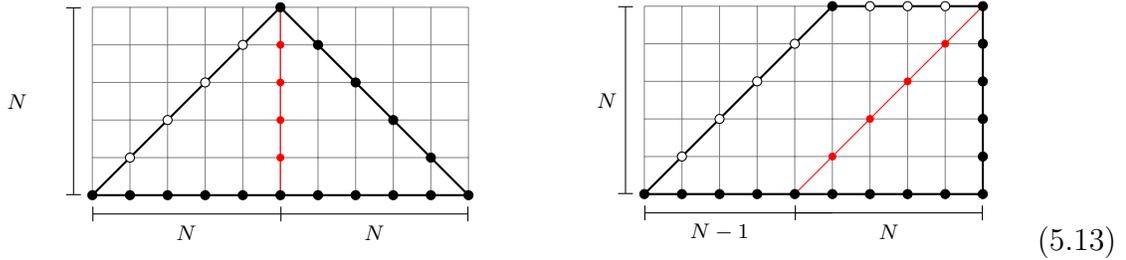
\label{GTPPN}
\begin{tikzpicture}[x=.5cm,y=.5cm]
	\draw[step=.5cm,gray,very thin] (4,0) grid (14,5);
	
	\draw[ligne] (4,0)--(14,0)--(9,5)--(4,0);
	\draw[red] (9,0)--(9,5);
	
	\foreach \x in {4,5,...,14}
	\node[bd] at (\x,0) {};
	\foreach \x in {1,2,3,4}
	\node[wd] at (4+\x,\x) {};
	\node[bd] at (9,5) {};
	
	\foreach \x in {1,2,3,4}
	\node[bd] at (14-\x,\x) {};
	\foreach \x in {1,2,3,4}
	\node[rd] at (9,\x) {};
	
	\node[] at (6.5,-1) {{\scriptsize $N$}};	
	\node[] at (11.5,-1) {{\scriptsize $N$}};
	\node[] at (2,2.5) {{\scriptsize $N$}};
	
	\draw[|-|] (3.5,0)--(3.5,5);
	\draw[|-] (4,-.5)--(9,-.5);
	\draw[|-|] (9,-.5)--(14,-.5);
\end{tikzpicture}\hspace{1.5cm}
\begin{tikzpicture}[x=.5cm,y=.5cm]
	\draw[step=.5cm,gray,very thin] (0,0) grid (9,5);
	
	\draw[ligne] (0,0)--(9,0)--(9,5)--(5,5)--(0,0);	\draw[red] (9,5)--(4,0);
	
	\foreach \x in {0,1,...,9}
	\node[bd] at (\x,0) {};
	\node[bd] at (9,5) {};
	\foreach \x in {1,2,3}
	\node[wd] at (9-\x,5) {};
	\node[bd] at (5,5) {};
	\foreach \x in {1,2,3,4}
	\node[wd] at (5-\x,5-\x) {};
	
	\foreach \x in {1,2,3,4}
	\node[rd] at (4+\x,\x) {};
	\foreach \x in {1,2,3,4}
	\node[bd] at (9,\x) {};
	
	\node[] at (2,-1) {{\scriptsize $N-1$}};	
	\node[] at (6.5,-1) {{\scriptsize $N$}};	
	\node[] at (-1,2.5) {{\scriptsize $N$}};
	
	\draw[|-|] (-.5,0)--(-.5,5);
	\draw[|-] (0,-.5)--(5,-.5);
	\draw[|-|] (4,-.5)--(9,-.5);
\end{tikzpicture}
\ee
The two representations are related by pruning the horizontal edge, i.e. in the left polygon we move a single segment from the bottom edge anti-clockwise through the polygon to give the top horizontal edge in the right polygon. Three edges are created in the process, realizing Hanany-Witten brane creation.
Further decoupling using the other curve gives $T_N$.

The magnetic quiver of $P_N$ can be obtained from that of $G_N$ by turning on FI deformations at the $U(N)$ node in the tail and the green node of the same rank (equivalently, we could use the $U(2N-1)$ node in the tail and the two color nodes). 
Subsequently, the magnetic quiver of $T_N$ may be obtained by turning on an FI parameter either at the $U(N-1)$ node in the long tail and the central $U(N-1)$ node (red), or at the two $U(N)$ nodes (blue): 
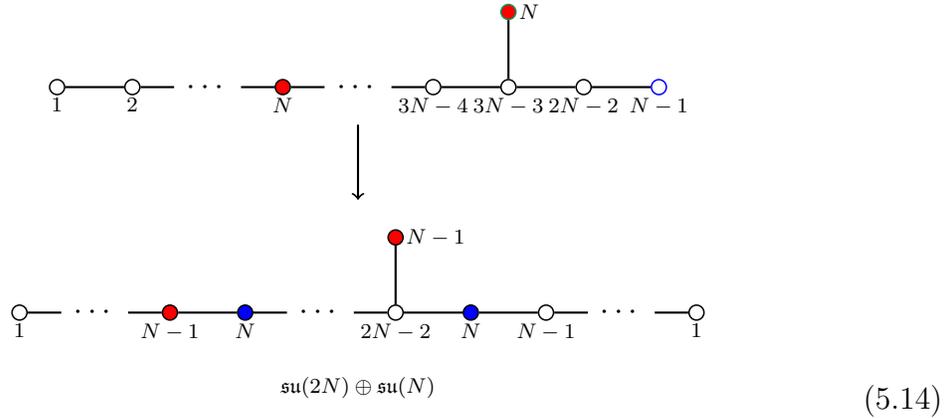
\be\label{gntopn} 
\begin{tikzpicture}[cross/.style={path picture={ 
  \draw[red]
(path picture bounding box.south east) -- (path picture bounding box.north west) (path picture bounding box.south west) -- (path picture bounding box.north east);
}}]
	\node (g1) at (1,2) [gauge,label={[yshift=3.2]below:{\scriptsize $1$}}] {};
	\node (g2) at (2,2) [gauge,label={[yshift=3.2]below:{\scriptsize $2$}}] {};
	\node (space1) at (2.7,2) {};
	\node (dots) at (3,2) {$\cdots$};
	\node (space2) at (3.3,2) {};
	\node (g7) at (4,2) [gauger,label={[yshift=3.2]below:{\scriptsize $N$}}] {};
	\node (space3) at (4.7,2) {};
	\node (dots) at (5,2) {$\cdots$};
	\node (space4) at (5.3,2) {};
	\node (g8) at (6,2) [gauge,label={[yshift=3.2]below:{\scriptsize $3N-4$}}] {};
	\node (g9) at (7,2) [gauge,label={[yshift=3.2]below:{\scriptsize $3N-3$}}] {};
	\node (g10) at (8,2) [gauge,label={[yshift=3.2]below:{\scriptsize $2N-2$}}] {};
	\node (g11) at (9,2) [gauge,draw=blue,label={[yshift=3.2]below:{\scriptsize $N-1$}}] {};
	\node (g12) at (7,3) [gauger,draw=ForestGreen,label={[xshift=-3.2]right:{\scriptsize $N$}}] {};
	\draw[gline] (g1)--(g2)--(space1);
	\draw[gline] (space2)--(g7)--(space3);
	\draw[gline] (space4)--(g8)--(g9)--(g10)--(g11);
	\draw[gline] (g9)--(g12);

%
	
	\draw[->,thick] (5,1.5)--(5,.5);
	
	\node (g1) at (.5,-1) [gauge,label={[yshift=3.2]below:{\scriptsize $1$}}] {};
	\node (space1) at (1.2,-1) {};
	\node (dots) at (1.5,-1) {$\cdots$};
	\node (space2) at (1.8,-1) {};	
	\node (g2) at (2.5,-1) [gauger,label={[yshift=3.2]below:{\scriptsize $N-1$}}] {};
	\node (g7) at (3.5,-1) [gauge,fill=blue,label={[yshift=3.2]below:{\scriptsize $N$}}] {};
	\node (space3) at (4.2,-1) {};
	\node (dots) at (4.5,-1) {$\cdots$};
	\node (space4) at (4.8,-1) {};	
	\node (g8) at (5.5,-1) [gauge,label={[yshift=3.2]below:{\scriptsize $2N-2$}}] {};
	\node (g9) at (6.5,-1) [gauge,fill=blue,label={[yshift=3.2]below:{\scriptsize $N$}}] {};
	\node (g10) at (7.5,-1) [gauge,label={[yshift=3.2]below:{\scriptsize $N-1$}}] {};
	\node (space5) at (8.2,-1) {};
	\node (dots) at (8.5,-1) {$\cdots$};
	\node (space6) at (8.8,-1) {};
	\node (g11) at (9.5,-1) [gauge,label={[yshift=3.2]below:{\scriptsize $1$}}] {};
	\node (g13) at (5.5,0) [gauger,label={[xshift=-3.2]right:{\scriptsize $N-1$}}] {};
	\draw[gline] (g1)--(space1);
	\draw[gline] (space2)--(g2)--(g7)--(space3);
	\draw[gline] (space4)--(g8)--(g9)--(g10)--(space5);
	\draw[gline] (space6)--(g11);
	\draw[gline] (g8)--(g13);
	
	\node at (5,-2) {{\scriptsize $\su (2N) \oplus \su (N)$}};
\end{tikzpicture}
\ee  
From either of these transitions we recover the magnetic quiver of $T_N$.

\subsection{IR Descriptions}

$T_N$ has a Lagrangian description as
\be 
\label{TNLagrange}
[N]-SU(N-1)_0-SU(N-2)- \cdots -SU(2)-[2]\,.
\ee 
In the toric polygon, we can go to this weak coupling description by successively introducing a set of $N-2$ internal edges as follows
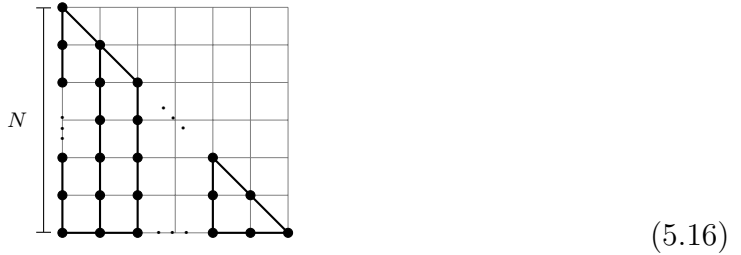
\be \label{TNruled}
\begin{tikzpicture}[x=.5cm,y=.5cm]
	\draw[step=.5cm,gray,very thin] (0,0) grid (6,6);
	
	\draw[ligne,black] (0,2)--(0,0)--(2,0);
	\draw[ligne,black] (4,0)--(6,0)--(4,2);
	\draw[ligne,black] (2,4)--(0,6)--(0,4);
	\foreach \x in {1,2,4}
	\draw[ligne,black] (\x,0)--(\x,6-\x);
	
	\foreach \x in {0,1,2,4,5,6}
	\node[bd] at (0,\x) {};
	\foreach \x in {1,2,4,5,6}
	\node[bd] at (\x,0) {};
	\foreach \x in {1,2,4,5}
	\node[bd] at (0+\x,6-\x) {};
	\foreach \x in {1,2,3,4}
	\node[bd] at (1,\x) {};
	\foreach \x in {1,2,3}
	\node[bd] at (2,\x) {};
	\node[bd] at (4,1) {};

	\node[] at (3,0) {$\dots$};
	\node[] at (0,3) {$\vdots$};
	\node[rotate=-45] at (3,3) {$\dots$};
	
	\node[] at (-1.2,3) {{\scriptsize $N$}};
	
	\draw[|-|] (-.5,0)--(-.5,6);
\end{tikzpicture}
\ee
The left-most ruling corresponds to turning on a SYM term for the $SU(N-1)$ gauge group. The magnetic quiver of this theory is obtained from that of $T_N$ by an FI deformation at two $U(1)$ nodes
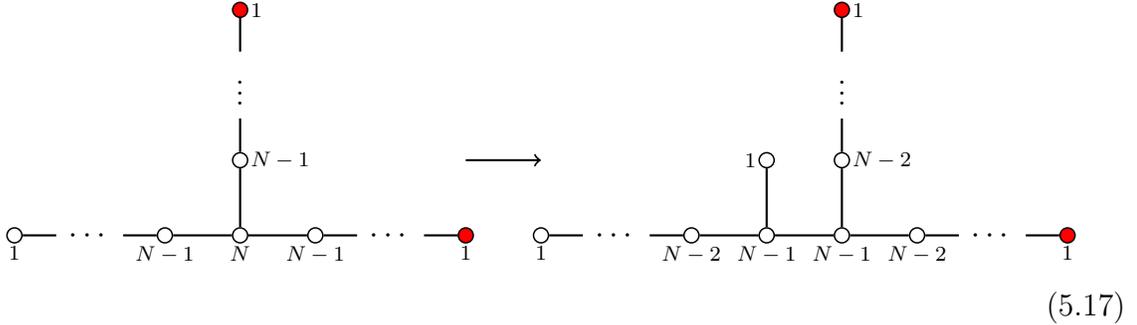
\be 
\label{MQTNLagrange}
\begin{tikzpicture}
	\node (g2) at (1,0) [gauge,label={[yshift=3.2]below:{\scriptsize $1$}}] {};
	\node (space1) at (1.7,0) {};
	\node (dots) at (2,0) {$\cdots$};
	\node (space2) at (2.3,0) {};
	\node (g3) at (3,0) [gauge,label={[yshift=3.2]below:{\scriptsize $N-1$}}] {};
	\node (g4) at (4,0) [gauge,label={[yshift=3.2]below:{\scriptsize $N$}}] {};
	\node (g5) at (5,0) [gauge,label={[yshift=3.2]below:{\scriptsize $N-1$}}] {};
	\node (space3) at (5.7,0) {};
	\node (dots) at (6,0) {$\cdots$};
	\node (space4) at (6.3,0) {};
	\node (g6) at (7,0) [gauger,label={[yshift=3.2]below:{\scriptsize $1$}}] {};
	\node (g8) at (4,1) [gauge,label={[xshift=-3.2]right:{\scriptsize $N-1$}}] {};
	\node (space5) at (4,1.7) {};
	\node (dots) at (4,2) {$\vdots$};
	\node (space6) at (4,2.3) {};
	\node (g9) at (4,3) [gauger,label={[xshift=-3.2]right:{\scriptsize $1$}}] {};
	\draw[gline] (g2)--(space1);
	\draw[gline] (space2)--(g3)--(g4)--(g5)--(space3);
	\draw[gline] (space4)--(g6);
	\draw[gline] (g4)--(g8)--(space5);
	\draw[gline] (space6)--(g9);
	
	\draw[->,thick] (7,1)--(8,1);
	
\node (g2) at (8,0) [gauge,label={[yshift=3.2]below:{\scriptsize $1$}}] {};
\node (space1) at (8.7,0) {};
\node (dots) at (9,0) {$\cdots$};
\node (space2) at (9.3,0) {};
\node (g3) at (10,0) [gauge,label={[yshift=3.2]below:{\scriptsize $N-2$}}] {};
\node (g4) at (11,0) [gauge,label={[yshift=3.2]below:{\scriptsize $N-1$}}] {};
\node (g5) at (12,0) [gauge,label={[yshift=3.2]below:{\scriptsize $N-1$}}] {};
\node (g11) at (13,0) [gauge,label={[yshift=3.2]below:{\scriptsize $N-2$}}] {};
\node (space3) at (13.7,0) {};
\node (dots) at (14,0) {$\cdots$};
\node (space4) at (14.3,0) {};
\node (g6) at (15,0) [gauger,label={[yshift=3.2]below:{\scriptsize $1$}}] {};
\node (g8) at (12,1) [gauge,label={[xshift=-3.2]right:{\scriptsize $N-2$}}] {};
\node (space5) at (12,1.7) {};
\node (dots) at (12,2) {$\vdots$};
\node (space6) at (12,2.3) {};
\node (g9) at (12,3) [gauger,label={[xshift=-3.2]right:{\scriptsize $1$}}] {};
\node (g12) at (11,1) [gauge,label={[xshift=3.2]left:{\scriptsize $1$}}] {};
\draw[gline] (g2)--(space1);
\draw[gline] (space2)--(g3)--(g4)--(g5)--(g11)--(space3);
\draw[gline] (space4)--(g6);
\draw[gline] (g5)--(g8)--(space5);
\draw[gline] (space6)--(g9);
\draw[gline] (g4)--(g12);	
\end{tikzpicture}
\ee 
The second ruling from left in \eqref{TNruled} is associated to the $SU(N-2)$ gauge group. The magnetic quiver of this theory is obtained from the FI deformation indicated in the right-hand magnetic quiver of \eqref{MQTNLagrange}. 
By iterating this procedure of turning on FI parameters at the $U(1)$ nodes in the symmetric tails, after $N-2$ times, we arrive at the following magnetic quiver for the IR Lagrangian theory \eqref{TNLagrange}:
\be 
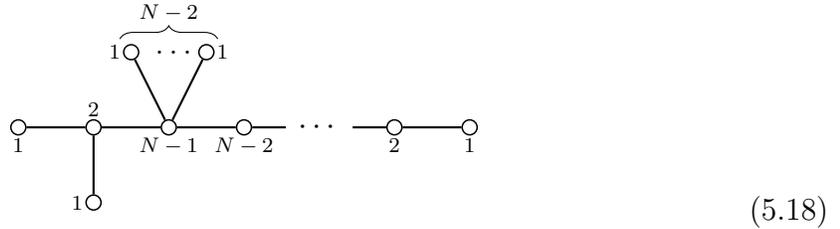
\begin{tikzpicture}
	\node (g1) at (0,0) [gauge,label={[yshift=3.2]below:{\scriptsize $1$}}] {};
	\node (g2) at (1,0) [gauge,label={[yshift=-3.2]above:{\scriptsize $2$}}] {};
	\node (g3) at (2,0) [gauge,label={[yshift=3.2]below:{\scriptsize $N-1$}}] {};
	\node (g4) at (3,0) [gauge,label={[yshift=3.2]below:{\scriptsize $N-2$}}] {};
	\node (space3) at (3.7,0) {};
	\node (dots) at (4,0) {$\cdots$};
	\node (space4) at (4.3,0) {};
	\node (g5) at (5,0) [gauge,label={[yshift=3.2]below:{\scriptsize $2$}}] {};
	\node (g6) at (6,0) [gauge,label={[yshift=3.2]below:{\scriptsize $1$}}] {};
	\node (g7) at (1,-1) [gauge,label={[xshift=3.2]left:{\scriptsize $1$}}] {};
	\node (g8) at (1.5,1) [gauge,label={[xshift=3.2]left:{\scriptsize $1$}}] {};
	\node (dots) at (2.1,1) {$\dots$};
	\node (g9) at (2.5,1) [gauge,label={[xshift=-3.2]right:{\scriptsize $1$}}] {};
	\draw[gline] (g1)--(g2)--(g3)--(g4)--(space3);
	\draw[gline] (space4)--(g5)--(g6);
	\draw[gline] (g3)--(g8);
	\draw[gline] (g3)--(g9);
	\draw[gline] (g2)--(g7);
	
	\draw [decorate,decoration={brace,amplitude=5},yshift=5]
(1.35,1)--(2.65,1) node [midway,yshift=10] {\scriptsize $N-2$};
	
\end{tikzpicture}
\ee 
with flavor symmetry algebra $\su(N) \oplus \su(2)^2$ for $N \neq 3$ and, for $N=3$, the balanced nodes form $\widetilde D_5$ .

In \cite{Eckhard:2020jyr} a new weakly coupled gauge theory description for $T_4$ was found, with three $SU(2)$ gauge groups with a trifundamental half-hypermultiplet in the representation $\half (\mathbf{2},\mathbf{2},\mathbf{2})$ among them, as well as 2 fundamentals associated to each gauge group. The corresponding partial resolution is
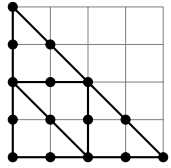
\be
\begin{tikzpicture}[x=.5cm,y=.5cm]
	\draw[step=.5cm,gray,very thin] (0,0) grid (4,4);
	
	\draw[ligne,black] (0,0)--(4,0)--(0,4)--(0,0);
	
	\foreach \x in {0,1,...,4}
	\node[bd] at (0,\x) {};
	\foreach \x in {1,2,...,4}
	\node[bd] at (\x,0) {};
	\foreach \x in {1,2,...,3}
	\node[bd] at (0+\x,4-\x) {};
	\node[bd] at (1,1) {};
	\node[bd] at (2,1) {};
	\node[bd] at (1,2) {};
	
	\draw[ligne,black] (2,0)--(2,2)--(0,2)--(2,0);
\end{tikzpicture}
\ee
This ruling structure is associated to three transitions in the magnetic quiver, namely
\be 
\begin{tikzpicture}
	\node (g1) at (0,0) [gauge,label={[yshift=3.2]below:{\scriptsize $1$}}] {};
	\node (g2) at (1,0) [gauger,label={[yshift=3.2]below:{\scriptsize $2$}}] {};
	\node (g3) at (2,0) [gauge,label={[yshift=3.2]below:{\scriptsize $3$}}] {};
	\node (g4) at (3,0) [gauge,label={[yshift=3.2]below:{\scriptsize $4$}}] {};
	\node (g5) at (4,0) [gauge,label={[yshift=3.2]below:{\scriptsize $3$}}] {};
	\node (g6) at (5,0) [gauge,label={[yshift=3.2]below:{\scriptsize $2$}}] {};
	\node (g7) at (6,0) [gauge,label={[yshift=3.2]below:{\scriptsize $1$}}] {};
	\node (g8) at (3,1) [gauge,label={[xshift=-3.2]right:{\scriptsize $3$}}] {};
	\node (g9) at (3,2) [gauger,label={[xshift=-3.2]right:{\scriptsize $2$}}] {};
	\node (g10) at (3,3) [gauge,label={[xshift=-3.2]right:{\scriptsize $1$}}] {};
	\draw[gline] (g1)--(g2)--(g3)--(g4)--(g5)--(g6)--(g7);
	\draw[gline] (g4)--(g8)--(g9)--(g10);
	
	\draw[->,thick] (7,0)--(8,0);
	
	\node (g1) at (9,0) [gauge,label={[yshift=3.2]below:{\scriptsize $1$}}] {};
	\node (g2) at (10,0) [gauge,label={[yshift=3.2]below:{\scriptsize $2$}}] {};
	\node (g3) at (11,0) [gauge,label={[yshift=3.2]below:{\scriptsize $3$}}] {};
	\node (g4) at (12,0) [gauger,label={[yshift=3.2]below:{\scriptsize $2$}}] {};
	\node (g5) at (13,0) [gauge,label={[yshift=3.2]below:{\scriptsize $1$}}] {};
	\node (g6) at (10,1) [gauge,label={[xshift=3.2]left:{\scriptsize $1$}}] {};
	\node (g7) at (11,1) [gauger,label={[xshift=3.2]left:{\scriptsize $2$}}] {};
	\node (g8) at (11,2) [gauge,label={[xshift=-3.2]right:{\scriptsize $1$}}] {};
	\node (g9) at (12,1) [gauge,label={[xshift=-3.2]right:{\scriptsize $1$}}] {};
	\draw[gline] (g1)--(g2)--(g3)--(g4)--(g5);
	\draw[gline] (g2)--(g6);
	\draw[gline] (g3)--(g7)--(g8);
	\draw[gline] (g7)--(g9);
	
	\draw[->,thick] (11,-1)--(11,-2);

	\node (g1) at (9.5,-4) [gauge,label={[yshift=3.2]below:{\scriptsize $1$}}] {};
	\node (g2) at (10.5,-4) [gauger,label={[xshift=-3.2, yshift=3.2]below right:{\scriptsize $2$}}] {};
	\node (g3) at (11.5,-4) [gauger,label={[xshift=3.2, yshift=3.2]below left:{\scriptsize $2$}}] {};
	\node (g4) at (12.5,-4) [gauge,label={[yshift=3.2]below:{\scriptsize $1$}}] {};
	\node (g5) at (10.5,-3) [gauge,label={[xshift=3.2]left:{\scriptsize $1$}}] {};
	\node (g6) at (10.5,-5) [gauge,label={[xshift=3.2]left:{\scriptsize $1$}}] {};
	\node (g7) at (11.5,-3) [gauge,label={[xshift=-3.2]right:{\scriptsize $1$}}] {};
	\node (g8) at (11.5,-5) [gauge,label={[xshift=-3.2]right:{\scriptsize $1$}}] {};
	\draw[gline] (g1)--(g2)--(g3)--(g4);
	\draw[gline] (g5)--(g2)--(g6);
	\draw[gline] (g7)--(g3)--(g8);	
	
	\draw[->,thick] (8,-4)--(7,-4);
	
	\node (g1) at (2,-4) [gauge,label={[xshift=3.2]left:{\scriptsize $1$}}] {};
	\node (g2) at (3,-4) [gauge,label={[yshift=2]below:{\scriptsize $2$}}] {};
	\node (g3) at (4,-4) [gauge,label={[xshift=-3.2]right:{\scriptsize $1$}}] {};
	\node (g4) at (2.5,-3.13) [gauge,label={[yshift=-3.2]above:{\scriptsize $1$}}] {};
	\node (g5) at (2.5,-4.87) [gauge,label={[yshift=3.2]below:{\scriptsize $1$}}] {};
	\node (g6) at (3.5,-3.13) [gauge,label={[yshift=-3.2]above:{\scriptsize $1$}}] {};
	\node (g7) at (3.5,-4.87) [gauge,label={[yshift=3.2]below:{\scriptsize $1$}}] {};
	\draw[gline] (g1)--(g2)--(g3);
	\draw[gline] (g4)--(g2)--(g5);
	\draw[gline] (g6)--(g2)--(g7);	
	
	\node at (3,-6) {{\scriptsize $\su (2)^6$}};	
\end{tikzpicture}
\ee

\subsection{New Lagrangians from FI Deformations}
\label{sec:newlag}

In this section we will see how the analysis of FI deformations can be helpful to guess new low-energy lagrangian descriptions for 5d SCFTs. We will focus on the case of $P_N$ theory, for which we find a new linear unitary quiver. The basic idea is to turn on FI deformations (in one-to-one correspondence with the number of simple factors in the gauge group) until we find the 3d mirror of a gauge theory. Notice that this argument alone does not allow us to determine the value of the CS levels or $\theta$ angles. 

We proceed inductively, starting from the magnetic quiver of $P_N$ theory given in \eqref{gntopn}. We first activate FI parameters at the $U(1)$ and $U(2)$ nodes on the right. 

\be\label{firstdef} 
\begin{tikzpicture}
	
	\node (g1) at (.5,2) [gauge,label={[yshift=3.2]below:{\scriptsize $1$}}] {};
	\node (space1) at (1.2,2) {};
	\node (dots) at (1.5,2) {$\cdots$};
	\node (space2) at (1.8,2) {};	
	\node (g2) at (2.5,2) [gauge,label={[yshift=3.2]below:{\scriptsize $N-1$}}] {};
	\node (g7) at (3.5,2) [gauge,label={[yshift=3.2]below:{\scriptsize $N$}}] {};
	\node (space3) at (4.2,2) {};
	\node (dots) at (4.5,2) {$\cdots$};
	\node (space4) at (4.8,2) {};	
	\node (g8) at (5.5,2) [gauge,label={[yshift=3.2]below:{\scriptsize $2N-2$}}] {};
	\node (g9) at (6.5,2) [gauge,label={[yshift=3.2]below:{\scriptsize $N$}}] {};
	\node (g10) at (7.5,2) [gauge,label={[yshift=3.2]below:{\scriptsize $N-1$}}] {};
	\node (space5) at (8.2,2) {};
	\node (dots) at (8.5,2) {$\cdots$};
	\node (space6) at (8.8,2) {};
	\node (g11) at (9.5,2) [gauger,label={[yshift=3.2]below:{\scriptsize $2$}}] {};
	\node (g12) at (10.5,2) [gauger,label={[yshift=3.2]below:{\scriptsize $1$}}] {};
	\node (g13) at (5.5,3) [gauge,label={[xshift=-3.2]right:{\scriptsize $N-1$}}] {};
	\draw[gline] (g1)--(space1);
	\draw[gline] (space2)--(g2)--(g7)--(space3);
	\draw[gline] (space4)--(g8)--(g9)--(g10)--(space5);
	\draw[gline] (space6)--(g11)--(g12);
	\draw[gline] (g8)--(g13);
	
   \draw[->,thick] (5.5,1)--(5.5,0);
  
	\node (h5) at (1.5,-2) [gauge,label={[yshift=3.2]below:{\scriptsize $1$}}] {};
	\node (h6) at (2.5,-2) [gauge,label={[yshift=3.2]below:{\scriptsize $2$}}] {};
	\node (space3) at (3.2,-2) {};
	\node (dots) at (3.5,-2) {$\cdots$};
	\node (space4) at (3.8,-2) {};
	\node (h7) at (4.5,-2) [gauge,label={[yshift=3.2]below:{\scriptsize $2N-4$}}] {};	
	\node (h8) at (5.5,-2) [gauge,label={[yshift=3.2]below:{\scriptsize $2N-4$}}] {};
	\node (h9) at (6.5,-2) [gauge,label={[yshift=3.2]below:{\scriptsize $2N-4$}}] {};
	\node (h10) at (7.5,-2) [gauge,label={[yshift=3.2]below:{\scriptsize $N-2$}}] {};
	\node (space5) at (8.2,-2) {};
	\node (dots) at (8.5,-2) {$\cdots$};
	\node (space6) at (8.8,-2) {};
	\node (h11) at (9.5,-2) [gauge,label={[yshift=3.2]below:{\scriptsize $2$}}] {};
	\node (h12) at (10.5,-2) [gauge,label={[yshift=3.2]below:{\scriptsize $1$}}] {};
	\node (h13) at (6.5,-1) [gauge,label={[xshift=-3.2]right:{\scriptsize $N-2$}}] {};
	\node (h14) at (4.5,-1) [gauge,label={[xshift=-3.2]left:{\scriptsize $1$}}] {};
	\draw[gline] (h5)--(h6)--(space3);
	\draw[gline] (space4)--(h7)--(h8)--(h9)--(h10)--(space5);
	\draw[gline] (space6)--(h11)--(h12);
	\draw[gline] (h9)--(h13);
	\draw[gline] (h7)--(h14); 
   
\end{tikzpicture}
\ee  
This deformation leads to the second quiver in \eqref{firstdef}, which describes (i.e. is the 3d mirror of) an $SU(2N-4)$ vector multiplet coupled to $2N-2$ flavors and to an SCFT whose 3d mirror is given by the quiver 
\be\label{newcftmirr}
\begin{tikzpicture} 
	\node (h5) at (1.5,-2) [gauge,label={[yshift=3.2]below:{\scriptsize $1$}}] {};
	\node (h6) at (2.5,-2) [gauge,label={[yshift=3.2]below:{\scriptsize $2$}}] {};
	\node (space3) at (3.2,-2) {};
	\node (dots) at (3.5,-2) {$\cdots$};
	\node (space4) at (3.8,-2) {};
	\node (h7) at (4.5,-2) [gauge,label={[yshift=3.2]below:{\scriptsize $2N-4$}}] {};	
	\node (h8) at (5.5,-2) [gauge,label={[yshift=3.2]below:{\scriptsize $N-2$}}] {};
	\node (space5) at (6.2,-2) {};
	\node (dots) at (6.5,-2) {$\cdots$};
	\node (space6) at (6.8,-2) {};
	\node (h9) at (7.5,-2) [gauge,label={[yshift=3.2]below:{\scriptsize $2$}}] {};
	\node (h10) at (8.5,-2) [gauge,label={[yshift=3.2]below:{\scriptsize $1$}}] {};
	\node (h11) at (4.5,-1) [gauge,label={[xshift=-3.2]right:{\scriptsize $N-2$}}] {};
	\draw[gline] (h5)--(h6)--(space3);
	\draw[gline] (space4)--(h7)--(h8)--(space5);
	\draw[gline] (space6)--(h9)--(h10);
	\draw[gline] (h7)--(h11); 
   
\end{tikzpicture}
\ee
Notice that the central node is underbalanced and we can use the Seiberg-like duality of \cite{Gaiotto:2008ak}: 
\be\label{dualseib} U(n)\; \text{w/ $2n-1$ flavors}\;\;\simeq\;\;  U(n-1)\;\text{w/ $2n-1$ flavors}+\text{1 free hyper}.\ee
With this modification both the $U(N-2)$ above and the $U(2N-3)$ on the left become underbalanced and we can apply the duality again. Once we get rid of all the underbalanced nodes with this procedure, we find $4N-8$ decoupled hypermultiplets, organized into 2 flavors of $SU(2N-4)$, and an interacting SCFT which can be identified with $P_{N-2}$. We therefore end up with the theory 
\be\label{firstdef2} P_{N-2}-SU(2N-4)-\boxed{2N}.\ee 
By iterating the above argument, we conclude that for $N$ odd ($N=2k+1$) after $k$ FI deformations we end up with the quiver 
\be\label{oddseclast}P_3-SU(6)-\dots-SU(2N-4)-\boxed{2N}.\ee 
Setting $N=3$ in (\ref{firstdef}) we recognize the magnetic quiver of the rank-1 $E_7$ theory, which flows to $SU(2)$ with 6 flavors and therefore the final answer is 
\be\label{oddfinal}SU(2)-SU(6)-\dots-SU(2N-4)-\boxed{2N}.\ee 
For $N$ even ($N=2k$) instead, after $k-1$ FI deformations we get 
\be\label{evenseclast}P_4-SU(8)-\dots-SU(2N-4)-\boxed{2N}.\ee 
Let us now focus on the FI deformation of $P_4$. For $N=4$ we recognize the second quiver in \eqref{firstdef} as the 3d mirror of $SU(4)$ with 8 fundamentals and 1 anti-symmetric hypermultiplet. Overall, the theory after $k$ FI deformations becomes 
\be\label{evenfinal}\boxed{1\;AS}-SU(4)-SU(8)-\dots-SU(2N-4)-\boxed{2N}.\ee 

We can substantiate that the new Lagrangians in \eqref{oddfinal} and \eqref{evenfinal} indeed UV complete to $P_N$ by constructing pre-GTPs realizing the gauge theories, and showing that they can be put into the form \eqref{GTPPN}. The pre-GTP of the quiver gauge theory in \eqref{oddfinal} with (CS levels 0) is shown to the left
\be
\begin{tikzpicture}[x=.5cm,y=.5cm]
	\draw[step=.5cm,gray,very thin] (0,-10) grid (4,4);
	
	\draw[ligne] (0,0)--(1,-1)--(2.25,-4.75);
	\draw[ligne] (2.75,-6.25)--(4,-10)--(4,-4);
	\draw[ligne] (4,-2)--(4,4)--(2.5,2.5);
	\draw[ligne] (1.5,1.5)--(0,0);	
	
	\draw (1,-1)--(1,1);
	\draw (3,-7)--(3,3);
	
	\foreach \x in {0,1,2,3}
	\node[bd] at (1+\x,-1-3*\x) {};
	\foreach \x in {1,2,3,4,5,9,10,11,12,13,14}
	\node[bd] at (4,-10+\x) {};
	\foreach \x in {0,1,3}
	\node[bd] at (\x,\x) {};
	
	\node[] at (5.2,-3) {{\scriptsize $2N$}};	
	\node[] at (-1.6,2) {{\scriptsize $\frac{N+1}{2}$}};
	\node[] at (2,-11) {{\scriptsize $\frac{N+1}{2}$}};	
	\node[] at (-1.6,-5.5) {{\scriptsize $\frac{3N-1}{2}$}};
	
	\draw[|-|] (4.5,-10)--(4.5,4);
	\draw[|-] (-.5,4)--(-.5,0);
	\draw[|-|] (-.5,0)--(-.5,-10);
	\draw[|-|] (0,-10.5)--(4,-10.5);
	
	\node[] at (4,-2.8) {$\vdots$};
	\node[] at (2.2,0) {$\dots$};
	\node[rotate=-45] at (2.2,2.2) {$\vdots$};
	\node[rotate=18.5] at (2.4,-5.2) {$\vdots$};
	
	\node at (-1.1,0) {{\scriptsize $\mathbf{v}_0$}};
	
	\draw[->,thick] (6.5,-3)--(8.5,-3) node[midway, above] {\scriptsize $\mathfrak{T}_\text{odd}$};
	
	\draw[step=.5cm,gray,very thin] (11,-10) grid (18,4);
	\draw[ligne] (18,4)--(15.5,1.5);
	\draw[ligne] (13.5,-.5)--(11,-3)--(13.5,-5.5);
	\draw[ligne] (15.5,-7.5)--(18,-10)--(18,-4);
	\draw[ligne] (18,-2)--(18,4);
	
	\foreach \x in {0,1,...,5,9,10,...,14}
	\node[bd] at (18,\x-10) {};
	\foreach \x in {0,1,2,5,6,7}
	\node[bd] at (11+\x,\x-3) {};
	\foreach \x in {1,2,5,6}
	\node[wd] at (11+\x,-\x-3) {};
	
	\node[] at (19.2,-3) {{\scriptsize $2N$}};	
	\node[] at (9.6,.5) {{\scriptsize $N$}};
	\node[] at (9.6,-6.5) {{\scriptsize $N$}};	
	
	\draw[|-|] (18.5,-10)--(18.5,4);
	\draw[|-] (10.5,4)--(10.5,-3);
	\draw[|-|] (10.5,-3)--(10.5,-10);
	
	\node[] at (18,-2.8) {$\vdots$};
	\node[rotate=-45] at (14.6,.6) {$\vdots$};
	\node[rotate=45] at (14.4,-6.4) {$\vdots$};
\end{tikzpicture}
\ee
where the $SU(2)$ gauge node sits along the left-most ruling with no matter multiplets. The ranks of the gauge groups increase in steps of 4 to the right, until $SU(2N-4)$ with its $2N$ hypers sitting along the right-hand vertical edge. The $(1,1)$ and $(1,-3)$ edges furnish the bi-fundamental matter.
In the notation of \cite{vanBeest:2020kou}, applying the edge-moves given by
\be
\mathfrak{T}_\text{odd}=\mathfrak{M}^-_1\mathfrak{M}^-_2 \dots \mathfrak{M}^-_{\frac{N-1}{2}}\,,
\ee 
renders the polygon convex, giving exactly the GTP of $P_N$ in \eqref{GTPPN} (left).
The pre-GTP of the gauge theory with anti-symmetric matter in \eqref{evenfinal} (and CS levels 0) is shown to the left
\be
\begin{tikzpicture}[x=.5cm,y=.5cm]
	\draw[step=.5cm,gray,very thin] (0,-11) grid (5,5);
	
	\draw[ligne] (0,0)--(2,-2)--(3.25,-5.75);
	\draw[ligne] (3.75,-7.25)--(5,-11)--(5,-4);
	\draw[ligne] (5,-2)--(5,5)--(3.5,3.5);
	\draw[ligne] (2.5,2.5)--(0,0);	
	
	\draw (2,-2)--(2,2);
	\draw (4,-8)--(4,4);
	
	\foreach \x in {0,1,2,3}
	\node[bd] at (2+\x,-2-3*\x) {};
	\foreach \x in {0,1,...,5,9,10,...,15}
	\node[bd] at (5,-10+\x) {};
	\foreach \x in {0,1,2,4}
	\node[bd] at (\x,\x) {};
	\node[wd] at (1,-1) {};
	
	\node[] at (6.2,-3) {{\scriptsize $2N$}};	
	\node[] at (-1.6,3.5) {{\scriptsize $\frac{N}{2}+1$}};
	\node[] at (2.5,-12) {{\scriptsize $\frac{N}{2}+1$}};	
	\node[] at (-1.6,-6.5) {{\scriptsize $\frac{3N}{2}-1$}};
	
	\draw[|-|] (5.5,-11)--(5.5,5);
	\draw[|-] (-.5,5)--(-.5,0);
	\draw[|-|] (-.5,0)--(-.5,-11);
	\draw[|-|] (0,-11.5)--(5,-11.5);
	
	\node[] at (5,-2.8) {$\vdots$};
	\node[] at (3.2,0) {$\dots$};
	\node[rotate=-45] at (3.2,3.2) {$\vdots$};
	\node[rotate=18.5] at (3.4,-6.2) {$\vdots$};
	
	\node at (-1.1,0) {{\scriptsize $\mathbf{v}_0$}};
	
	\draw[->,thick] (7.5,-3)--(9.5,-3) node[midway, above] {\scriptsize $\mathfrak{T}_\text{even}$};
	
	\draw[step=.5cm,gray,very thin] (12,-11) grid (20,5);
	\draw[ligne] (20,5)--(16.5,1.5);
	\draw[ligne] (14.5,-.5)--(12,-3)--(14.5,-5.5);
	\draw[ligne] (16.5,-7.5)--(20,-11)--(20,-4);
	\draw[ligne] (20,-2)--(20,5);
	
	\foreach \x in {-1,0,...,5,9,10,...,15}
	\node[bd] at (20,\x-10) {};
	\foreach \x in {0,1,2,5,6,7}
	\node[bd] at (12+\x,\x-3) {};
	\foreach \x in {1,2,5,6,7}
	\node[wd] at (12+\x,-\x-3) {};
	
	\node[] at (21.2,-3) {{\scriptsize $2N$}};	
	\node[] at (10.6,.5) {{\scriptsize $N$}};
	\node[] at (10.6,-6.5) {{\scriptsize $N$}};	
	
	\draw[|-|] (20.5,-11)--(20.5,5);
	\draw[|-] (11.5,5)--(11.5,-3);
	\draw[|-|] (11.5,-3)--(11.5,-11);
	
	\node[] at (20,-2.8) {$\vdots$};
	\node[rotate=-45] at (15.6,.6) {$\vdots$};
	\node[rotate=45] at (15.4,-6.4) {$\vdots$};
\end{tikzpicture}
\ee
where the left-hand sector with the white vertex supplies the anti-symmetric matter, and the $SU(4)$ sits along the left-most ruling. Again, the ranks of the gauge groups increase in steps of 4 to the right until $SU(2N-4)$, which has $2N$ hypers sitting on the vertical edge. Applying the following edge-moves,
\be
\mathfrak{T}_\text{even}=\mathfrak{M}^-_1\mathfrak{M}^-_2 \dots \mathfrak{M}^-_{\frac{N}{2}-1}\,,
\ee 
gives the GTP of $P_N$.

\subsubsection*{Acknowledgements}
We thank Antoine Bourget, Stefano Cremonesi, Julius Grimminger, Amihay Hanany and Sakura Sch\"afer-Nameki for discussions. The work of SG is supported in part by the ERC Consolidator Grant number 682608 ``Higgs bundles: Supersymmetric Gauge Theories and Geometry (HIGGSBNDL)" and partly by the INFN grant ``Per attività di formazione per sostenere progetti di ricerca'' (GRANT 73/STRONGQFT).
The work of MvB is in part supported by the ERC Consolidator Grant number 682608 ``Higgs bundles: Supersymmetric Gauge Theories and Geometry (HIGGSBNDL)".


\bibliographystyle{ytphys}
\bibliography{ref}

\end{document}